\documentclass[aps,prl,amssymb,groupedaddress,nofootinbib]{revtex4}

\usepackage[english]{babel}
\usepackage{amsfonts}
\usepackage{amssymb}
\usepackage{epsfig}
\usepackage{amsmath}
\usepackage{fancyhdr}
\usepackage{textcomp}
\usepackage{setspace}

\begin{document}

\title{General CMB and Primordial Trispectrum Estimation}

\author{D.M.~Regan}

\author{E.P.S.~Shellard}

\author{J.R.~Fergusson}

\affiliation{Centre for Theoretical Cosmology,\\
Department of Applied Mathematics and Theoretical Physics,\\
University of Cambridge,
Wilberforce Road, Cambridge CB3 0WA, United Kingdom}

\date{\today}

\pacs{1}

\begin{abstract}
In this paper we present trispectrum estimation methods which can be applied
to general non-separable primordial and CMB trispectra.  We review the relationship
between the reduced CMB trispectrum and  the reduced primordial trispectrum.
We present a general optimal  estimator for the connected part of the trispectrum,
for which we derive a quadratic term to incorporate the effects of inhomogeneous
noise and  masking.  We describe a general algorithm for creating simulated maps
with given arbitrary (and independent) power spectra, bispectra and trispectra.
We propose a universal definition of the trispectrum parameter $T_{NL}$,
 so that the integrated trispectrum on the observational domain can be consistently
 compared between theoretical models.   We define a shape function for the
 primordial trispectrum, together with a shape correlator and a useful parametrisation
 for visualizing the trispectrum; these methods might also be applied to the late-time trispectrum for large scale structure. We derive separable analytic CMB solutions in the
 large-angle limit for constant and local models.
 We present separable mode decompositions which can be used to describe any primordial
 or CMB trispectra on their respective wavenumber or multipole domains.   By extracting
 coefficients of these separable basis functions from an observational map, we are able
 to present an efficient estimator for any given theoretical model with a nonseparable
 trispectrum.   The estimator  has two manifestations,  comparing the theoretical and
 observed coefficients at either primordial or late times, thus encompassing a wider
 range of models, such as secondary anisotropies, lensing and cosmic strings.
 We show that these mode decomposition methods are numerically tractable with
 order $l^5$ operations for the CMB estimator and approximately order $l^6$ for the
 general primordial estimator (reducing to order $l^3$ in both cases for a special class of models). We also demonstrate how the trispectrum can be reconstructed
 from observational maps using these methods.
\end{abstract}
\maketitle

\section{I. Introduction}
Single field slow-roll inflationary fluctuations in the standard picture of cosmology predict a nearly scale invariant spectrum of adiabatic perturbations with a nearly Gaussian distribution. Hence it can be described very accurately by its angular power spectrum. These predictions agree well with measurements of the cosmic microwave background (CMB) and large scale structure, such as those provided by WMAP and SDSS. However, it remains possible that there exists a mechanism for generating large non-Gaussianities in the early Universe. Measurements of such non-Gaussianities open up the opportunity of investigating the physics of the early universe including different inflationary models and competing alternative scenarios. In order to study such observations, higher order correlators, beyond the two-point function, offer possibly the best prospects. General methods for comparing the three point correlator, dubbed the bispectrum, were developed in \cite{Ferg1,Ferg2,Ferg3}. In those papers an integrated measure of the bispectrum was defined, as well as a set of formalisms for comparing, evolving and constraining the bispectrum in the case of both the primordial and CMB three-point correlators. In this paper we will generalise many of these methods to the four-point correlator which is denoted the trispectrum. We will emphasise the application of these methods to the primordial and CMB trispectra. The primary motivation for this paper is to develop formalisms to bring observations to bear on this broader class of cosmological models. We will demonstrate that despite the complexity of trispectrum estimation, these methods are numerically tractable given 
present resources, even at Planck satellite resolution.  
\par
In order to get large non-Gaussianity we must move away from the standard single field slow-roll inflation~\cite{Chen3}. Multifield inflation allows the possibility for superhorizon evolution. Non-Gaussianities are generated when this evolution is nonlinear. We can consider superhorizon behaviour as occurring in patches separated by horizons which evolve independently of each other. This locality in position space translates to non-locality in momentum space and indicates that for such models we expect the signal to peak for $k_4\ll k_1,k_2,k_3$. This forms the so-called local model. Such models have been investigated in the context of the trispectrum in \cite{Seery1,Seery2,Seery3,Adshead,Bartolo,Valen,Rodriguez,Huang1,Lehners}. Since subhorizon modes oscillate and so average out, the only chance to have large non-Gaussianity in single field inflationary models is when all modes have similar wavelengths and exit at the same time. A non-standard kinetic term allows for such a possibility. Since the signal peaks when the modes have similar wavelengths this class of forms are known as equilateral models and have been investigated using the trispectrum in \cite{aChen,bArroja,cChen4,dArroja2,eArroja3,fSenatZal,gGao}. It should be noted that this amplification of nonlinear effects around the time the modes exit the horizon is not possible for slow-roll single field inflation. It has also been shown in \cite{Huang2,Izumi} that a large trispectrum may be generated in the ghost inflation model. These models are so-called as they are based on the idea of a ghost condensate, i.e. a kind of fluid with equation of state $p=-\rho$, that can fill the universe, and which provides an alternative method of realising de Sitter phases in the early universe. Of course there are other methods to generate non-Gaussianity such as having sharp features in the potential or a non-Bunch-Davies vacuum. Also there are models which have features that resemble the aforementioned forms in different regimes, e.g. quasi-single field inflation~\cite{Chen2}, or have mixed contributions, e.g. in multifield DBI inflation~\cite{RenauxDBI}.
\par
One of the motivations for studying the four-point correlator is that it may be possible that the bispectrum is suppressed but still have a large trispectrum. In particular, this behaviour may be realised in quasi-single field inflation~\cite{Chen2} or in the curvaton model~\cite{Sasaki}. It also occurs in the case of cosmic strings where the bispectrum is suppressed by symmetry considerations \cite{Hindmarsh2009,Regan}. The effects of non-Gaussianity  could also be detectable in a wide range of astrophysical measurements, such as cluster abundances and the large scale clustering of highly biased tracers. In \cite{Sefusatti} the possibility of using the galaxy bispectrum to constrain the local form of the trispectrum has been reviewed. 
\par
The trispectrum, $T(k_1,k_2,k_3,k_4)$, is generally parametrised using the variable $\tau_{NL}$ which schematically is given by the ratio $\tau_{NL}\approx T(k,k,k,k)/P(k)^3$. Standard slow-roll inflation predicts $\tau_{NL}\lesssim r/50$ where $r<1$ is the tensor to scalar ratio~\cite{Seery1}. Such a low signal would be undetectable since it is below the level of non-Gaussian contamination that would be expected from secondary anisotropies $\tau_{NL}\approx \mathcal{O}(1)$. Using the analysis of N-point probability distribution of the CMB anisotropies~\cite{vielva}, where a local non-linear perturbative model $\Phi=\Phi_L+f_{NL}(\Phi^2_L-\langle \Phi_L^2\rangle)+g_{NL}\Phi_L^3+\mathcal{O}(\Phi_L^4)$ is used to characterise the large scale anistropies, the constraint $-5.6\times 10^5<g_{NL}<6.4\times 10^5$ was obtained\footnote{It should be noted that for single field local inflation $\tau_{NL}^{\rm{loc}}=\left(\frac{5}{6}f_{NL}\right)^2$. Since $f_{NL}$ is constrained by the bispectrum, $g_{NL}$ is the quantity that is constrained by the trispectrum directly in this case.}. For the more general case, there is only a weak experimental bound imposed on non-Gaussianity by the trispectrum, which is roughly $|\tau_{NL}|\lesssim 10^8$~\cite{Lyth}. In~\cite{Cooray,Cooray2} an improved constraint on $\tau_{NL}$ was presented using estimators to allow a joint fit of $f_{NL}$ and $g_{NL}$ using the trispectrum of WMAP5 data. However, the analysis therein included an incomplete formula for the CMB trispectrum due to local non-Gaussianity\footnote{The formula for the reduced local CMB trispectrum has been used in place of the full local CMB trispectrum,
which appears to simplify the analysis.}. Nonetheless, the approach indicates that vast improvements to trispectrum constraints should be achievable in the near future. In fact, it is expected that the Planck satellite will be sensitive to a value of $|\tau_{NL}|\sim 560$~\cite{Kogo}.
\par
The analysis of the trispectrum is a computationally intensive operation. In fact only the trispectrum induced by the local shape has been constrained so far by CMB data. The local form is an example of a separable shape - a notion which we will define more concretely in this paper. Essentially, since the primordial trispectrum is a six dimensional quantity, separability means the trispectrum is the product of one dimensional functions of each of these variables. Exploiting this separability reduces the problem from one of $\mathcal{O}(l_{\rm{max}}^7)$ operations to a more manageable $\mathcal{O}(l_{\rm{max}}^5)$. In special cases we get a further reduction to $\mathcal{O}(l_{\rm{max}}^3)$. 
\par
In the next section we shall describe the CMB trispectrum and its relation to the primordial equivalent. We will make use of a particular parametrisation of the reduced primordial trispectrum and exploit a Legendre series expansion in terms of one of these parameters to write an expression for the reduced CMB trispectrum which is valid in general. We will also outline a general correlation method for comparing different trispectra. In this section we will also give a formula for the kurtosis in terms of the multipoles. In section III we define a shape function which is a scale invariant form of the trispectrum. Using this function we define a shape correlator that is expected to predict closely the correlation between the respective trispectra. Next, we show how to decompose this shape in order to provide a method for visualising trispectra. We apply this visualisation to the case of the local and equilateral models which we describe in section IV. We also present the Sachs Wolfe limit ($l<100$) for the local and constant models. In section V we describe how to form a mode expansion for general non-separable shapes. This provides a rigorous method to find a separable approximation to any shape and therefore makes analysis of the trispectrum far more tractable. This expansion can be performed for both the primordial and CMB trispectra. Of immediate relevance in terms of Planck is to find a general measure for the size of the trisectrum. This is addressed in section VI in both the primordial and CMB cases. It is clearly desirable to be able to reconstruct the underlying trispectrum given the data. As we shall describe in section VII this is a computationally intensive task, but it is tractable. We will observe here that there is a degeneracy in reconstruction of the primordial trispectrum, implying that only the zeroth Legendre mode is recoverable. Finally in section VIII we outline a method for performing CMB map simulations for given general bispectra and trispectra.

\section{II. The CMB Trispectrum}
\subsection{Definition of the primordial and CMB trispectra}
\begin{figure}[htp]
\centering 
\includegraphics[width=102mm]{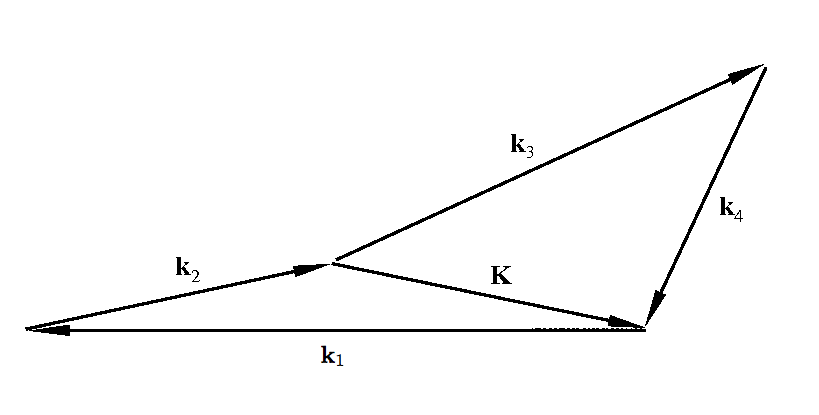}
\caption{Quadrilateral defined by the four wave-vectors ${\bf k}_i$.  The diagonal is represented
for ${\bf K}$.}
\label{fig:Tquad}
\end{figure}

We are concerned with the analysis of the four-point function induced by a non-Gaussian primordial gravitational potential $\Phi(\mathbf{k})$ in the CMB temperature fluctuation field. The temperature anisotropies
may be represented using the $a_{lm}$ coefficients of a spherical harmonic decomposition of the cosmic microwave sky,
\begin{eqnarray}\label{DeltaT}
\frac{\Delta T}{T}(\hat{\mathbf{n}})=\sum_{l,m} a_{lm}Y_{lm}(\hat{\mathbf{n}}).
\end{eqnarray}
The primordial potential $\Phi$ induces the multipoles $a_{lm}$ via a convolution with the transfer functions $\Delta_l(k)$ through the relation
\begin{eqnarray}
a_{lm}=4\pi (-i)^l \int \frac{d^3k}{(2\pi)^3} \Delta_l(k) \Phi(\mathbf{k}) Y_{lm}(\hat{\mathbf{k}}).
\end{eqnarray}
The connected part of the four point correlator of the $a_{lm}$ gives us the trispectrum. In particular, 
\begin{eqnarray}\label{Tconnls}
T_{l_1 m_1 l_2 m_2 l_3 m_3 l_4 m_4}&=&\langle a_{l_1 m_1}	 a_{l_2 m_2} a_{l_3 m_3} a_{l_4 m_4}	\rangle_c \nonumber\\
&=& (4\pi)^4 (-i)^{\sum_i l_i}\int  \frac{d^3 k_1 d^3 k_2 d^3 k_3 d^3 k_4}{(2\pi)^{12}}\Delta_{l_1}(k_1)\Delta_{l_2}(k_2)\Delta_{l_3}(k_3)\Delta_{l_4}(k_4)\times \nonumber\\
&&\langle \Phi(\mathbf{k_1})\Phi(\mathbf{k_2})\Phi(\mathbf{k_3})\Phi(\mathbf{k_4}) \rangle_c Y_{l_1 m_1}(\hat{\mathbf{k_1}}) Y_{l_2 m_2}(\hat{\mathbf{k_2}}) Y_{l_3 m_3}(\hat{\mathbf{k_3}}) Y_{l_4 m_4}(\hat{\mathbf{k_4}}),
\end{eqnarray}
where $k_i=|\mathbf{k}_i|$ and the subscript $c$ is used to denote the connected component. Naively, we would define the primordial trispectrum as 
\begin{eqnarray*}
\langle \Phi(\mathbf{k_1})\Phi(\mathbf{k_2})\Phi(\mathbf{k_3})\Phi(\mathbf{k_4}) \rangle_c=(2\pi)^3 \delta(\mathbf{k_1+k_2+k_3+k_4}) T'_{\Phi}(\mathbf{k}_1,\mathbf{k}_2,\mathbf{k}_3,\mathbf{k}_4).
\end{eqnarray*}
Here, the four wave-vectors form a quadrilateral as shown in Figure~\ref{fig:Tquad}.  
However, a more useful definition is to write
\begin{eqnarray}\label{Tconn}
\langle \Phi(\mathbf{k_1})\Phi(\mathbf{k_2})\Phi(\mathbf{k_3})\Phi(\mathbf{k_4}) \rangle_c=(2\pi)^3 \int d^3 K \delta (\mathbf{k_1+k_2+K})\delta (\mathbf{k_3+k_4-K})T_{\Phi}(\mathbf{k}_1,\mathbf{k}_2,\mathbf{k}_3,\mathbf{k}_4;\mathbf{K}).
\end{eqnarray}
Here the delta function indicates that the diagonal $\mathbf{K}$ makes triangles with $(\mathbf{k_1, k_2})$ and $(\mathbf{k_3, k_4})$, respectively. Of course there are symmetries implicit in this definition of $T_{\Phi}$ - namely, that we may form triangles with different combinations of the vectors. In particular,
\begin{eqnarray}
T_{\Phi}(\mathbf{k}_1,\mathbf{k}_2,\mathbf{k}_3,\mathbf{k}_4;\mathbf{K})=&&P_{\Phi}(\mathbf{k}_1,\mathbf{k}_2,\mathbf{k}_3,\mathbf{k}_4;\mathbf{K})+\int d^3 K' [\delta (\mathbf{k_3-k_2-K+K'})P_{\Phi}(\mathbf{k}_1,\mathbf{k}_3,\mathbf{k}_2,\mathbf{k}_4;\mathbf{K'})\nonumber \\
&&+\delta (\mathbf{k_4-k_2-K+K'})P_{\Phi}(\mathbf{k}_1,\mathbf{k}_4,\mathbf{k}_3,\mathbf{k}_2;\mathbf{K'}) ]
\end{eqnarray}
where $P_{\Phi}$ are constructed using a reduced trispectrum $\mathcal{T}_{\Phi}$ via
\begin{eqnarray}
P_{\Phi}(\mathbf{k}_1,\mathbf{k}_2,\mathbf{k}_3,\mathbf{k}_4;\mathbf{K})&=&\mathcal{T}_{\Phi}(\mathbf{k}_1,\mathbf{k}_2,\mathbf{k}_3,\mathbf{k}_4;\mathbf{K})+\mathcal{T}_{\Phi}(\mathbf{k}_2,\mathbf{k}_1,\mathbf{k}_3,\mathbf{k}_4;\mathbf{K})\nonumber\\
&+&\mathcal{T}_{\Phi}(\mathbf{k}_1,\mathbf{k}_2,\mathbf{k}_4,\mathbf{k}_3;\mathbf{K})+\mathcal{T}_{\Phi}(\mathbf{k}_2,\mathbf{k}_1,\mathbf{k}_4,\mathbf{k}_3;\mathbf{K}).
\end{eqnarray}
Therefore, we need only consider the reduced trispectrum $\mathcal{T}$ from one particular arrangement of the vectors and form the other contributions by permuting the symbols. 

The CMB trispectrum may also be written in a rotationally invariant way as
\begin{eqnarray}\label{Ttot}
T_{l_1 m_1 l_2 m_2 l_3 m_3 l_4 m_4}=\sum_{LM} (-1)^M \left( \begin{array}{ccc}
l_1 & l_2 & L \\
m_1 & m_2 & -M \end{array} \right) \left( \begin{array}{ccc}
l_3 & l_4 & L \\
m_3 & m_4 & M \end{array} \right) T^{l_1 l_2}_{l_3 l_4}(L).
\end{eqnarray}
The Wigner $3j$ symbols impose the triangle conditions on the multipole combinations $(l_1, l_2 ,L)$ and $(l_3,l_4 ,L)$.
As in the case of the primordial trispectrum, there are implicition symmertries in this definition. In a similar manner to the primordial case we can write
\begin{eqnarray}\label{Prel1}
T_{l_1 m_1 l_2 m_2 l_3 m_3 l_4 m_4}=\sum_{LM} (-1)^M \left( \begin{array}{ccc}
l_1 & l_2 & L \\
m_1 & m_2 & -M \end{array} \right) \left( \begin{array}{ccc}
l_3 & l_4 & L \\
m_3 & m_4 & M \end{array} \right) P^{l_1 l_2}_{l_3 l_4}(L) + (l_2\leftrightarrow l_3)+ (l_2\leftrightarrow l_4),
\end{eqnarray}
with
\begin{eqnarray}\label{Prel2}
P^{l_1 l_2}_{l_3 l_4}(L) =\mathcal{T}^{l_1 l_2}_{l_3 l_4}(L)+(-1)^{l_1+l_2+L} \mathcal{T}^{l_2 l_1}_{l_3 l_4}(L)+(-1)^{l_3+l_4+L} \mathcal{T}^{l_1 l_2}_{l_4 l_3}(L)+(-1)^{l_1+l_2+l_3+l_4} \mathcal{T}^{l_2 l_1}_{l_4 l_3}(L).
\end{eqnarray}
where the factors of powers of $(-1)$ are induced by identities of the Wigner $3j$ symbol. Therefore, we again need only consider the reduced trispectrum $\mathcal{T}$ from one particular arrangement of the multipoles. Indeed we need only find the reduced CMB trispectrum induced by the reduced primordial trispectrum. In particular, we denote
\begin{eqnarray}\label{RedTrisp}
\mathcal{T}_{l_1 m_1 l_2 m_2 l_3 m_3 l_4 m_4}=\sum_{LM} (-1)^M \left( \begin{array}{ccc}
l_1 & l_2 & L \\
m_1 & m_2 & -M \end{array} \right) \left( \begin{array}{ccc}
l_3 & l_4 & L \\
m_3 & m_4 & M \end{array} \right) \mathcal{T}^{l_1 l_2}_{l_3 l_4}(L),
\end{eqnarray}
and observe that
\begin{eqnarray}\label{TotalRedTrisp}
T_{l_1 m_1 l_2 m_2 l_3 m_3 l_4 m_4}&=&\mathcal{T}_{l_1 m_1 l_2 m_2 l_3 m_3 l_4 m_4}+\mathcal{T}_{l_2 m_2 l_1 m_1 l_3 m_3 l_4 m_4}+\mathcal{T}_{l_1 m_1 l_2 m_2 l_4 m_4 l_3 m_3 }+\mathcal{T}_{ l_2 m_2 l_1 m_1 l_4 m_4 l_3 m_3 }\nonumber\\
&+&\mathcal{T}_{l_1 m_1 l_3 m_3  l_2 m_2 l_4 m_4}+\mathcal{T}_{ l_3 m_3 l_1 m_1 l_2 m_2 l_4 m_4}+\mathcal{T}_{l_1 m_1  l_3 m_3 l_4 m_4 l_2 m_2}+\mathcal{T}_{l_3 m_3 l_1 m_1   l_4 m_4 l_2 m_2}\nonumber\\
&+&\mathcal{T}_{l_1 m_1  l_4 m_4 l_2 m_2 l_3 m_3}+\mathcal{T}_{ l_4 m_4 l_1 m_1 l_2 m_2 l_3 m_3}+\mathcal{T}_{l_1 m_1  l_4 m_4  l_3 m_3 l_2 m_2}+\mathcal{T}_{ l_4 m_4 l_1 m_1 l_3 m_3  l_2 m_2}.
\end{eqnarray}

\subsection{Relation between the primordial and CMB trispectra}
In order to relate the above definitions for the  primordial and CMB trispectra we use the following identities
\begin{eqnarray}\label{Identities}
\delta(\mathbf{k})&=&\frac{1}{(2\pi)^3}\int e^{i \mathbf{r.k}}d^3 r,\nonumber\\
e^{i \mathbf{r.k}}&=& 4\pi \sum_{l,m} i^l j_l(k r) Y_{l m}(\hat{\mathbf{k}}) Y_{lm}^*(\hat{\mathbf{r}}),\nonumber\\
Y_{l -m}&=&(-1)^m Y_{lm}^*.
\end{eqnarray}
We find using these identities with equations~\eqref{Tconnls} and~\eqref{Tconn}
\begin{eqnarray}
\mathcal{T}_{l_1 m_1 l_2 m_2 l_3 m_3 l_4 m_4}&=&\left(\frac{2}{\pi}\right)^5 (-i)^{\sum l_i}\int \left(\Pi_{i=1}^4d^3 k_i \Delta_{l_i}(k_i)  Y_{l_i m_i}(\hat{\mathbf{k}}_i) \right)d^3 K  \mathcal{T}_{\Phi}(\mathbf{k}_1,\mathbf{k}_2,\mathbf{k}_3,\mathbf{k}_4;\mathbf{K})\nonumber \\
&&\times \sum_{l_i',L' ,L''} \sum_{m_i',M',M''}\int d^3 r_1 d^3 r_2 i^{\sum_{i'} l_i' +L'-L''} [j_{l_1'}(k_1 r_1) Y_{l_1' m_1'}(\hat{\mathbf{k}}_1) Y^*_{l_1' m_1'}(\hat{\mathbf{r}}_1)  ] \nonumber \\
&&\times [j_{l_2'}(k_2 r_1) Y_{l_2' m_2'}(\hat{\mathbf{k}}_2) Y^*_{l_2' m_2'}(\hat{\mathbf{r}}_1)  ] [j_{l_3'}(k_3 r_2) Y_{l_3' m_3'}(\hat{\mathbf{k}}_2) Y^*_{l_3' m_3'}(\hat{\mathbf{r}}_2)  ] [j_{l_4'}(k_4 r_2) Y_{l_4' m_4'}(\hat{\mathbf{k}}_4) Y^*_{l_4' m_4'}(\hat{\mathbf{r}}_2)  ] \nonumber\\
&&\times [j_{L'}(K r_1) Y_{L' M'}(\hat{\mathbf{K}}) Y^*_{L' M'}(\hat{\mathbf{r}}_1)  ] [j_{L''}(K r_2) Y^*_{L'' M''}(\hat{\mathbf{K}}) Y_{L'' M''}(\hat{\mathbf{r}}_2)  ]. 
\end{eqnarray}
where $\hat{\mathbf{k}}_i$ represents the unit vector in the direction $\mathbf{k}_i$.
\begin{figure}[htp]
\centering 
\includegraphics[width=102mm]{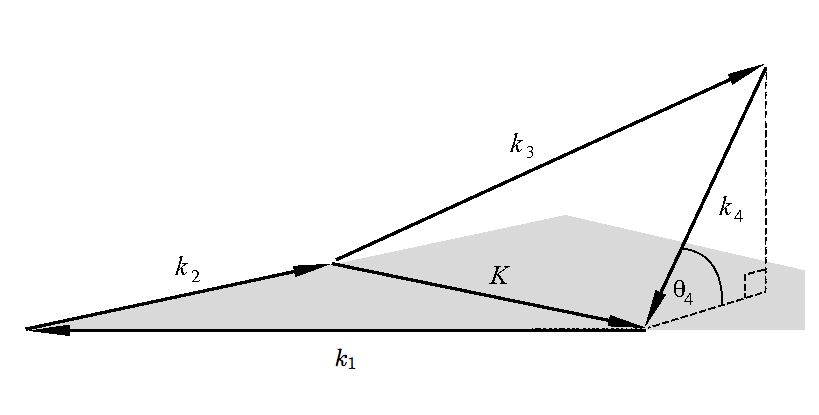}
\caption{Quadrilateral defined by the four wavenumbers $ k_i$, the diagonal $K$, and the 
angle $\theta_4$ out of the plane of the first triangle.}
\label{fig:Tquad2}
\end{figure}

To calculate further, we must choose an appropriate parametrisation for $\mathcal{T}_{\Phi}$. 
We note that the primordial trispectrum shape has $6$ degrees of freedom. We could define the quadrilateral uniquely by the lengths of the four sides $k_i = |{\bf k}_i|$,
together with the two diagonals $K = |{\bf K}|$ and $\tilde K = |\tilde{\bf K}|$.  However, we find it 
more convenient to represent the sixth degree of freedom with the angle $\theta_4$ which represents
the deviation of the quadrilateral from planarity (as illustrated in Figure~\ref{fig:Tquad2}).  Many well-motivated primordial models, such as
the local and equilateral cases we shall discuss, are planar (i.e. $\theta_4=0$). So we choose the independent parameters to identify the shape to be $(k_1,k_2, k_3,k_4, K, \theta_4)$, that is, 
 $\mathcal{T}_{\Phi}=\mathcal{T}_{\Phi}(k_1,k_2, k_3,k_4; K, \theta_4)$. With this parametrisation, we have the following identities,
\begin{eqnarray}\label{Gaunt}
\int d\Omega_{\hat{r}} Y_{l_1 m_1}(\hat{\mathbf{r}})Y_{l_2 m_2}(\hat{\mathbf{r}})Y_{l_3 m_3}(\hat{\mathbf{r}})&=&\sqrt{\frac{(2l_1+1)(2l_2+1)(2l_3+1)}{4\pi}}\left( \begin{array}{ccc}
l_1 & l_2 & l_3 \\
0 & 0 & 0 \end{array} \right)\left( \begin{array}{ccc}
l_1 & l_2 & l_3 \\
m_1 & m_2 & m_3 \end{array} \right),\nonumber\\
\int d \Omega_{\hat{r}}Y_{lm}(\hat{\mathbf{r}})Y_{l' m'}^*(\hat{\mathbf{r}})&=&\delta_{l l'}\delta_{m m'}
\end{eqnarray}
and~\eqref{Identities} we find
\begin{eqnarray}
\mathcal{T}_{l_1 m_1 l_2 m_2 l_3 m_3 l_4 m_4}&=&\left(\frac{2}{\pi}\right)^5 \sum_{L',M'}\sum_{l_4',m_4'} (-1)^{M'}\int (k_1 k_2 k_3 k_4 K)^2 dk_1 dk_2 dk_3 dk_4 dK r_1^2 dr_1 r_2^2 dr_2 j_L(K r_1) j_L(K r_2)\nonumber\\
&&\times[j_{l_1}(k_1 r_1)\Delta_{l_1}(k_1)][j_{l_2}(k_2 r_1)\Delta_{l_2}(k_2)][j_{l_3}(k_3 r_2)\Delta_{l_3}(k_3)][j_{l_4}(k_4 r_2)\Delta_{l_4}(k_4)] h_{l_1 l_2 L'} h_{l_3 l_4' L'} (-1)^{m_4'}\nonumber\\
&&\times \left( \begin{array}{ccc}
l_1 & l_2 & L' \\
m_1 & m_2 & -M' \end{array} \right)\left( \begin{array}{ccc}
l_3 & l_4' & L' \\
m_3 & -m_4' & M' \end{array} \right)\int d\Omega_{\hat{\mathbf{k}}_4} \mathcal{T}_{\Phi}(k_1,k_2, k_3,k_4; K, \theta_4)Y_{l_4 m_4}(\hat{\mathbf{k}}_4)Y_{l_4' m_4'}(\hat{\mathbf{k}}_4)
\end{eqnarray}
where we write
\begin{eqnarray}
h_{l_1 l_2 L'}=\sqrt{\frac{(2l_1+1)(2l_2+1)(2L'+1)}{4\pi}}\left( \begin{array}{ccc}
l_1 & l_2 & L' \\
0 & 0 & 0 \end{array} \right). 
\end{eqnarray}
Next, we note that inverting equation~\eqref{RedTrisp} gives the expression
\begin{eqnarray}
\mathcal{T}^{l_1 l_2}_{l_3 l_4}(L)=\sum_{m_i,M}(2L+1)(-1)^M \left( \begin{array}{ccc}
l_1 & l_2 & L \\
m_1 & m_2 & M \end{array} \right)\left( \begin{array}{ccc}
l_3 & l_4 & L \\
m_3 & m_4 & -M \end{array} \right) \mathcal{T}_{l_1 m_1 l_2 m_2 l_3 m_3 l_4 m_4}.
\end{eqnarray}
The sum over $m_1, m_2$ is proportional to
\begin{eqnarray}\label{Orthog1}
\sum_{m_1 ,m_2} (2L+1) \left( \begin{array}{ccc}
l_1 & l_2 & L \\
m_1 & m_2 & M \end{array} \right)\left( \begin{array}{ccc}
l_1 & l_2 & L' \\
m_1 & m_2 & -M' \end{array} \right)=\delta_{L,L'} \delta_{M,-M'}
\end{eqnarray}
and therefore the sum over $L', M'$ implies $L'=L$ and $M'=-M$. The sum over $m_3, M$ is then proportional to
\begin{eqnarray}\label{Orthog2}
\sum_{m_3 ,M} \left( \begin{array}{ccc}
l_3 & l_4 & L \\
m_3 & m_4 & -M \end{array} \right)\left( \begin{array}{ccc}
l_3 & l_4' & L \\
m_3 &- m_4' & -M \end{array} \right)=\frac{1}{2l_4'+1}\delta_{l_4,l_4'} \delta_{m_4,-m_4'}.
\end{eqnarray}
Combining these we find that
\begin{eqnarray}\label{TrispRed1}
\mathcal{T}^{l_1 l_2}_{l_3 l_4}(L)&=& h_{l_1 l_2 L} h_{l_3 l_4 L} \left(\frac{2}{\pi}\right)^5 \int (k_1 k_2 k_3 k_4 K)^2 dk_1 dk_2 dk_3 dk_4 dK r_1^2 dr_1 r_2^2 dr_2 j_L(K r_1) j_L(K r_2)\nonumber\\
&&\times[j_{l_1}(k_1 r_1)\Delta_{l_1}(k_1)][j_{l_2}(k_2 r_1)\Delta_{l_2}(k_2)][j_{l_3}(k_3 r_2)\Delta_{l_3}(k_3)][j_{l_4}(k_4 r_2)\Delta_{l_4}(k_4)] \nonumber\\
&&\times \frac{1}{2 l_4+1}\sum_{m_4=-l_4}^{l_4} \int d\Omega_{\hat{\mathbf{k}}_4} \mathcal{T}_{\Phi}(k_1,k_2, k_3,k_4; K, \theta_4)Y_{l_4 m_4}(\hat{\mathbf{k}}_4)Y_{l_4 m_4}^*(\hat{\mathbf{k}}_4)
\end{eqnarray}
We can decompose this expression further by expanding the primordial trispectrum as a Legendre series. In particular, we write
\begin{eqnarray}\label{expansion}
\mathcal{T}_{\Phi}(k_1,k_2, k_3,k_4; K, \theta_4)=\sum_{n=0}^{\infty}\mathcal{T}_{\Phi,n}(k_1,k_2, k_3,k_4; K) P_n(\cos\theta_4).
\end{eqnarray}
This is an expansion about the $n=0$ planar mode which, as we have noted, is sufficient for describing many well-motivated models. Noting that $P_n=\sqrt{\frac{4\pi}{2 n+1}}Y_{n 0}$,  our expression for the CMB trispectrum becomes
\begin{eqnarray*}
\mathcal{T}^{l_1 l_2}_{l_3 l_4}(L)&=& h_{l_1 l_2 L} h_{l_3 l_4 L} \left(\frac{2}{\pi}\right)^5 \int (k_1 k_2 k_3 k_4 K)^2 dk_1 dk_2 dk_3 dk_4 dK r_1^2 dr_1 r_2^2 dr_2 j_L(K r_1) j_L(K r_2)\nonumber\\
&&\times[j_{l_1}(k_1 r_1)\Delta_{l_1}(k_1)][j_{l_2}(k_2 r_1)\Delta_{l_2}(k_2)][j_{l_3}(k_3 r_2)\Delta_{l_3}(k_3)][j_{l_4}(k_4 r_2)\Delta_{l_4}(k_4)] \nonumber\\
&&\times \sum_{m_4=-l_4}^{l_4}\sum_{n=0}^{\infty} (-1)^{m_4}    \left( \begin{array}{ccc}
l_4 & l_4 & n \\
0 &0 & 0 \end{array} \right)    \left( \begin{array}{ccc}
l_4 & l_4 & n \\
m_4 & -m_4 & 0 \end{array} \right)\mathcal{T}_{\Phi,n}(k_1,k_2, k_3,k_4; K). 
\end{eqnarray*}
This expression may be further simplified by noting
\begin{eqnarray*}
\sum_{m_4} (-1)^{m_4}  \left( \begin{array}{ccc}
l_4 & l_4 & n \\
m_4 & -m_4 & 0 \end{array} \right)=(-1)^{l_4} \sqrt{2 l_4+1} \delta_{n 0}
\end{eqnarray*}
and
\begin{eqnarray*}
  \left( \begin{array}{ccc}
l_4 & l_4 & 0 \\
0 &0 & 0 \end{array} \right) = (-1)^{l_4}\frac{1}{ \sqrt{2 l_4+1}},
\end{eqnarray*}
which together imply that the final line reduces to $\mathcal{T}_{\Phi,0}$.
 In particular, we have
\begin{eqnarray}\label{TrispRed2}
\mathcal{T}^{l_1 l_2}_{l_3 l_4}(L)&=& h_{l_1 l_2 L} h_{l_3 l_4 L} \left(\frac{2}{\pi}\right)^5 \int (k_1 k_2 k_3 k_4 K)^2 dk_1 dk_2 dk_3 dk_4 dK r_1^2 dr_1 r_2^2 dr_2 j_L(K r_1) j_L(K r_2)\nonumber\\
&&\times[j_{l_1}(k_1 r_1)\Delta_{l_1}(k_1)][j_{l_2}(k_2 r_1)\Delta_{l_2}(k_2)][j_{l_3}(k_3 r_2)\Delta_{l_3}(k_3)][j_{l_4}(k_4 r_2)\Delta_{l_4}(k_4)] \mathcal{T}_{\Phi,0}(k_1,k_2, k_3,k_4; K). 
\end{eqnarray}
The reduction to the $n=0$ mode in~\eqref{TrispRed2} shows clearly that the CMB only probes and constrains the planar component of the primordial trispectrum $T_{\Phi}$. In order to test theories which have general non-planar $n>0$ contributions we will have to use 3D data, such as $21$cm surveys or large-scale galaxy distributions (as we shall discuss later).
\par
From equation~\eqref{TrispRed1} it is clear that the definition reduced trispectrum \cite{Hu} includes an unnecessary geometrical factor $h_{l_1 l_2 L} h_{l_3 l_4 L}$ and we therefore advocate the use of the true reduced trispectrum,
\begin{eqnarray}\label{ExtraTrispRed}
t^{l_1 l_2}_{l_3 l_4}(L)=\frac{\mathcal{T}^{l_1 l_2}_{l_3 l_4}(L)}{h_{l_1 l_2 L}h_{l_3 l_4 L}},
\end{eqnarray}
by analogy with the reduced bispectrum $b_{l_1 l_2 l_3}=B_{l_1 l_2 l_3}/h_{l_1 l_2 l_3}$, where $B_{l_1 l_2 l_3}$ represents the angle-averaged bispectrum. To prevent confusion, however, we refer to $t^{l_1 l_2}_{l_3 l_4}(L)$ as the `extra'-reduced trispectrum.

\subsection{Relationship between the primordial trispectrum and other probes}
As is clear from equation~\eqref{TrispRed2} the CMB trispectrum depends only on the zeroth Legendre mode of the primordial trispectrum. Therefore, in order to break this degeneracy other probes of non-Gaussianity should be considered. As has been discussed in \cite{SefZal} the matter density perturbations are related to the primordial fluctuations by the Poisson equation via the expression
\begin{eqnarray}
\delta_{\mathbf{k}}(a)=M(k;a) \Phi_{\mathbf{k}},
\end{eqnarray}
where $a$ is the scale factor and $M(k;a)$ is given by
\begin{eqnarray}
M(k;a)=-\frac{3}{5}\frac{k^2 T(k)}{\Omega_m H_0^2}D_+(a),
\end{eqnarray}
where $T(k)$ is the transfer function, $D_+(a)$ is the growth factor in linear perturbation theory, $\Omega_m$ is the present value of the dark matter density and $H_0$ is the present value of the Hubble constant.
Therefore, the primordial contribution to the $n$- point connected correlation function of matter density perturbations at a given value of the scale factor is given by
\begin{eqnarray}
\langle \delta_{\mathbf{k}_1}(a) \delta_{\mathbf{k}_2}(a)\dots\delta_{\mathbf{k}_n} (a) \rangle_c =M(k_1;a)M(k_2;a)\dots M(k_n;a)\langle \Phi_{\mathbf{k}_1} \Phi_{\mathbf{k}_2}\dots\Phi_{\mathbf{k}_n}  \rangle_c.
\end{eqnarray}
 Possible probes of the matter density perturbations include galaxy surveys and the Lyman alpha forest, i.e. the sum of absorption lines from the Ly-$\alpha$ transition of the neutral hydrogen in the spectra of distant galaxies and quasars. There are three sources of non-Gaussianity in such surveys \cite{ReviewLFSS}: one primordial, one due to gravitational instability and the last due to nonlinear bias. $21$cm observations offer another probe of non-Gaussianity which are less subject to the unknown galaxy bias, especially at high redshift. However, uncertainties in the neutral fraction replaces the uncertainties in the bias in this case. There are also complications due to redshift space distortions arising from peculiar velocities. Despite these drawbacks, recent advances in this area suggest that probes of the matter density perturbations potentially represent a powerful tool to detect non-Gaussianity and possibly break the degeneracy implicit in trispectrum measurements using the CMB. The study of such data involves using the full Legendre expansion of the primordial trispectrum as in equation~\eqref{expansion}. In the remainder of this paper we proceed to investigate the CMB trispectrum. However, many of the results presented here are straightforwardly extended to alternative probes of non-Gaussianity as discussed here.

\subsection{Ideal Estimator}
Unfortunately the trispectrum signal, like the bispectrum, is too weak for us to measure individual multipoles directly. Therefore, in order to compare theory with observations it is necessary to use an estimator that sums over all multipoles. Estimators can be thought of as performing a least squares fit of the trispectrum predicted by theory, $\langle a_{l_1m_1}a_{l_2m_2}a_{l_3m_3}a_{l_4m_4}\rangle_c$, to the trispectrum obtained from observations. The trispectrum from observations is given by $(a_{l_1 m_1}^{\rm{obs}}a_{l_2 m_2}^{\rm{obs}}a_{l_3 m_3}^{\rm{obs}}a_{l_4 m_4}^{\rm{obs}})_c$ where we subtract the unconnected or Gaussian part, denoted $\rm{uc}$, from the four point function, $a_{l_1 m_1}^{\rm{obs}}a_{l_2 m_2}^{\rm{obs}}a_{l_3 m_3}^{\rm{obs}}a_{l_4 m_4}^{\rm{obs}}$. This unconnected part is related to the observed
angular power spectrum $C_{l}^{\rm{obs}}$ by
\begin{eqnarray}
(a_{l_1 m_1}^{\rm{obs}}a_{l_2 m_2}^{\rm{obs}}a_{l_3 m_3}^{\rm{obs}}a_{l_4 m_4}^{\rm{obs}})_{\rm{uc}}=&&(-1)^{m_1+m_3}C_{l_1}^{\rm{obs}}C_{l_3}^{\rm{obs}}\delta_{l_1, l_2}\delta_{m_1, -m_2}\delta_{l_3, l_4}\delta_{m_3, -m_4}+(-1)^{m_1+m_2}C_{l_1}^{\rm{obs}}C_{l_2}^{\rm{obs}}\nonumber\\
&&\times\Big(\delta_{l_1, l_3}\delta_{m_1, -m_3}\delta_{l_2, l_4}\delta_{m_2, -m_4}
+\delta_{l_1, l_4}\delta_{m_1, -m_4}\delta_{l_2, l_3}\delta_{m_2, -m_3}\Big).
\end{eqnarray}
We define the estimator to be
\begin{eqnarray}\label{Estimator}
\mathcal{E}=\frac{1}{N_T}\sum_{l_i m_i}\frac{\langle a_{l_1m_1}a_{l_2m_2}a_{l_3m_3}a_{l_4m_4}\rangle_c \left(a_{l_1 m_1}^{\rm{obs}}a_{l_2 m_2}^{\rm{obs}}a_{l_3 m_3}^{\rm{obs}}a_{l_4 m_4}^{\rm{obs}}\right)_c}{C_{l_1}C_{l_2}C_{l_3}C_{l_4}}
\end{eqnarray}
where the normalisation factor $N_T$ is given by (see Appendix A)
\begin{eqnarray}\label{Normal}
N_T=\sum_{l_i, L}\frac{T^{l_1 l_2}_{l_3 l_4}(L)T^{l_1 l_2}_{l_3 l_4}(L)}{(2L+1)C_{l_1}C_{l_2}C_{l_3}C_{l_4}}.
\end{eqnarray}
As is clear from the earlier discussion, assuming isotropy for a given theoretical model, we need only calculate the reduced trispectrum, $\mathcal{T}^{l_1 l_2}_{l_3 l_4}(L)$,  rather than the more challenging full trispectrum $\langle a_{l_1m_1}a_{l_2m_2}a_{l_3m_3}a_{l_4m_4}\rangle_c$. 
\par
This estimator naturally defines a correlator for testing whether two competing trispectra could be differentiated by an ideal experiment. Replacing the observed trispectrum with one calculated from a competing theory we have,
\begin{eqnarray}
\mathcal{C}(T,T')&=&\frac{1}{N_T}\sum_{l_i, m_i}\frac{\langle a_{l_1m_1}a_{l_2m_2}a_{l_3m_3}a_{l_4m_4}\rangle_c \langle a_{l_1m_1}'a_{l_2m_2}' a_{l_3m_3}' a_{l_4m_4}'\rangle_c }{C_{l_1}C_{l_2}C_{l_3}C_{l_4}}\nonumber\\
&=&\frac{1}{N_T}\sum_{l_i, L}\frac{T^{l_1 l_2}_{l_3 l_4}(L)T'^{l_1 l_2}_{l_3 l_4}(L)}{(2L+1)C_{l_1}C_{l_2}C_{l_3}C_{l_4}},
\end{eqnarray}
where now the normalisation $N_T$ is defined as follows,
\begin{eqnarray}
N_T=\sqrt{\sum_{l_i, L}\frac{T^{l_1 l_2}_{l_3 l_4}(L)T^{l_1 l_2}_{l_3 l_4}(L)}{(2L+1)C_{l_1}C_{l_2}C_{l_3}C_{l_4}}}\sqrt{\sum_{l_i, L}\frac{T'^{l_1 l_2}_{l_3 l_4}(L)T'^{l_1 l_2}_{l_3 l_4}(L)}{(2L+1)C'_{l_1}C'_{l_2}C'_{l_3}C'_{l_4}}}.
\end{eqnarray}
An alternative correlator between two trispectra, which is easier to solve numerically, is found by replacing the trispectra by the respective reduced trispectra in the above definitions. Therefore, when comparing two trispectra we shall use this latter definition, $\mathcal{C(T,T')}$. The exact relation between the two correlators can be deduced from Appendix B in ref.~\cite{Hu}.

\subsection{General Estimator}
The above estimator is applicable for general trispectra in the limit where non-Gaussianity is small and the observed map is free of instrument noise and foreground contamination. Of course, this is an idealised case and we need to consider taking into account the effect of sky cuts and inhomogeneous noise. Here we follow the approach of~\cite{Amendola} (an approach that is further elucidated in~\cite{Babichopt} and~\cite{Komatsu2}). As we prove in Appendix B the appropriate form of the optimal estimator becomes
\begin{eqnarray}\label{Estimator2}
\mathcal{E}^{\rm{general}}&=&\frac{f_{\rm{sky}}}{\tilde{N}}\sum_{l_i m_i} \langle a_{l_1m_1}a_{l_2m_2}a_{l_3m_3}a_{l_4m_4}\rangle_c  \Bigg[(C^{-1} a^{\rm{obs}})_{l_1 m_1}(C^{-1} a^{\rm{obs}})_{l_2 m_2} (C^{-1} a^{\rm{obs}})_{l_3 m_3} (C^{-1} a^{\rm{obs}})_{l_4 m_4}\nonumber\\
&&-6(C^{-1})_{l_1 m_1,l_2 m_2} (C^{-1} a^{\rm{obs}})_{l_3 m_3} (C^{-1} a^{\rm{obs}})_{l_4 m_4}+3 (C^{-1})_{l_1 m_1,l_2 m_2}(C^{-1})_{l_3 m_3,l_4 m_4} \Bigg],
\end{eqnarray}
where
\begin{eqnarray*}
\tilde{N}= \sum_{l_i m_i}  \langle a_{l_1m_1}a_{l_2m_2}a_{l_3m_3}a_{l_4m_4}\rangle_c  (C^{-1})_{l_1 m_1,l_1' m_1'} (C^{-1})_{l_2 m_2,l_2' m_2'} (C^{-1})_{l_3 m_3,l_3' m_3'} (C^{-1})_{l_4 m_4,l_4' m_4'} \langle a_{l_1' m_1'}a_{l_2' m_2'}a_{l_3' m_3'}a_{l_4' m_4'}\rangle_c,
\end{eqnarray*}
$f_{\rm{sky}}$ is the fraction of the sky outside the mask, and where the covariance matrix $C$ is now non-diagonal due to mode-mode coupling introduced by the mask and anisotropic noise. Due to the breaking of isotropy extra terms have been added in order to maintain the optimality of the estimator. The optimal estimator, in the case that the covariance matrix is diagonal, reads
\begin{eqnarray}
\mathcal{E}&=&\frac{f_{\rm{sky}}}{N_T}\sum_{l_i m_i}  \frac{\langle a_{l_1m_1}a_{l_2m_2}a_{l_3m_3}a_{l_4m_4}\rangle_c }{C_{l_1}C_{l_2}C_{l_3}C_{l_4}}\Big[a^{\rm{obs}}_{l_1 m_1}a^{\rm{obs}}_{l_2 m_2}a^{\rm{obs}}_{l_3 m_3}a^{\rm{obs}}_{l_4 m_4}-6 (-1)^{m_1}C_{l_1}\delta_{l_1 l_2}\delta_{m_1 -m_2}a^{\rm{obs}}_{l_3 m_3}a^{\rm{obs}}_{l_4 m_4}\nonumber\\
&&+3(-1)^{m_1+m_3}\delta_{l_1 l_2}\delta_{m_1 -m_2} \delta_{l_3 l_4}\delta_{m_3 -m_4}C_{l_1}C_{l_3}\Big],
\end{eqnarray}
where $N_T$ is given by equation~\eqref{Normal}. We note also that the average of this estimator is
\begin{eqnarray}
\langle\mathcal{E}\rangle&=&\frac{f_{\rm{sky}}}{N_T}\sum_{l_i m_i}  \frac{\langle a_{l_1m_1}a_{l_2m_2}a_{l_3m_3}a_{l_4m_4}\rangle_c \langle a^{\rm{obs}}_{l_1 m_1}a^{\rm{obs}}_{l_2 m_2}a^{\rm{obs}}_{l_3 m_3}a^{\rm{obs}}_{l_4 m_4}\rangle_c}{C_{l_1}C_{l_2}C_{l_3}C_{l_4}},
\end{eqnarray}
as expected.

In the remainder of this paper we shall refer to the ideal estimator unless otherwise stated. However, this formula is important for the general implementation of the formalisms introduced here.

\subsection{Kurtosis as a measure of non-Gaussianity}
As an aside,  we note that the use of non-optimal estimators may also provide useful information, e.g. as a reality check on these complex calculations. The kurtosis of the one point temperature distribution offers such an estimator. The kurtosis, $g_2$, is defined as
\begin{eqnarray}\label{Kurtosis1} 
g_2=\frac{\Bigg\langle\left(\frac{\Delta T}{T}(\hat{n}) \right)^4\Bigg\rangle}{\left(\Bigg\langle \left(\frac{\Delta T}{T}(\hat{n}) \right)^2\Bigg\rangle\right)^2}-3.
\end{eqnarray}
As we show in Appendix C (where we also include a discussion on the skewness for completeness) the kurtosis may be written in the following form
\begin{eqnarray}\label{Kurtosis2}
g_2=\frac{48\pi\sum_{l_i,L} h_{l_1 l_2 L}^2 h_{l_3 l_4 L}^2t^{l_1 l_2}_{l_3 l_4}(L)/(2L+1)}{\left(\sum_l (2 l+1)C_l\right)^2}.
\end{eqnarray}
The calculation of this quantity is relatively straightforward compared to the full estimator due to the absence of Wigner 6j symbols in the expression.

\section{III. The Shape of Primordial Trispectra}

\subsection{Shape function}
It is known from CMB observations that the power spectrum is nearly scale-invariant. Analysis of the bispectrum is performed using the shape function, which is a scale invariant form of the bispectrum. To parallel this analysis we wish to write a scale invariant form of the trispectrum (or in particular the trispectrum modes). Therefore, we need to eliminate a $k^9$ scaling. Motivated by~\eqref{TrispRed2} we define this shape by
\begin{eqnarray}\label{Shape}
S_T(k_1,k_2,k_3,k_4,K)&=&\frac{(k_1 k_2 k_3 k_4)^2 K}{\Delta_{\Phi}^3 N} T_{\Phi,0}(k_1,k_2,k_3,k_4;K),
\end{eqnarray}
where $N$ is an appropriate normalisation factor. For clarity in what follows we note that we shall use the symbol $S_{\mathcal{T}}$ when referring to the shape induced by the reduced primordial trispectrum. Of course, this choice of the shape function is not unique. Another choice of shape function is
\begin{eqnarray}\label{Shape2}
\tilde{S}_T(k_1,k_2,k_3,k_4,K)&=&\frac{(k_1 k_2 k_3 k_4)^{9/4}}{\Delta_{\Phi}^3 N} T_{\Phi,0}(k_1,k_2,k_3,k_4;K),
\end{eqnarray}
which has the advantage of remaining independent of the diagonal $K$ if the underlying trispectrum has this property. Such a class of models are discussed further in Appendix D. Nonetheless we proceed with $S_{T}$ as our choice of shape function in this paper, leaving further investigation of this issue to a future publication~\cite{Regan2}. We should also notice that, since our analysis here is focused on the CMB, we have only included the zeroth mode of the Legendre expansion as indicated by~\eqref{TrispRed2}. However, for more general probes of non-Gaussianity, as discussed in Section II, the full Legendre expansion described by equation~\eqref{expansion} is required. In such a case the analysis outlined here can be applied mode-by-mode. Due to orthogonality of the Legendre modes, extending the study is a trivial task.
\par
 If we rewrite the reduced CMB trispectrum in terms of the shape function, $S_{\mathcal{T}}$, we have
\begin{eqnarray} 
\mathcal{T}^{l_1 l_2}_{l_3 l_4}(L)&=&N  h_{l_1 l_2 L} h_{l_3 l_4 L} \left(\frac{2}{\pi}\right)^5 \int d\mathcal{V}_k  S_{\mathcal{T}} (k_1,k_2,k_3,k_4,K)K\nonumber\\
&&\times \Delta_{l_1}(k_1)\Delta_{l_2}(k_2)\Delta_{l_3}(k_3)\Delta_{l_4}(k_4) I^G_{l_1 l_2 l_3 l_4 L}(k_1,k_2,k_3,k_4,K)
\end{eqnarray}
where the integral $I^G$ is given by
\begin{eqnarray}
 I^G_{l_1 l_2 l_3 l_4 L}(k_1,k_2,k_3,k_4,K)&=&\int r_1^2 r_2^2 dr_1 dr_2 j_L(K r_1) j_L (K r_2) j_{l_1}(k_1 r_1) j_{l_2}(k_2 r_1) j_{l_3}(k_3 r_2) j_{l_4}(k_4 r_2)
 \end{eqnarray}
and $d \mathcal{V}_k$ corresponds to the area inside the region $k_i,K/2\in[0,k_{\rm{max}}]$ allowed by the triangle conditions. Therefore the shape function is the signal that is evolved via the transfer functions to give the trispectrum today. Essentially, $I^G$ acts like a window function on all the shapes as it projects from $k$ to $l-$space, that is, it will tend to smear out their sharper distinguishing features. This means that the shape function $S_{\mathcal{T}}$, especially in the scale invariant case, can be thought of as the primordial counterpart of the reduced CMB trispectrum $\mathcal{T}^{l_1 l_2}_{l_3 l_4}(L)$ before projection.

\subsection{Shape correlators}
We wish to construct a primordial shape correlator that predicts the value of the CMB correlator $\mathcal{C(T,T')}$. To this end we should consider something of the form
\begin{eqnarray}
F(S_{\mathcal{T}},S_{\mathcal{T}'})=\int d\mathcal{V}_k  S_{\mathcal{T}}(k_1,k_2,k_3,k_4,K) S_{\mathcal{T}'}(k_1,k_2,k_3,k_4,K) \omega(k_1,k_2,k_3,k_4,K)
\end{eqnarray}
where $\omega$ is an appropriate weight function. With this choice of weight the primordial shape correlator then takes the form
\begin{eqnarray}
\overline{\mathcal{C}}(S_{\mathcal{T}},S_{\mathcal{T}'})=\frac{F(S_{\mathcal{T}},S_{\mathcal{T}'})}{\sqrt{F(S_{\mathcal{T}},S_{\mathcal{T}})F(S_{\mathcal{T}'},S_{\mathcal{T}'})}}.
\end{eqnarray}
\par
The question now is what weight function should we choose? Our goal is to choose $S^2 \omega$ in $k-$ space such that it produces the same scaling as the estimator $T^2/((2L+1)C^4)$ in $l-$ space. Let's consider the simplest case where $k=k_1=k_2=k_3=k_4=K$ and $l=l_1=l_2=l_3=l_4=L$. For primordial trispectra which are scale invariant, then
\begin{eqnarray}
S_{\mathcal{T}}^2(k,k,k,k,k)\,\omega(k,k,k,k,k)\propto \omega(k,k,k,k,k).
\end{eqnarray}
If we work in the large angle approximation, and assume $l+1\approx l$, then we know $C_l\propto l^{-2}$ and from the analytic solution for the local model which we will describe below (see equations\eqref{LocA} and~\eqref{LocB}) we have
\begin{eqnarray}
T^{l l}_{l l}(l)\propto h_{l l l}^2 \frac{1}{l^6}.
\end{eqnarray}
Now $h_{ l l l}\propto l^{3/2} \left( \begin{array}{ccc}
l & l & l \\
0 & 0 & 0 \end{array} \right)$ and the Wigner $3J$ symbol has an exact solution for which
\begin{eqnarray}
\left( \begin{array}{ccc}
l & l & l \\
0 & 0 & 0 \end{array} \right)\approx (-1)^{3l/2}\frac{1}{\sqrt{3 l+1}}\sqrt{\frac{l!^3}{3l!}}\frac{(3l/2)!}{(l/2)!^3}\approx (-1)^{3l/2}\sqrt{\frac{2}{\sqrt{3}\pi}}\frac{1}{l}.
\end{eqnarray}
Therefore $T^{l l}_{l l}(l)\propto  l^{-5}$ and so
\begin{eqnarray}
\frac{T^{l l}_{l l}(l)^2}{(2l+1)C_l^4 }\propto l^{-3}.
\end{eqnarray}
Hence we find that we should choose a weight function $\omega(k,k,k,k,k)\propto k^{-3}$. The particular choice of $\omega$ may significantly improve forecasting accuracy - by, for instance, using a phenomenological window function to incorporate damping due to photon diffusion or smoothing due to projection from $k$- to $l$- space, but it does not impact important qualitative insights. A specific choice of weight function, motivated by the choice of weight function for the bispectrum, is the following
\begin{eqnarray}\label{Weightk}
w(k_1,k_2,k_3,k_4,K)=\frac{K}{(k_1+k_2+K)^2(k_3+k_4+K)^2}.
\end{eqnarray}

\subsection{Shape Decomposition}
\begin{figure}[htp]
\centering 
\includegraphics[width=172mm]{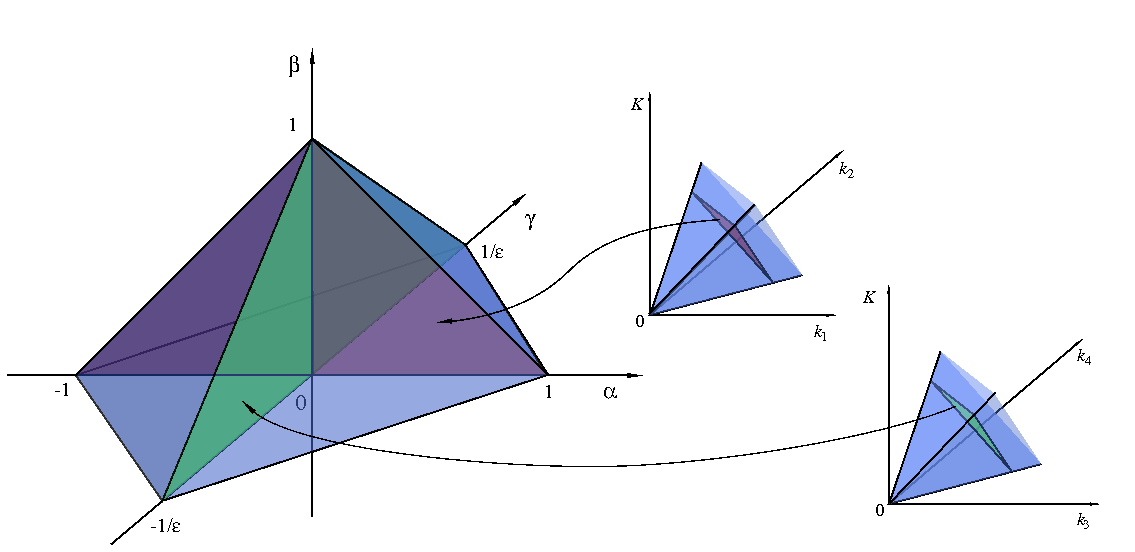}
\caption{Three-dimensional shape function domain for a fixed value of $\epsilon$, i.e. for a particular ratio of the perimeters of the triangles created by the wavenumbers $(k_1,k_2,K)$ and $(k_3,k_4,K)$ respectively. Note that the triangle conditions on these two wavenumber sets restrict them to the two tetrahedral domains illustrated (right), slices through which are mapped as shown into the full domain, a rectangular pyramid (left).}
\label{fig:shape}
\end{figure}

Given the strong observational limits on the scalar tilt we expect all shape functions to exhibit behaviour close to scale-invariance, so that $S_{\mathcal{T}}(k_1,k_2,k_3,k_4,K)$ will depend only weakly on the overall magnitude of the summed wavenumbers. Here we choose to parametrise the magnitude of the wavenumbers with the quantity
\begin{eqnarray} 
k=\frac{1}{2}(k_1+k_2+K).
\end{eqnarray}
$k$ is the semi-perimeter of the triangle formed by the vectors $\mathbf{k}_1,\mathbf{k}_2,\mathbf{K}$.
Due to the scaling behaviour the form of the shape function on a cross-section is essentially independent of $k$, so that 
\begin{eqnarray}
S_{\mathcal{T}}(k_1,k_2,k_3,k_4,K)=f(k)\overline{S}_{\mathcal{T}}(\hat{k}_1,\hat{k}_2,\hat{k}_3,\hat{k}_4,\hat{K})
\end{eqnarray}
where $\hat{k}_i=k_i/k$ and $\hat{K}=K/k$. Since we are restricted to the region where the wavenumbers $(k_1,k_2,K)$ and $(k_3,k_4,K)$ form triangles by momentum conservation, we will reparametrise the allowed region to separate out the overall scale $k$ from the behaviour on a cross-sectional slice. This four-dimensional slice is spanned by the remaining coordinates. Concentrating on each triangle individually we reparametrise in a similar fashion to the analysis done in~\cite{Ferg2}. For triangle $(k_1,k_2,K)$ we have
\begin{eqnarray}
K&=&k(1-\beta)\nonumber\\
k_1&=&\frac{k}{2}(1+\alpha+\beta)\nonumber\\
k_2&=&\frac{k}{2}(1-\alpha+\beta)
\end{eqnarray}
while for triangle $(k_3,k_4,K)$ we have
\begin{eqnarray}
K&=&\epsilon k(1-\delta)\nonumber\\
k_3&=&\frac{\epsilon k}{2}(1+\gamma+\delta)\nonumber\\
k_4&=&\frac{\epsilon k}{2}(1-\gamma+\delta)
\end{eqnarray}
where $\epsilon$ parametrises the ratio of the perimeters of the two triangles, i.e. $\epsilon=\frac{k_3+k_4+K}{k_1+k_2+K}$. We consider $1\leq\epsilon<\infty$. The different expressions for $K$ imply that
\begin{eqnarray}
1-\beta=\epsilon (1-\delta).
\end{eqnarray}
The conditions for triangle $(k_1,k_2,K)$ that $0\leq k_1,k_2,K\leq k$ imply that $0\leq \beta \leq 1$ and $-(1-\beta)\leq \alpha \leq 1-\beta$, while the conditions for triangle $(k_3,k_4,K)$ that $0\leq k_1,k_2,K\leq \epsilon k$, along with the relationship between $\delta$ and $\beta$ and the requirement that $\epsilon\geq 1$, imply that 
$-(1-\beta)/\epsilon \leq\gamma\leq (1-\beta)/\epsilon$. In summary, we have the following domains,
\begin{eqnarray}
0\leq k<\infty,\quad  1\leq\epsilon<\infty,\quad 0\leq \beta \leq 1,\quad -(1-\beta)\leq \alpha \leq 1-\beta,\quad -\frac{1-\beta}{\epsilon} \leq\gamma\leq \frac{1-\beta}{\epsilon}.
\end{eqnarray}
With this parametrisation we can re-write the shape function and the volume element respectively as
\begin{eqnarray}
S_{\mathcal{T}}(k_1,k_2,k_3,k_4,K)=f(k)\overline{S}_{\mathcal{T}}(\alpha,\beta,\gamma,\epsilon),\quad d\mathcal{V}_{k}=d k_1 d k_2 dk_3 dk_4 dK= \epsilon k^4 dk d\alpha d\beta d\gamma d\epsilon.
\end{eqnarray}
In order to represent the shape function graphically we can choose fixed values of $\epsilon$ in which case the shape $\overline{S}$ becomes three dimensional. The particular three dimensional domain is shown in Figure~\ref{fig:shape}. From the image we see how the particular triangles created by the wavenumbers generate the three dimensional slice for each $\epsilon$. We can envisage  the four-dimensional shape by
imagining an orthogonal direction for $\epsilon$ out of the page, along which are located increasingly squeezed rectangular pyramids. 

\section{IV. Separable Shapes}

\subsection{Examples: Local, equilateral and constant models}
\begin{figure}[htp]
\centering 
\includegraphics[width=120mm]{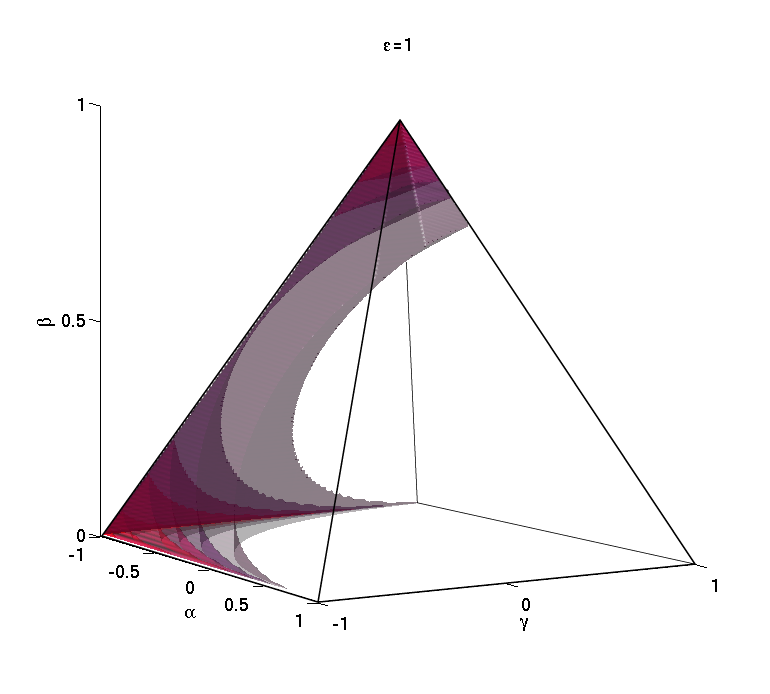}
\includegraphics[width=80mm]{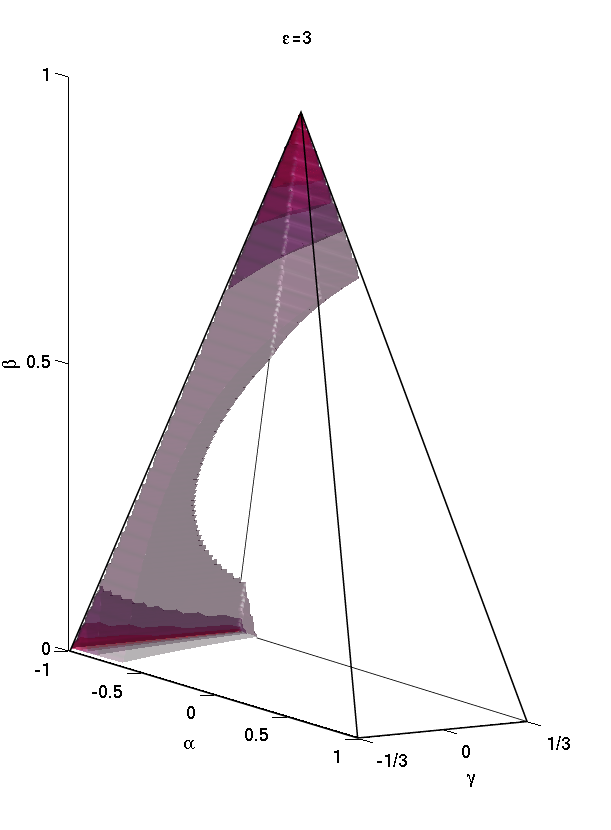}
\includegraphics[width=62mm]{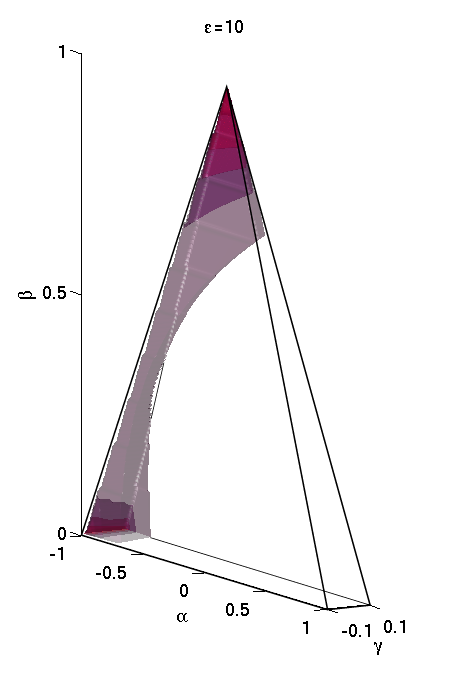}
\caption{Local $A$ model~\eqref{LocA}. The peak as $\beta\rightarrow 1$ corresponds to $K\rightarrow 0$, i.e. the `doubly-squeezed' limit. The other peak corresponds to $k_1\rightarrow 0, k_3\rightarrow (\epsilon-1)k$. As $\epsilon$ rises above unity (i.e. for triangle $(k_3,k_4,K)$ bigger than triangle $(k_1,k_2,K)$) we expect this peak to become suppressed as observed.}
\label{fig:LocA}
\end{figure}

\begin{figure}[htp]
\centering 
\includegraphics[width=120mm]{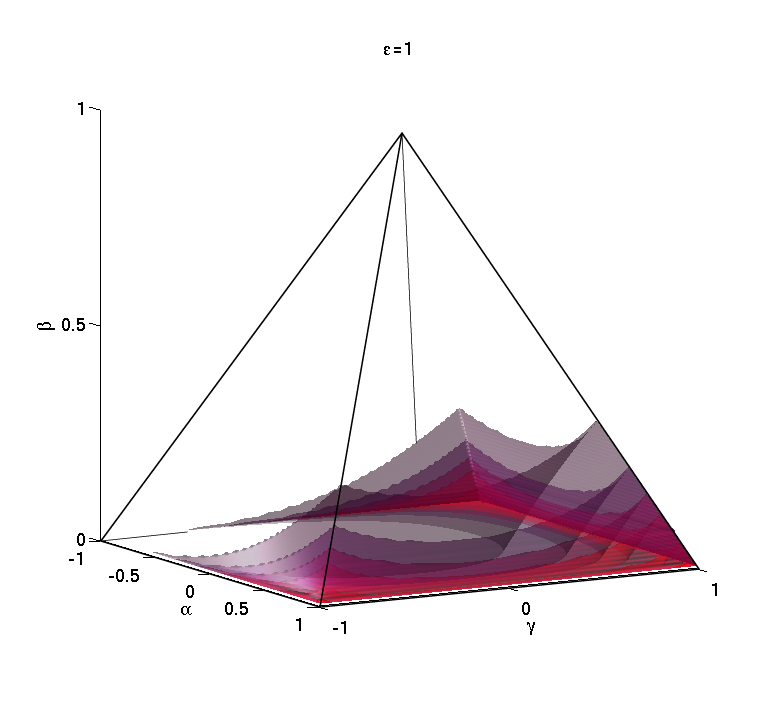}
\includegraphics[width=80mm]{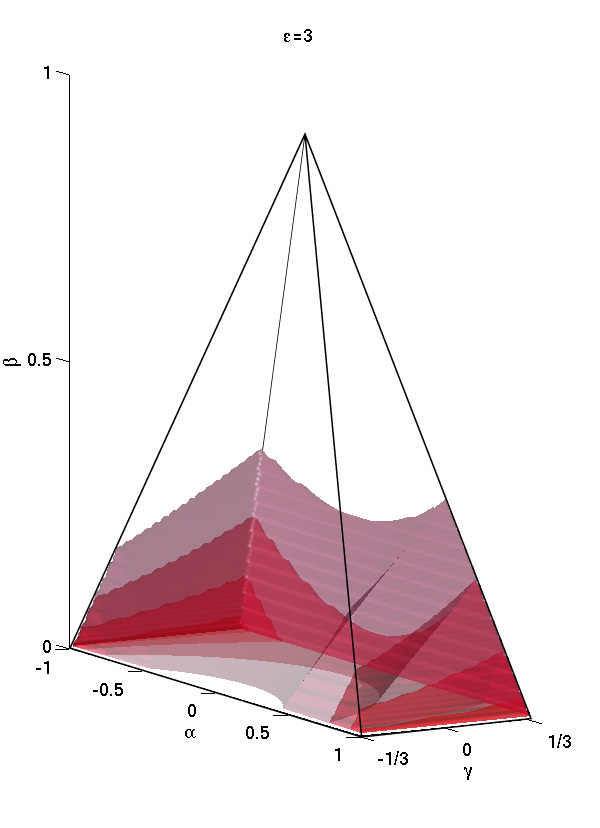}
\includegraphics[width=70mm]{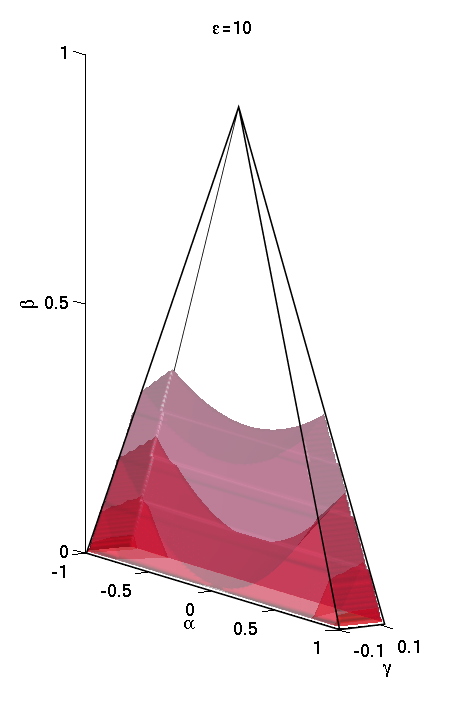}
\caption{Local $B$ model~\eqref{LocB}. The peaks at $\alpha=1$ correspond to $k_2\rightarrow 0$, while the peaks at $\alpha=-1$ correspond to $k_1\rightarrow 0$. For $\epsilon=1$ (i.e. equal triangle sizes $(k_1,k_2,K)$ and $(k_3,k_4,K)$) we see a more confined peaking at $\gamma=-1,\gamma=1$, i.e. $k_3\rightarrow 0, k_4\rightarrow 0$ respectively. We observe the peaking of the local B models to be somewhat orthogonal to the local A model.}
\label{fig:LocB}
\end{figure}

The local model is given by the reduced primordial trispectrum 
\begin{eqnarray}
\mathcal{T}^{\rm{loc}}_{\Phi}(\mathbf{k}_1,\mathbf{k}_2,\mathbf{k}_3,\mathbf{k}_4;\mathbf{K})=\mathcal{T}^{\rm{loc}}_{\Phi A}(\mathbf{k}_1,\mathbf{k}_2,\mathbf{k}_3,\mathbf{k}_4;\mathbf{K})+\mathcal{T}^{\rm{loc}}_{ \Phi B} (\mathbf{k}_1,\mathbf{k}_2,\mathbf{k}_3,\mathbf{k}_4;\mathbf{K})
\end{eqnarray}
where
\begin{eqnarray}
\mathcal{T}^{\rm{loc}}_{\Phi A}(\mathbf{k}_1,\mathbf{k}_2,\mathbf{k}_3,\mathbf{k}_4;\mathbf{K})&=&\frac{25}{9}\tau_{NL} P_{\Phi}(K)P_{\Phi}(k_1)P_{\Phi}(k_3)\\
\mathcal{T}^{\rm{loc}}_{ \Phi B} (\mathbf{k}_1,\mathbf{k}_2,\mathbf{k}_3,\mathbf{k}_4;\mathbf{K})&=&g_{NL}\left[P_{\Phi}(k_2)P_{\Phi}(k_3)P_{\Phi}(k_4)+P_{\Phi}(k_1)P_{\Phi}(k_2)P_{\Phi}(k_4)\right].
\end{eqnarray}
For single field inflation we have $\tau_{NL}=(\frac{6}{5}f_{NL})^2$. This relationship breaks down for multifield inflation (see \cite{Byrnes}). 
We can see clearly here that the local trispectrum is independent of the angle $\theta_4$, i.e. the zeroth mode of the local trispectrum is exactly the full local trispectrum. The primordial shapes for each of these expressions may be shown visually using the prescription described in the previous section and they are shown in Figures~\ref{fig:LocA} and~\ref{fig:LocB}. As expected for the local model the signal peaks in the corners. However, as is easily observable the `peaking' behaviour is somewhat orthogonal between the two models. Working in the Sachs-Wolfe approximation, where we replace the transfer function with a Bessel function,
\begin{eqnarray}
\Delta_l(k)=\frac{1}{3}j_l((\tau_0-\tau_{\rm{dec}})k),
\end{eqnarray}
the integral for the reduced trispectrum can be expressed in closed form. Setting $P_{\Phi}(k)=\Delta_{\Phi}k^{-3}$ we find
\begin{eqnarray}\label{LocA}
\mathcal{T}^{l_1 l_2\rm{loc}}_{l_3 l_4, A}(L)&=&\frac{25\tau_{NL}}{9} \frac{\Delta_{\Phi}^3}{3^4} \left(\frac{2}{\pi}\right)^5 h_{l_1 l_2 L}h_{l_3 l_4 L}\int r_1^2 d r_1 r_2^2 d r_2 K^{-1}dK k_1^{-1}dk_1
k_3^{-1}dk_3 I_{l_2}(2,r_1) I_{l_4}(2,r_1)\nonumber\\
&=&\frac{25\tau_{NL}}{9}  \frac{\Delta_{\Phi}^3}{3^4} \left(\frac{2}{\pi}\right)^5 h_{l_1 l_2 L}h_{l_3 l_4 L} \left( \frac{\pi}{2}\right)^2 I_L(-1,1) I_{l_1}(-1,1) I_{l_3}(-1,1)\nonumber\\
&=&\frac{25\tau_{NL}}{36} \pi^2\frac{\Delta_{\Phi}^3}{3^4} \left(\frac{2}{\pi}\right)^5 h_{l_1 l_2 L}h_{l_3 l_4 L} \left( \frac{1}{2 L(L+1)2 l_1(l_1+1)2 l_3(l_3+1)} \right)
\end{eqnarray}
where
\begin{eqnarray}
I_l (p,x)=\int k^p dk j_l(k)j_l(x k),
\end{eqnarray}
and we have used
\begin{eqnarray}\label{IdentitySpecial}
I_l(2,r)&=& \frac{\pi}{2 r^2}\delta(r-1) \\
I_l(-1,1)&=&\frac{1}{2l(l+1)}.
\end{eqnarray}
Similarly,
\begin{eqnarray}\label{LocB}
\mathcal{T}^{l_1 l_2 \rm{loc}}_{l_3 l_4, B}(L)&=&
 g_{NL}\frac{\pi^2}{4}\frac{\Delta_{\Phi}^3}{3^4} \left(\frac{2}{\pi}\right)^5 h_{l_1 l_2 L}h_{l_3 l_4 L} \left( \frac{1}{2 l_2(l_2+1)2 l_3(l_3+1)2 l_4(l_4+1)}+(l_1\leftrightarrow l_3)\right)
\end{eqnarray}

Next we propose a $constant$ model for the primordial trispectrum, analogous to the 
simplest model for the bispectrum. This is given by
\begin{eqnarray}\label{Constant}
\frac{1}{\Delta_{\Phi}^3 N}(k_1 k_2 k_3 k_4)^2 K\mathcal{T}_{\Phi,0}(k_1,k_2,k_3,k_4;{K})=S_{\mathcal{T}}(k_1,k_2,k_3,k_4,K)=1
\end{eqnarray}
with $N$ the normalisation factor of equation~\eqref{Constant} and the choice $\Delta_{\Phi}^3$ motivated by comparison to the local model.
Again, as for the local model, the primordial trispectrum is already a zero mode quantity with respect to angle $\theta_4$, i.e. $T=T_{,0}$. Using the Sachs Wolfe approximation the integral~\eqref{TrispRed2} can be written as
\begin{eqnarray}
\mathcal{T}^{l_1 l_2\rm{const}}_{l_3 l_4}(L)=\frac{\Delta_{\Phi}^3 N}{3^4 }\left(\frac{2}{\pi}\right)^5 h_{l_1 l_2 L}h_{l_3 l_4 L}\int dx x^2  dr_1 r_1^3 I_{l_1}(0,r_1) I_{l_2}(0,r_1) I_{l_3}(0,r_1 x) I_{l_4}(0,r_1 x) I_L(1,x)
\end{eqnarray}
where we write $r_2/r_1=x$. Now we can evaluate
\begin{eqnarray}
I_l(0,x)&=&\frac{\pi}{2(2l+1)} x^{-(l+1)}\qquad \mbox{for}\,\, x>1 \nonumber \\
&=&\frac{\pi}{2(2l+1)} x^{l}\qquad\qquad \,\mbox{for}\,\, x<1 \\
I_L(1,x)&=&\frac{\pi \Gamma(L+1)}{2\Gamma(L+3/2)} x^{-(L+2)}   {}_2F_1(\frac{1}{2},L+1;L+\frac{3}{2};x^{-2}) \qquad \, \mbox{for}\,\, x>1 \nonumber \\
&=&\frac{\pi \Gamma(L+1)}{2\Gamma(L+3/2)} x^{L}   {}_2F_1(\frac{1}{2},L+1;L+\frac{3}{2};x^{2})\qquad\qquad \quad \,\mbox{for}\,\, x<1
\end{eqnarray}
where $_2F_1$ is a generalised hypergeometric function. We can write $_2F_1$  in terms of a series expansion in the form
\begin{eqnarray}
_2F_1(a,b;c;z)=\sum_{n=0}^{\infty}\frac{(a)_n (b)_n}{(c)_n}\frac{z^n}{n!}.
\end{eqnarray}
where $(p)_n=\Gamma(p+n)/\Gamma(p)$.
The conditions for convergence, namely that this series converges for $c$ a non-negative integer with $|z|<1$, are satisfied in this case. Using this decomposition we find 
\begin{eqnarray}
\mathcal{T}^{l_1 l_2\rm{const}}_{l_3 l_4}(L)&=&\frac{\Delta_{\Phi}^3 N}{3^4 } h_{l_1 l_2 L}h_{l_3 l_4 L}\frac{1}{{\pi}}\frac{1}{(2l_1+1)(2l_2+1)(2l_3+1)(2l_4+1)}\sum_{n=0}^{\infty}\frac{\Gamma(1/2+n) \Gamma(L+1+n)}{\Gamma(L+3/2+n)}\frac{1}{n!} \nonumber\\
&&\times \Bigg[ \frac{1}{2n+3+l_3+l_4+L}\left( \frac{1}{\sum l_i+4}-\frac{1}{A_1}\right)+\frac{1}{2n+1+l_1+l_2+L}\left( \frac{1}{\sum l_i}+\frac{1}{A_1}\right) \nonumber\\
&& + \frac{1}{2n+3+l_1+l_2+L}\left( \frac{1}{\sum l_i+4}-\frac{1}{A_2}\right)+\frac{1}{2n+1+l_3+l_4+L}\left( \frac{1}{\sum l_i}+\frac{1}{A_2}\right)\Bigg]
\end{eqnarray}
Notice that this sum is still finite if the denominators $A_1=l_1+l_2-l_3-l_4+2$ or $A_2=l_3+l_4-l_1-l_2+2$ are zero since in those cases the respective numerators vanish. Alternatively we can integrate over the hypergeometric function directly and write the solution in the following closed form
\begin{eqnarray}
\mathcal{T}^{l_1 l_2\rm{const}}_{l_3 l_4}(L)&=&\frac{\Delta_{\Phi}^3 N}{3^4} h_{l_1 l_2 L}h_{l_3 l_4 L}\frac{1}{2{\pi}}\frac{1}{(2l_1+1)(2l_2+1)(2l_3+1)(2l_4+1)} B(L+1,1/2)\nonumber \\
&&\times\Bigg[\left( \frac{1}{\sum l_i+4}-\frac{1}{A_1}\right)B\left(\frac{C_1+2}{2},1\right) {}_3 F_2\left(\{\frac{C_1+2}{2},\frac{1}{2},L+1 \};\{ L+\frac{3}{2},\frac{C_1+4}{2}  \};1\right) \nonumber \\
&&+\left( \frac{1}{\sum l_i}+\frac{1}{A_1}\right)B\left(\frac{C_2}{2},1\right) {}_3 F_2\left(\{\frac{C_2}{2},\frac{1}{2},L+1 \};\{ L+\frac{3}{2},\frac{C_2+2}{2}  \};1\right) \nonumber\\
&&+\left( \frac{1}{\sum l_i+4}-\frac{1}{A_2}\right) B\left(\frac{C_2+2}{2},1\right) {}_3 F_2\left(\{\frac{C_2+2}{2},\frac{1}{2},L+1 \};\{ L+\frac{3}{2},\frac{C_2+4}{2}  \};1\right) \nonumber\\
&&+\left( \frac{1}{\sum l_i}+\frac{1}{A_2}\right)B\left(\frac{C_1}{2},1\right) {}_3 F_2\left(\{\frac{C_1}{2},\frac{1}{2},L+1 \};\{ L+\frac{3}{2},\frac{C_1+2}{2}  \};1\right) \Bigg]
\end{eqnarray}
where $B(x,y)$ denotes the beta function and $C_1=1+l_3+l_4+L , C_2=1+l_1+l_2+L$.
\par
\begin{figure}[htp]
\centering 
\includegraphics[width=120mm]{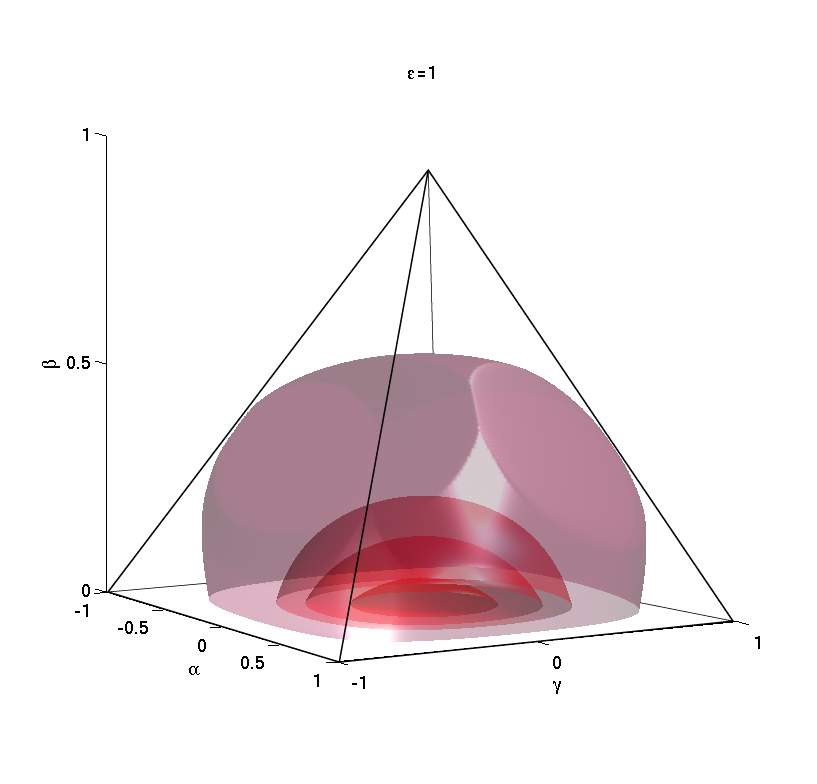}
\includegraphics[width=80mm]{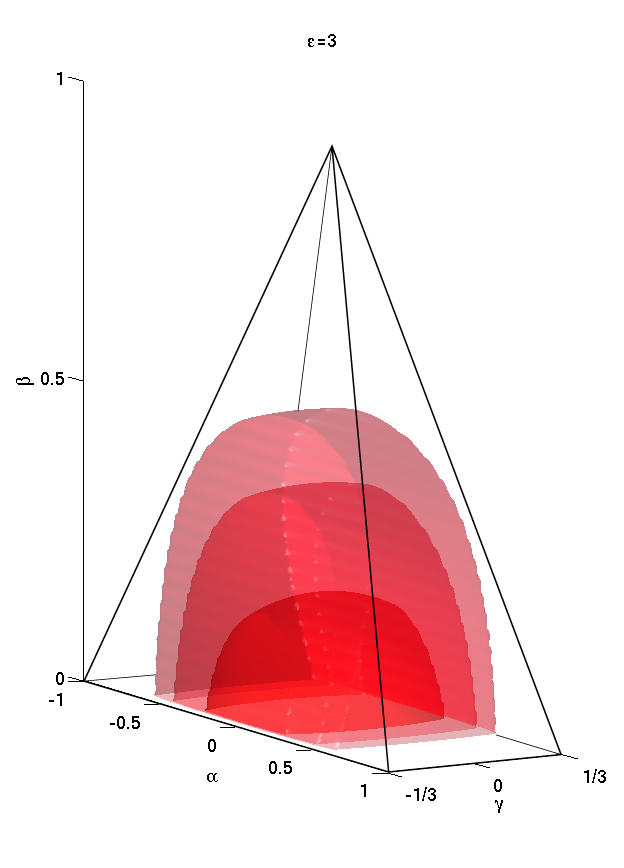}
\includegraphics[width=70mm]{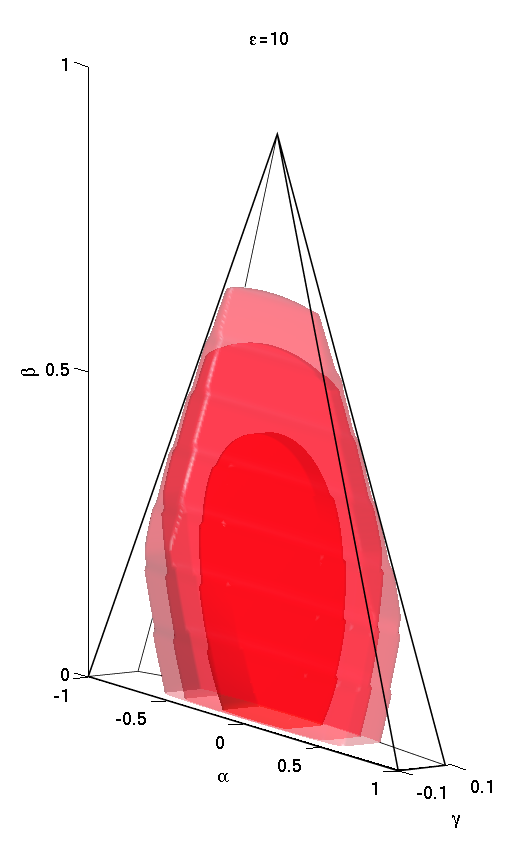}
\caption{Equilateral $c_1$ model~\eqref{c1}. The signal peaks at $\epsilon=1$ towards $\gamma=0,\alpha=0,\beta=0$, i.e. at $k_1=k_2=k_3=k_4=K/2$. For $\epsilon>1$ the signal similarly peaks for $k_1=k_2$, $k_3=k_4$ but since the triangles $(k_1,k_2,K)$ and $(k_3,k_4,K)$ are now unequal the peak position is less sharp and shifts to smaller values of $K$.}
\label{fig:equilc1}
\end{figure}
\begin{figure}[htp]
\centering 
\includegraphics[width=120mm]{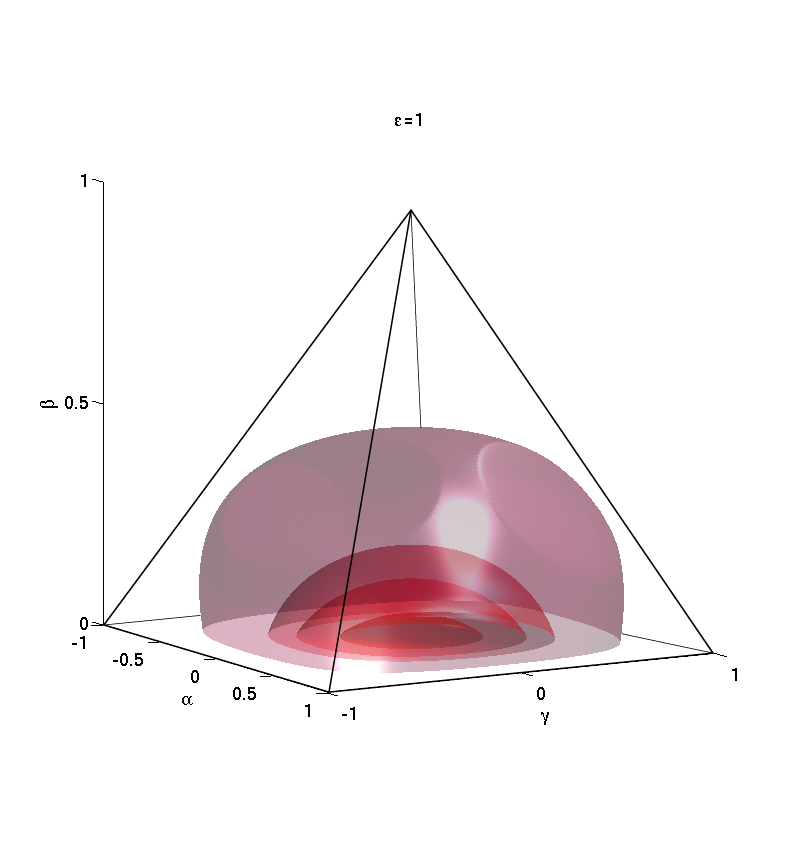}
\includegraphics[width=80mm]{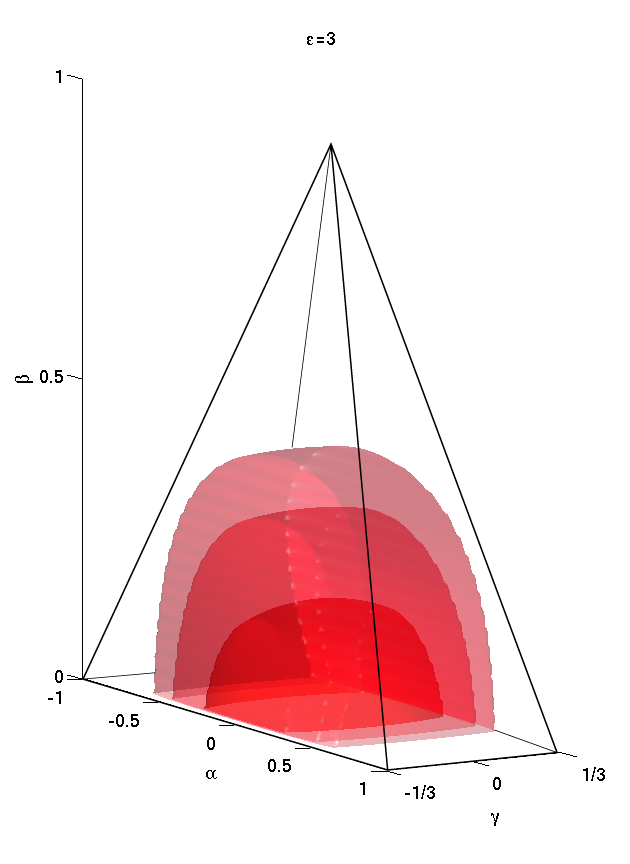}
\includegraphics[width=70mm]{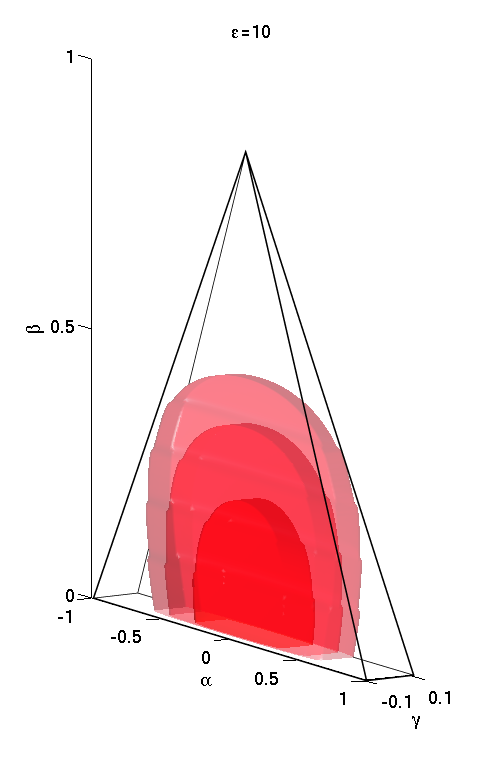}
\caption{Equilateral $s_1$ model~\eqref{s1}. As for the equilateral $c_1$ model the signal peaks for equal $k_i$ at $\epsilon$, while for $\epsilon>1$ the peak position becomes less sharp. As is clearly observable from the figures the equilateral $c_1$ and $s_1$ models are highly correlated.}
\label{fig:equils1}
\end{figure}

The equilateral shape has also received a lot of attention in the literature. As has been described in~\cite{aChen}, for the purposes of data analyses, there are two representative forms for the equilateral trispectra. These are given by the following shapes for the reduced trispectra
\begin{eqnarray}
S_{\mathcal{T}, c_1}^{\rm{equil}}(\mathbf{k}_1,\mathbf{k}_2,\mathbf{k}_3,\mathbf{k}_4;\mathbf{K})&\propto& K \frac{k_1 k_2 k_3 k_4}{(k_1 +k_2 +k_3 +k_4)^5}\label{c1}\\
S_{\mathcal{T}, s_1}^{\rm{equil}}(\mathbf{k}_1,\mathbf{k}_2,\mathbf{k}_3,\mathbf{k}_4;\mathbf{K})&\propto& \frac{k_1k_2k_3k_4 K^2}{(k_3+k_4 +K)^3}\bigg( \frac{1}{2(k_1+k_2+K)^3} +\frac{6(k_3+k_4 +K)^2}{(k_1+k_2+k_3+k_4)^5}\nonumber\\
&&+\frac{3(k_3+k_4 +K)}{(k_1+k_2+k_3+k_4)^4}+\frac{1}{(k_1+k_2+k_3+k_4)^3}\bigg)\label{s1}
\end{eqnarray}
where we use the notation $c_1$ and $s_1$ to correspond to~\cite{aChen}. These shapes are similar in most regions apart from the doubly squeezed limit ($k_3=k_4\rightarrow 0$). It has been observed that the first ansatz is factorisable by introducing the integral $1/M^n=(1/\Gamma(n))\int_0^{\infty} t^{n-1}e^{-M t} dt$ where $M=\sum k_i$. As we observe from Figures~\ref{fig:equilc1} and~\ref{fig:equils1} it is clear that the shapes for the two representative forms are highly correlated. Therefore, for the purposes of analysis of the equilateral model, it may only be necessary to consider the $c_1$ model.

\section{V. Mode decomposition}
Our goal is to represent an arbitrary non-separable reduced primordial trispectrum (zero mode) $\mathcal{T}_{\Phi,0}(k_1,k_2,k_3,k_4;K)$ or reduced CMB trispectrum $\mathcal{T}^{l_1 l_2}_{l_3 l_4}(L)$ on their respective wavenumber or multipole domains using a rapidly convergent mode expansion. We need to achieve this in a separable manner, in order to make tractable the five dimensional integrals ($\sim dk_1 dk_2 dk_3 dk_4 dK$) required for trispectrum estimation by breaking them down into products of one-dimensional integrals. In particular, this means that we wish to expand an arbitrary non-separable primordial (reduced) shape function in the form
\begin{eqnarray}\label{decomp}
S_{\mathcal{T}}(k_1,k_2,k_3,k_4,K)
&=&\sum_{p, r, s,u,v} \alpha_{p r s u v} q_p(k_1)q_r(k_2)q_s(k_3)q_u(k_4) r_v(K) 
\end{eqnarray}
where the $q_p, r_v$ are appropriate basis mode functions which are convergent and complete, that is, they span the space of all functions on the wavenumber domain. The differing notation, $q,r$ is due to the different ranges of the variables - $k_i\in [0,k_{\rm{max}}]$ but $K\in [0,2 k_{\rm{max}}]$. In the case of more general probes of non-Gaussianity this is easily extended to include the other Legendre modes of equation~\eqref{expansion} by writing
\begin{eqnarray}
S(k_1,k_2,k_3,k_4,K,\theta_4)=\sum_n S_{\mathcal{T} n}(k_1,k_2,k_3,k_4,K)P_n(\cos\theta_4)
\end{eqnarray}
where $S$ is the shape function applied to the full Legendre expansion~\eqref{expansion}. The shape function of the nth Legendre mode, $S_{\mathcal{T}n}$, may be decomposed as in equation~\eqref{decomp}.

We will present one method for finding the basis functions $q,r$ below. We will achieve this objective in stages. First, we create examples of one dimensional mode functions which are orthogonal and well-behaved over the full wavenumber (or multipole) domain. We then construct five dimensional products of these wavefunctions $q_p(k_1)q_r(k_2)q_s(k_3)q_u(k_4) r_v(K) \rightarrow \mathcal{Q}_m$. This creates a complete basis for all possible reduced trispectra on the given domain. By orthonormalising these product basis functions $\mathcal{Q}_m\rightarrow \mathcal{R}_m$ we obtain a rapid and convenient method for calculating the expansion coefficients $ \alpha_{p r s u v}$
(or $ \alpha_m$). Here we use bounded symmetric polynomials as a concrete implementation of this methodology. Of course as outlined in the case of the bispectrum in~\cite{Ferg3} there are alternatives to using the polynomials $ \mathcal{Q}_m,\mathcal{R}_m$ but the shortcoming of these alternatives is that either (i) they can lead to overshooting at the domain boundaries or (ii) the choice may compromise separability. However, it is possible that an alternative to the polynomial expansion may be desirable to improve the rate of convergence. These should be able to conveniently represent functions in a separable form, and should be derived explicitly for the domain.

\subsection{A. Domain and weight functions}
\begin{figure}[htp]
\centering 
\includegraphics[width=82mm]{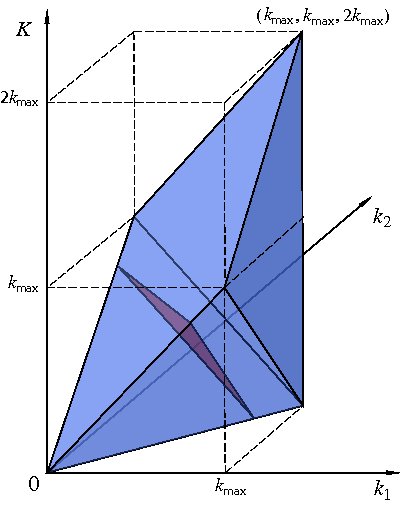}
\caption{`Tetrapiped' domain for allowed wavenumbers of the primordial reduced trispectrum $\mathcal{T}(k_1,k_2,k_3,k_4;K,\theta_4)$ imposed by the triangle created by $(k_1,k_2,K)$. There is a corresponding tetrapiped domain imposed by the triangle created by $(k_3,k_4,K)$. The region is an extension of the tetrapyd domain described in~\cite{Ferg3} due to the extended range of $K$. The same domain is valid for allowed multipole values $l_i,L$ in the case of the reduced CMB trispectrum $\mathcal{T}^{l_1 l_2}_{l_3 l_4}(L)$. The shaded area denotes the region described in Figure~\ref{fig:shape}.}
\label{fig:tetrapiped}
\end{figure}

In Fourier space, the primordial reduced trispectrum  zero mode $\mathcal{T}_{\Phi,0}(k_1,k_2,k_3,k_4;K)$ is defined when the wavevectors $\mathbf{k_1,k_2,K}$ and $\mathbf{k_3,k_4,K}$ close to form triangles subject to $\mathbf{k_1}+\mathbf{k_2}+\mathbf{K}=0=\mathbf{k_3}+\mathbf{k_4}-\mathbf{K}$. Each such triangle is uniquely defined by the lengths of the sides $k_1,k_2,K$ and $k_3,k_4,K$. In terms of these wavenumbers, the triangle conditions restricts the allowed combinations into a region defined by
\begin{eqnarray}\label{Domain1}
k_1\leq K+k_2\,\,\rm{for}\,\, k_1\geq k_2,K,\quad \rm{or}\quad k_2\leq K+k_1\,\,\rm{for}\,\, k_2\geq k_1,K,\quad \rm{or}\quad K\leq k_1+k_2\,\,\rm{for}\,\, K\geq k_1,k_2
\end{eqnarray}
and
\begin{eqnarray}\label{Domain2}
k_3\leq K+k_4\,\,\rm{for}\,\, k_3\geq k_4,K,\quad \rm{or}\quad k_4\leq K+k_3\,\,\rm{for}\,\, k_4\geq k_3,K,\quad \rm{or}\quad K\leq k_3+k_4\,\,\rm{for}\,\, K\geq k_3,k_4.
\end{eqnarray}
Since the wavenumber $K$ is common to both triangles the region is a product of the tetrahedral domains defined by the conditions~\eqref{Domain1} and~\eqref{Domain2}. Considering each region individually we note that they each describe a regular tetrahedron for $k_1+k_2+K<2k_{\rm{max}}$ (or $k_3+k_4+K<2k_{\rm{max}}$). However, motivated by issues of separability and observation, it is more natural to extend the domain out to values given by a maximum wavenumber $k_{\rm{max}}$. In particular, we have $k_i<k_{\rm{max}}$ and $K<2k_{\rm{max}}$. In each case the allowed region is a hexahedron formed by the intersection of a tetrahedron and a rectangular parallelepiped. For brevity we will denote this configuration as a $tetrapiped$. This region is an extension of the tetrapyd referred to in~\cite{Ferg3} due to the extended range of $K$ and is shown in Figure~\ref{fig:tetrapiped}.

\par
In order to integrate functions $f(k_1,k_2,k_3,k_4,K)$ over the tetrapiped domains, which we denote $\mathcal{V}_{\mathcal{T}}$, we note the presence of $K$ in both regions. We find that the integration is given explicitly by
\begin{eqnarray}
\mathcal{I}[f]&\equiv& \int_{\mathcal{V}_{\mathcal{T}}}f(k_1,k_2,k_3,k_4,K) \omega(k_1,k_2,k_3,k_4,K)d \mathcal{V}_{\mathcal{T}}\\
&=&k_{\rm{max}}^5\Big\{   \int_0^{1/2}dt\left(\int_0^t ds \int_{t-s}^{t+s}dx+\int_t^{1-t}  ds \int_{s-t}^{t+s}dx+\int_{1-t}^1 ds \int_{s-t}^1 dx\right)\nonumber\\
&&\times \left(\int_0^t dy \int_{t-y}^{t+y}dz+\int_t^{1-t}  dy \int_{y-t}^{t+y}dz+\int_{1-t}^1 dy \int_{y-t}^1 dz\right) FW \nonumber\\
&&+ \int_{1/2}^1 dt  F W\left(\int_0^{1-t} ds \int_{t-s}^{t+s}dx+\int^t_{1-t}  ds \int_{t-s}^{t+s}dx+\int_{t}^1 ds \int_{s-t}^1 dx\right) \nonumber\\ 
&&\times \left(\int_0^{1-t} dy \int_{t-y}^{t+y}dz+\int^t_{1-t}  dy \int_{t-y}^{t+y}dz+\int_{t}^1 dy \int_{y-t}^1 dz\right)+\int_1^2 dt \int_{t-1}^1 ds\int_{t-s}^1 dx \int_{t-1}^1 dy \int_{t-y}^1    dz F  W  \nonumber \Big\},
\end{eqnarray}
where  $\omega(k_1,k_2,k_3,k_4,K)$ is an appropriate weight function and we have made the transformation $t=K/k_{\rm{max}}, s=k_1/k_{\rm{max}},x=k_2/k_{\rm{max}}, y=k_3/k_{\rm{max}}, z=k_4/k_{\rm{max}}$ with $F(s,x,y,z,t)=f(k_{\rm{max}}\times(s,x,y,z,t))$ and $W(s,x,y,z,t)=\omega(k_{\rm{max}}\times(s,x,y,z,t))$. For integrals over the product of two functions $f$ and $g$ we can define the inner product $\langle f,g \rangle=\mathcal{I}[f g]$. This inner product essentially defines a Hilbert space of possible shape functions in the domain. The total volume of the domain is given by $\mathcal{I}[1]=k_{\rm{max}}^5/3$. Initially we will restrict attention to the case of weight $\omega=1$. However, it is important to incorporate a weight function for a variety of reasons which we will discuss later.
\par
Analysis of the CMB `extra'-reduced trispectrum, $t^{l_1 l_2}_{l_3 l_4}(L)$, is more straightforward than in the primordial case. This is because the CMB trispectrum, being defined on a two sphere, is an explicitly five dimensional quantity and therefore is defined completely in terms of the multipoles. We note here that the quantity $p^{l_1 l_2}_{l_3 l_4}(L)=t^{l_1 l_2}_{l_3 l_4}(L)+t^{l_2 l_1}_{l_3 l_4}(L)+t^{l_1 l_2}_{l_4 l_3}(L)+t^{l_2 l_1}_{l_4 l_3}(L)$ is probably a more elegant expression for this analysis since it is more symmetric, whilst being defined on the same domain and being subject to the same weighting over the domain. Nonetheless, we proceed in this paper with analysis of $t^{l_1 l_2}_{l_3 l_4}(L)$ leaving exploration of this minor issue to an upcoming paper~\cite{Regan2}. As for the primordial case, we extend the tetrahedral domains to include multipoles out to $l_i, L/2<l_{\rm{max}}$. The respective tetrapiped domains for the extra-reduced trispectrum becomes the discrete $\{l_1,l_2,l_3,l_4,L\}$ satisfying
\begin{eqnarray}
&&l_1,l_2,l_3,l_4,L/2 <l_{\rm{max}},\,\, l_i,L\in \mathbb{N},\nonumber\\
&&l_1\leq l_2+L \,\, \rm{for}\,\, l_1\geq l_1,L, +\,\,\rm{cyclic \,perms}\nonumber\\
&&l_3\leq l_4+L \,\, \rm{for}\,\, l_3\geq l_4,L, +\,\,\rm{cyclic \,perms}\nonumber\\
&&l_1+l_2+L=2 n_1,\quad l_3+l_4+L=2 n_2,\quad n_1,n_2\in \mathbb{N}.
\end{eqnarray}
In multipole space, we will be primarily dealing with a summation over all possible  $\{l_1,l_2,l_3,l_4,L\}$  combinations in the correlator $\mathcal{C(T,T')}$. The appropriate weight function in the sum from~\eqref{ExtraTrispRed} is then
\begin{eqnarray}
\omega(l_1,l_2,l_3,l_4,L)=h_{l_1 l_2 L}^2 h_{l_3 l_4 L}^2.
\end{eqnarray}
A straightforward continuum version of this can be deduced by comparison of this `weight' formula to the bispectrum multipole weight function in~\cite{Ferg3}. Similarly to that analysis, we should eliminate a scaling in this weight such that the overall weight becomes very nearly constant. We can do this by using a separable weight function as
\begin{eqnarray}
\omega_s(l_1 l_2 l_3 l_4 L)=\frac{\omega(l_1,l_2,l_3,l_4,L)}{(2 l_1+1)^{1/3}(2 l_2+1)^{1/3}(2 l_3+1)^{1/3}(2 l_4+1)^{1/3}(2 L+1)^{2/3}}.
\end{eqnarray}
We note that there is also a freedom to absorb an arbitrary separable function $v_l$ into the weight fucntions. If we define a new weight $\overline{\omega}$ in the estimator as
\begin{eqnarray}
\overline{\omega}_{l_1 l_2 l_3 l_4 L}=\omega_{l_1 l_2 l_3 l_4 L}/(v_{l_1}v_{l_2}v_{l_3}v_{l_4}v_L)^2,
\end{eqnarray}
then we must rescale the estimator functions by the factor $v_{l_1}v_{l_2}v_{l_3}v_{l_4}v_L$. The important point is to use both the weight $\overline{\omega}$ and the estimator rescaling throughout the analysis, including the generation of appropriate orthonormal mode functions.

\subsection{B. Orthogonal polynomials on the domain}
We now construct some concrete realisations of mode functions which are orthogonal on the domain $\mathcal{V}_{\mathcal{T}}$ and which have the form required for a separable expansion. First, we will generate one-dimensional orthogonal polynomials $q_p(s), r_v(t)$ for unit weight $\omega=1$. Considering functions $q_p(s)$ depending only on the $s$-coordinate\footnote{We can consider $s$ as corresponding to any of the $k_i$.}, we integrate over the $t,x,y,z$-directions to yield the weight functions $\overline{\omega}(s)$ for $s\in [0,1]$ (for simplicity we take $k_{\rm{max}}=1$):
\begin{eqnarray}
\overline{\omega}(s)=s-s^3+\frac{5}{12}s^4,\quad \rm{with}\quad \mathcal{I}[f]=\int_0^1 f(s) \overline{\omega}(s) ds.
\end{eqnarray}
Therefore, the moments for each power of $s$ become
\begin{eqnarray}
\overline{\omega}_n\equiv\mathcal{I}[s^n]&=&\frac{1}{n+2}+\frac{5}{12(n+5)}-\frac{1}{n+4}\nonumber\\
&=&\frac{5n^2+54 n+160}{12(n+2)(n+4)(n+5)}.
\end{eqnarray}
For functions $r_v(t)$ (where $t$ corresponds to the K coordinate), we integrate over the $s,x,y,z$- directions to yield the weight functions $\hat{\omega}(t)$ for $t\in[0,2]$:
\begin{eqnarray}
\hat{\omega}(t)&=&\left(\frac{t}{2}(4-3t)\right)^2\,\,\rm{for}\,\, t\in [0,1]\nonumber\\
&=&\frac{(t-2)^4}{4}\qquad\quad \rm{for}\,\, t\in [1,2], \quad \rm{with}\quad \mathcal{I}[f]=\int_0^2 f(t) \hat{\omega}(t) dt.
\end{eqnarray}
With this choice of weight the moments of each power of $t$ become
\begin{eqnarray}
\hat{\omega}_n\equiv\mathcal{I}[t^n]&=&\frac{n^2+15n+68}{4(n+3)(n+4)(n+5)}+\frac{768\times 2^n-744-474 n-131n^2-18n^3-n^4}{4(n+1)(n+2)(n+3)(n+4)(n+5)}.
\end{eqnarray}
From these moments we can create orthogonal polynomials using the generating functions,
\begin{eqnarray}
q_n(s)=\frac{1}{\mathcal{N}_1} \left| \begin{array}{ccccc}
1/3 & 73/360 & 1/7&\dots&\overline{\omega}_n \\
73/360 & 1/7&367/3360&\dots&\overline{\omega}_{n+1}\\
\dots&\dots&\dots&\dots&\dots\\
\overline{\omega}_{n-1}&\overline{\omega}_{n}&\overline{\omega}_{n+1}&\dots&\overline{\omega}_{2n-1}\\
1&s&s^2&\dots&s^n \end{array} \right|
\end{eqnarray}
and
\begin{eqnarray}
r_n(t)=\frac{1}{\mathcal{N}_2} \left| \begin{array}{ccccc}
1/3 & 7/30 & 4/21&\dots&\hat{\omega}_n \\
7/30 & 4/21&73/420&\dots&\hat{\omega}_{n+1}\\
\dots&\dots&\dots&\dots&\dots\\
\hat{\omega}_{n-1}&\hat{\omega}_{n}&\hat{\omega}_{n+1}&\dots&\hat{\omega}_{2n-1}\\
1&t&t^2&\dots&t^n \end{array} \right|,
\end{eqnarray}
where we choose the normalisation factors $\mathcal{N}_1,\mathcal{N}_2$ such that $\mathcal{I}[q_n^2]=1$ and $\mathcal{I}[r_n^2]=1$ for all $n\in \mathbb{N}$, that is such that the $q_n(s)$ (or $r_n(t)$) are orthonormal
\begin{eqnarray}
\langle q_n,q_p\rangle\equiv\mathcal{I}[q_n q_p]&=&\int_{\mathcal{V}_{\mathcal{T}}}q_n(s)q_p(s) d\mathcal{V}_{\mathcal{T}}=\delta_{np},\\
\langle r_v,r_u\rangle\equiv\mathcal{I}[r_v r_u]&=&\int_{\mathcal{V}_{\mathcal{T}}}r_v(t)r_v(t) d\mathcal{V}_{\mathcal{T}}=\delta_{vu}.
\end{eqnarray}
The first few orthonormal polynomials on the domain $\mathcal{V}_{\mathcal{T}}$ are explicitly
\begin{eqnarray}
q_0(s)&=&\sqrt{3}\nonumber\\
q_1(s)&=&7.16103(-0.608333+s)\nonumber\\
q_2(s)&=&7.76759-33.2061 s+29.0098 s^2,\nonumber\\
q_3(s)&=&-11.7911+93.1318s-194.111s^2+116.964 s^3 ,\dots
\end{eqnarray}
and
\begin{eqnarray}
r_0(t)&=&\sqrt{3}\nonumber\\
r_1(t)&=&6.06977(-0.7+t)\nonumber\\
r_2(t)&=&7.65066-24.1493t+16.1942t^2\nonumber\\
r_3(t)&=&-12.2182+63.4315 t-91.6438 t^2+38.7091 t^3,\dots
\end{eqnarray}
\par
\begin{figure}[htp]
\centering 
\includegraphics[width=152mm]{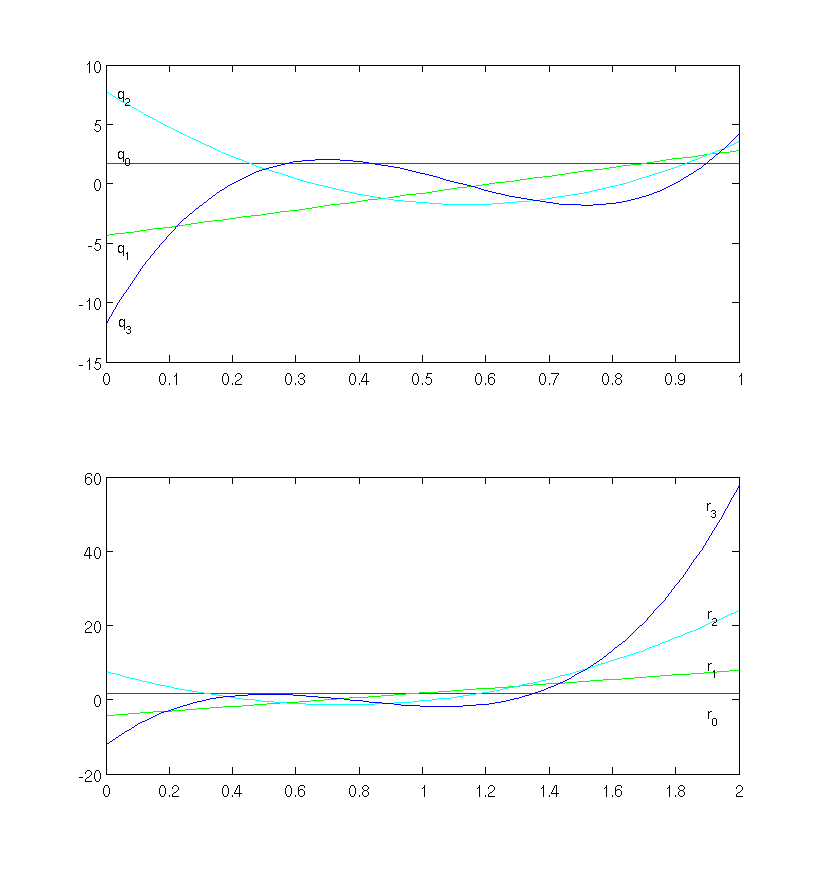}
\caption{The orthonormal one-dimensional eigenmodes $q_n, r_v$ plotted on their respective domains for $n,v=0,1,2,3$. The $q_n, r_v$ plotted are calculated for unit weight $\omega=1$ on the domain $\mathcal{V}_{\mathcal{T}}$. The shape of these eigenmodes alters for different choices of the weighting. }
\label{fig:orthoModes}
\end{figure}

We note that the $q_n,r_v$'s are only orthogonal in one dimension (e.g. $\langle q_n (s)r_v(t)\rangle\neq \delta_{n v}$ and $\langle q_n (s)q_m(x)\rangle\neq \delta_{n m}$ in general). However, as product functions of $t,s,x,y$ and $z$ they form an independent and well-behaved basis which we will use to construct orthonormal five-dimensional eigenfunctions. In practice the $q_n$'s and $r_v$'s remain the primary calculation tools, notably in performing separable integrations. In Figure~\ref{fig:orthoModes} we plot the first few $q_n$ and $r_v$'s.
\par
Now we turn to the polynomials $\overline{q}(s)$ and $\overline{r}(t)$ which are orthonormal on the multipole domain. Using the weight function $\omega=1$ we find the same polynomials as above. For the scaled weight function $\omega_s$ the polynomials will, of course, differ. While using either polynomial sets would suffice as independent basis functions on the multipole domain, the use of correctly weighted functions lead to improvements in the immediate orthogonality of the derived five dimensional polynomial sets. For definiteness we take $(t,s,x,y,z)\times l_{\rm{max}}=(L,l_1,l_2,l_3,l_4)$. The generating function is obtained as above but now using the moments $\overline{\omega}_n\equiv \mathcal{I}[s^n]=\int \omega(t,s,x,y,z)s^n d\mathcal{V}_{\mathcal{T}} $ and $\hat{\omega}_n\equiv \mathcal{I}[t^n]=\int \omega(t,s,x,y,z)t^n d\mathcal{V}_{\mathcal{T}}$.

\subsection{C. Five-dimensional basis functions}
We can represent arbitrary (reduced) trispectra Legendre modes on the domain $\mathcal{V}_{\mathcal{T}}$ using a suitable set of independent basis functions formed from products $q_p(s)q_r(x)q_s(y)q_u(z)r_v(t)$. (Here again we take $t=K/k_{\rm{max}},s=k_1/k_{\rm{max}},x=k_2/k_{\rm{max}},y=k_3/k_{\rm{max}},z=k_4/k_{\rm{max}}$ or $t=L/l_{\rm{max}}$, etc.) We denote the $5D$ basis function as
\begin{eqnarray}
\mathcal{Q}_m(t,s,x,y,z)= q_p(s)q_r(x)q_s(y)q_u(z) r_v(t). 
\end{eqnarray}
We can order these products linearly with a single index $m$ in a similar manner to that described in~\cite{Ferg3} for the bispectrum.
\par
While the $Q_m$'s by construction are an independent set of five-dimensional functions on the domain $\mathcal{V}_{\mathcal{T}}$, they are not in general orthogonal. To construct an orthonormal set $\mathcal{R}_m$ from the $\mathcal{Q}_m$ we perform an iterative Gram-Schmidt orthogonalisation process such that
\begin{eqnarray}
\langle \mathcal{R}_n \mathcal{R}_m\rangle=\delta_{n m}.
\end{eqnarray}
In particular, we form the Gram matrix $\Gamma=(\langle  \mathcal{Q}_n \mathcal{Q}_m \rangle )$ which needs to be factorised as $\Gamma=\Lambda^T \Lambda$ where $\Lambda=(\langle  \mathcal{Q}_n \mathcal{R}_m \rangle )$ is triangular. This process is described in more detail in~\cite{Ferg3}.

\subsection{D. Mode decomposition of the trispectrum}
Having formed the orthonormal basis $\{\mathcal{R}_m\}$ we consider an arbitrary primordial reduced trispectrum (zero mode) $\mathcal{T}_{,0}(k_1,k_2,k_3,k_4;K)$ described by the shape function $S_{\mathcal{T}}$ and decompose it as follows
\begin{eqnarray}
S_{\mathcal{T}}(k_1,k_2,k_3,k_4,K)=\sum_{m} \alpha_{m}^{\mathcal{R}}\mathcal{R}_m (t,s,x,y,z)
\end{eqnarray}
where the expansion coefficients $\alpha_{m}^{\mathcal{R}}$ are given by
\begin{eqnarray}
\alpha_{m}^{\mathcal{R}}=\langle \mathcal{R}_m,S_{\mathcal{T}}\rangle=\int_{\mathcal{V}_{\mathcal{T}}}\mathcal{R}_m S_{\mathcal{T}}\omega d\mathcal{V}_{\mathcal{T}}
\end{eqnarray}
and $k_{\rm{max}}(t,s,x,y,z)=(K,k_1,k_2,k_3,k_4)$ on the domain $\mathcal{V}_{\mathcal{T}}$ defined in~\eqref{Domain1} and \eqref{Domain2}. In practice, we must always work with partial sums up to a given $N=n_{\rm{max}}$ with
\begin{eqnarray}
S_{\mathcal{T}}^N=\sum_{m=0}^N  \alpha_{m}^{\mathcal{R}}\mathcal{R}_m (t,s,x,y,z), \quad S_{\mathcal{T}}=\lim_{N\rightarrow \infty}S_{\mathcal{T}}^N.
\end{eqnarray}
Given the complete orthonormal basis $\mathcal{R}_m$, Parseval's theorem for the integrated product of two functions implies
\begin{eqnarray}
\langle S_{\mathcal{T} },S'_{\mathcal{T} }\rangle=\int_{\mathcal{V}_{\mathcal{T}}} S_{\mathcal{T} }S'_{\mathcal{T} }\omega d\mathcal{V}_{\mathcal{T}}=\lim_{N\rightarrow \infty}\sum_{m=0}^N \alpha_{m}^{\mathcal{R}}\alpha_{m}^{\mathcal{R}'}
\end{eqnarray}
which for the square of a mode $S_{\mathcal{T} }$ yields the sum of the squares of the expansion coefficients, $\mathcal{I}[S_{\mathcal{T} }^2]=\sum_m \alpha_{m}^{\mathcal{R}\,\,2}$.
\par
In order to accomplish the goal of a general separable expansion, we must transform backwards from the orthogonal sum $\mathcal{R}_m$ into an expansion over the separable product functions $\mathcal{Q}_m$ through
\begin{eqnarray}
S_{\mathcal{T}}^N=\sum_{m=0}^N  \alpha_{m}^{\mathcal{Q}}\mathcal{Q}_m (t,s,x,y,z).
\end{eqnarray}
The $ \alpha_{m}^{\mathcal{Q}}$ can be obtained from the  $ \alpha_{m}^{\mathcal{R}}$ via
\begin{eqnarray}
 \alpha_{m}^{\mathcal{Q}}=\sum_{p=0}^N (\lambda^T)_{m p} \alpha_{p}^{\mathcal{R}}
\end{eqnarray}
where the transformation matrix $\lambda_{np}$ was defined above. Using the inverse relation $ \alpha_{m}^{\mathcal{R}}=\sum_{p=0}^N (\lambda^T)_{m p}^{-1} \alpha_{p}^{\mathcal{Q}}$ we find that the matrices $\Gamma$ and $\Lambda$ are related by
\begin{eqnarray}
\left(\gamma^{-1}\right)_{np}=\sum_{r}^N \left(\lambda^T\right)_{n r} \lambda_{r p}.
\end{eqnarray}
This implies that 
\begin{eqnarray}
\langle S_{\mathcal{T}}^N,S_{\mathcal{T}}^{N}\rangle =\sum_{m=0}^N \alpha_{m}^{\mathcal{R}\,\,2}=\sum_{m=0}^N \sum_{p=0}^N \alpha_{m}^{\mathcal{Q}}\gamma_{m p}\alpha_{p}^{\mathcal{Q}}.
\end{eqnarray}
As we have already noted the separable $\mathcal{Q}_m$ expansion is most useful for practical calculations. However, its coefficients must be constructed from the orthonormal $\mathcal{R}_m$.
\par
We can expand the CMB extra-reduced trispectrum $t^{l_1 l_2}_{l_3 l_4}(L)$ at late times using the same polynomials. However, as noted previously the CMB trispectrum is an explicitly five dimensional quantity and, as such we do not require the extra step of finding the zero mode of the Legendre series expansion. In particular, the appropriate expansion is of the form $t^{l_1 l_2}_{l_3 l_4}(L)=\sum_m \overline{\alpha}_m^{\mathcal{R}}\mathcal{R}_m(t,s,x,y,z)(=\sum_m \overline{\alpha}_m^{\mathcal{Q}}\mathcal{Q}_m(t,s,x,y,z))$.

\section{VI. Measures of $T_{NL}$}
\subsection{A. Primordial estimator}
We have obtained related mode expansions for a general primordial shape function, one with the orthonormal basis $\mathcal{R}_m$ and the other with the separable basis functions $\mathcal{Q}_m$. Substitution of the (reduced) separable form into the expression for the `extra'-reduced trispectrum~\eqref{ExtraTrispRed} offers an efficient route to its direct calculation through
\begin{eqnarray}
t^{l_1 l_2}_{l_3 l_4}(L)&=&N \Delta_{\Phi}^3 \left(\frac{2}{\pi}\right)^5  \int r_1^2 dr_1 r_2^2 dr_2 dk_1 dk_2 dk_3 dk_4 dK K\\
&&\times  \sum_{m}\alpha_{m}^{\mathcal{Q}} \mathcal{Q}_m (k_1,k_2,k_3,k_4,K) j_L(K r_1) j_L(K r_2)[j_{l_1}(k_1 r_1)\Delta_{l_1}(k_1)][j_{l_2}(k_2 r_1)\Delta_{l_2}(k_2)]\nonumber\\
&&\times[j_{l_3}(k_3 r_2)\Delta_{l_3}(k_3)][j_{l_4}(k_4 r_2)\Delta_{l_4}(k_4)] \nonumber\\
&=&N \Delta_{\Phi}^3 \sum_{m}     \alpha_{m}^{\mathcal{Q}} \int r_1^2 dr_1 r_2^2 dr_2\mathcal{Q}^{l_1 l_2 l_3 l_4 L}_{m} (r_1,r_2)\nonumber
\end{eqnarray}
where
\begin{eqnarray}
\mathcal{Q}^{l_1 l_2 l_3 l_4 L}_{m} (r_1,r_2)=     q_p^{l_1}(r_1)q_r^{l_2}(r_1)q_s^{l_3}(r_2)q_{u}^{l_4}(r_2)r_v^{L}(r_1,r_2)
\end{eqnarray}
with
\begin{eqnarray}
q_p^{l}(r)&=&\frac{2}{\pi}\int dk q_p(k) \Delta_{l}(k)j_{l}(k r),\nonumber\\
r_v^{L}(r_1,r_2)&=&\frac{2}{\pi}\int dK K r_v(K) j_L(K r_1) j_L(K r_2).
\end{eqnarray}
Next, we note from~\eqref{TotalRedTrisp} that
\begin{eqnarray}
\mathcal{T}_{l_1 m_1 l_2 m_2 l_3 m_3 l_4 m_4}=\sum_{L M}(-1)^M  \left( \begin{array}{ccc}
l_1 & l_2 & L \\
m_1 &m_2 & -M \end{array} \right)    \left( \begin{array}{ccc}
l_3 & l_4 & L \\
m_3 & m_4 & M \end{array} \right)\mathcal{T}^{l_1 l_2}_{l_3 l_4}(L).
\end{eqnarray}
Therefore, using these formulae in the estimator~\eqref{Estimator} (we omit the normalisation factor $N_T$ here and return to it later in the section) we find
\begin{eqnarray}\label{EstimatorPrim1}
\mathcal{E}&=&12 \sum_{l_i m_i} \mathcal{T}_{l_1 m_1 l_2 m_2 l_3 m_3 l_4 m_4}\frac{\left(a_{l_1 m_1}a_{l_2 m_2}a_{l_3 m_3}a_{l_4 m_4}\right)_c}{C_{l_1}C_{l_2}C_{l_3}C_{l_4}}\nonumber\\
\implies \mathcal{E}&=&\sum_{l_i m_i}\sum_{LM}12 N \Delta_{\Phi}^3\Bigg[\left(\int d\hat{n}_1 Y_{l_1 m_1}(\hat{n}_1)Y_{l_2 m_2}(\hat{n}_1)Y_{L M}^*(\hat{n}_1)\right)\left(\int d\hat{n}_2 Y_{l_3 m_3}(\hat{n}_2)Y_{l_4 m_4}(\hat{n}_2)Y_{L M}(\hat{n}_2)\right)\nonumber\\
&&\times \sum_{m}  \alpha_{ m}^{\mathcal{Q}} \int r_1^2 dr_1 r_2^2 dr_2\mathcal{Q}^{l_1 l_2 l_3 l_4 L}_{m} (r_1,r_2)\Big] \frac{\left(a_{l_1 m_1} a_{l_2 m_2} a_{l_3 m_3} a_{l_4 m_4} \right)_c}{C_{l_1}C_{l_2}C_{l_3}C_{l_4}}.
\end{eqnarray}
Now using the notation $\mathcal{E}=\mathcal{E}_{\rm{tot}}-\mathcal{E}_{\rm{uc}}$ where $\rm{tot}$ refers to $a_{l_1 m_1} a_{l_2 m_2} a_{l_3 m_3} a_{l_4 m_4}$ and $uc$ refers to $\left(a_{l_1 m_1} a_{l_2 m_2} a_{l_3 m_3} a_{l_4 m_4} \right)_{\rm{uc}}$ in place of $\left(a_{l_1 m_1} a_{l_2 m_2} a_{l_3 m_3} a_{l_4 m_4} \right)_c$, we find
\begin{eqnarray}
\mathcal{E}_{\rm{uc}}
&=&12 N\Delta_{\Phi}^3\sum_{ m}\alpha_{m}^{\mathcal{Q}}\int d\hat{n}_1 d\hat{n}_2\int dr_1 dr_2 r_1^2 r_2^2 N_v(\hat{n}_1,\hat{n}_2,r_1,r_2) \Bigg(M^{\rm{uc}}_{p r}(\hat{n}_1,\hat{n}_1,r_1,r_1)M^{\rm{uc}}_{s u }(\hat{n}_2,\hat{n}_2,r_2,r_2)\nonumber\\
&&+M^{\rm{uc}}_{p s}(\hat{n}_1,\hat{n}_2,r_1,r_2)M^{\rm{uc}}_{r u }(\hat{n}_1,\hat{n}_2,r_1,r_2)+M^{\rm{uc}}_{p u }(\hat{n}_1,\hat{n}_2,r_1,r_2)M^{\rm{uc}}_{r s}(\hat{n}_1,\hat{n}_2,r_1,r_2)\Bigg)
\end{eqnarray}
and
\begin{eqnarray}
\mathcal{E}_{\rm{tot}}&=&12 N\Delta_{\Phi}^3\sum_{ m}\alpha_{m}^{\mathcal{Q}}\int d\hat{n}_1 d\hat{n}_2\int dr_1 dr_2 r_1^2 r_2^2 M_{p}(\hat{n}_1,r_1)M_{r}(\hat{n}_1,r_1)M_{s}(\hat{n}_2,r_2)M_{u }(\hat{n}_2,r_2)N_v(\hat{n}_1,\hat{n}_2,r_1,r_2),\nonumber\\
\end{eqnarray}
where
\begin{eqnarray}
M^{\rm{uc}}_{p s}(\hat{n}_1,\hat{n}_2,r_1,r_2)&=&\sum_{l_1 m_1}
\frac{Y_{l_1 m_1}(\hat{n}_1)Y_{l_1 m_1}^*(\hat{n}_2)q_p^{l_1}(r_1) q_s^{l_1}(r_2)}{C_{l_1}} ,\nonumber\\
M_{p}(\hat{n}_1,r_1)&=&\sum_{l_1 m_1}\frac{Y_{l_1 m_1}(\hat{n}_1) a_{l_1 m_1}q_p^{l_1}(r_1)}{C_{l_1}},\nonumber\\
N_v(\hat{n}_1,\hat{n}_2,r_1,r_2)&=&\sum_{L M}Y_{L M}^* (\hat{n}_1) Y_{LM}(\hat{n}_2) r_v^L (r_1,r_2).
\end{eqnarray}
We can summarise these results (substituting back in $N_T$) as
\begin{eqnarray}\label{EstimatorPrim2}
\mathcal{E}&=&\frac{12 N\Delta_{\Phi}^3}{N_T}\sum_{m}\alpha_{m}^{\mathcal{Q}}\int d\hat{n}_1 d\hat{n}_2\int dr_1 dr_2 r_1^2 r_2^2 \mathcal{M}_{m}^{\mathcal{Q}}(\hat{n}_1,\hat{n}_2,r_1,r_2)\nonumber\\
&=&\frac{N\Delta_{\Phi}^3}{N_T}\sum_{m}\alpha_{m}^{\mathcal{Q}}\beta_{m}^{\mathcal{Q}}.
\end{eqnarray}
with
\begin{eqnarray}
\beta_{m}^{\mathcal{Q}}=12\int d\hat{n}_1 d\hat{n}_2\int dr_1 dr_2 r_1^2 r_2^2 \mathcal{M}_{m}^{\mathcal{Q}}(\hat{n}_1,\hat{n}_2,r_1,r_2)
\end{eqnarray}
and the form of $\mathcal{M}_{m}^{\mathcal{Q}}$ inferred from the above equations. The estimator has been reduced entirely to tractable integrals and sums which can be performed relatively quickly.
\par
We can estimate the computational time needed to find this estimator. The multipole summation needed for each basis is $\mathcal{O}(l_{\rm{max}})$ (since the sum over the $m$'s can be precomputed). The integral $\sim \int d^2 \hat{n}$ is an $\mathcal{O}(l_{\rm{max}}^2)$ calculation, while the line of sight integral $\sim \int d r$ is conservatively estimated as an $\mathcal{O}(100)$ operation. Therefore, in total the estimated number of operations is $\mathcal{O}(10 000)\times \mathcal{O}(l_{\rm{max}}^5)$.
\par
In the case that the primordial trispectrum is independent of the diagonal $K$, the estimated number of operations reduces to $\mathcal{O}(100)\times \mathcal{O}(l_{\rm{max}}^3)$ as outlined in Appendix D.

\subsection{B. CMB estimator}
In the case of a precomputed CMB trispectrum or a late times source of non-Gaussianity in the CMB, such as gravitational lensing or active models such as cosmic strings, we wish to find a late-time CMB estimator. For the late-time analysis we wish to expand the estimator functions using the separable $\overline{\mathcal{Q}}_m(l_1,l_2,l_3,l_4,L)$ mode functions created out of the $\overline{q}_p(l)$ and $\overline{r}_v(L)$ polynomials (Note that we denote the multipole modes with a bar to distinguish from the primordial equivalents, and also that we have no need for a subscript for the zeroth Legendre mode since the CMB trispectrum is an explicitly five dimensional quantity as described earlier). In order to effectively expand in mode functions modulated by the $C_l$'s we choose to decompose the estimator functions directly as
\begin{eqnarray}\label{DecompExtra}
\frac{v_{l_1}v_{l_2}v_{l_3}v_{l_4}v_{L}}{\sqrt{C_{l_1}C_{l_2}C_{l_3}C_{l_4}}}t^{l_1 l_2}_{l_3 l_4}(L)=\sum_m \overline{\alpha}^{\mathcal{Q}}_m  \overline{\mathcal{Q}}_m 
\end{eqnarray}
where the separable $v_l$ incorporates the freedom to make the weight function $\omega$ even more scale invariant. The estimator expansion with $C_l$ in~\eqref{EstimatorPrim1} is appropriate for primordial models, but it is expected that flatter choices will be more suitable for late-time anisotropy, such as from cosmic strings.
\par
Substituting this mode expansion into the estimator~\eqref{Estimator} (where again we omit the normalisation factor $N_T$ and return to it later in the section) we find,
\begin{eqnarray}
\mathcal{E}&=&12\sum_{l_i m_i}\sum_{L M}\sum_n \overline{\alpha}^{\mathcal{Q}}_n \overline{q}_p(l_1)\overline{q}_r(l_2)\overline{q}_s(l_3)\overline{q}_u(l_4)\overline{r}_v(L)\int d^2 \hat{n}_1 Y_{l_1 m_1}(\hat{n}_1)Y_{l_2 m_2}(\hat{n}_1)Y^*_{L M}(\hat{n}_1)\nonumber\\
&&\times\int d^2 \hat{n}_2 Y_{l_3 m_3}(\hat{n}_2)Y_{l_4 m_4}(\hat{n}_2)Y_{L M}(\hat{n}_2) \frac{\left(a_{l_1 m_1}a_{l_2 m_2}a_{l_3 m_3}a_{l_4 m_4}\right)_c}{v_{l_1}v_{l_2}v_{l_3}v_{l_4}v_{L}\sqrt{C_{l_1}C_{l_2}C_{l_3}C_{l_4}}}.
\end{eqnarray}
After some algebra we find
\begin{eqnarray}
\mathcal{E}^{\rm{tot}}&=&12 N\Delta_{\Phi}^3\sum_n \overline{\alpha}_n^{\mathcal{Q}} \int d^2 \hat{n}_1 d^2 \hat{n}_2\overline{M}_p(\hat{n}_1) \overline{M}_r(\hat{n}_1) \overline{M}_s(\hat{n}_2) \overline{M}_u(\hat{n}_2)\overline{N}_v(\hat{n}_1,\hat{n}_2),\\
\mathcal{E}^{\rm{uc}}&=&12 N\Delta_{\Phi}^3\sum_n \overline{\alpha}_n^{\mathcal{Q}}\int d^2 \hat{n}_1 d^2 \hat{n}_2 \nonumber\\
&&\times\left(\overline{M}^{\rm{uc}}_{p r}(\hat{n}_1,\hat{n}_1) \overline{M}^{\rm{uc}}_{s u}(\hat{n}_2,\hat{n}_2) +\overline{M}^{\rm{uc}}_{p s}(\hat{n}_1,\hat{n}_2) \overline{M}^{\rm{uc}}_{r u}(\hat{n}_1,\hat{n}_2)+\overline{M}^{\rm{uc}}_{p u}(\hat{n}_1,\hat{n}_2) \overline{M}^{\rm{uc}}_{r s}(\hat{n}_1,\hat{n}_2) \right)  \overline{N}_v(\hat{n}_1,\hat{n}_2),\nonumber
\end{eqnarray}
where
\begin{eqnarray}
\overline{M}_p(\hat{n}_1)&=&\sum_{l_1 m_1}\frac{a_{l_1 m_1}Y_{l_1 m_1}(\hat{n}_1) \overline{q}_p(l_1)}{v_{l_1}\sqrt{C_{l_1}}},\nonumber\\
\overline{M}^{\rm{uc}}_{p s}(\hat{n}_1,\hat{n}_2) &=&\sum_{l_1 m_1}\frac{Y_{l_1 m_1}(\hat{n}_1)Y_{l_1 m_1}^*(\hat{n}_2)   \overline{q}_p(l_1) \overline{q}_s(l_1)  }{v_{l_1}^2} ,\nonumber\\
 \overline{N}_v(\hat{n}_1,\hat{n}_2)&=& \sum_{L M}\frac{Y_{L M}^* (\hat{n}_1) Y_{LM}(\hat{n}_2) \overline{r}_v(L)}{v_L}.
\end{eqnarray}
Again we can summarise these results (substituting back in $N_T$) as
\begin{eqnarray}
\mathcal{E}=\frac{N\Delta_{\Phi}^3}{N_T}\sum_n  \overline{\alpha}_n^{\mathcal{Q}}  \overline{\beta}_n^{\mathcal{Q}} 
\end{eqnarray}
with
\begin{eqnarray}
 \overline{\beta}_n^{\mathcal{Q}} =12\int d^2 \hat{n}_1 d^2 \hat{n}_2 \overline{\mathcal{M}}_n^{\mathcal{Q}}(\hat{n}_1,\hat{n}_2),
\end{eqnarray}
and the form of $\overline{\mathcal{M}}_n^{\mathcal{Q}}(\hat{n}_1,\hat{n}_2)$ can be deduced from the equations for $\mathcal{E}^{\rm{tot}}$ and $\mathcal{E}^{\rm{uc}}$.
\par
Since there are no line of sight integrals ($\sim \int d r$) for this estimator the number of operations required in this case is $\mathcal{O}(l_{\rm{max}}^5)$ suggesting that the late-time estimator is much more computationally efficient than the primordial version.
\par
Similarly to the primordial case, there is a reduction in complexity to $\mathcal{O}(l_{\rm{max}}^3)$ in the case that the late-time extra-reduced trispectrum is independent of the diagonal $L$. This is explained further in Appendix D.

\subsection{C. $T_{NL}$ Estimator}
As with shortcomings of normalising the quantity $f_{NL}$ of the bispectrum that was addressed in~\cite{Ferg3} the current method~\cite{aChen}  of normalising the level of non-Gaussianity due to the trispectrum, $t_{NL}$ poses problems. In particular, the level of non-Gaussianity is found by normalising the shape function against a central point. More specifically we can identify this method as setting $S_{\mathcal{T}}(k,k,k,k,k)=1$ and identifying the normalisation $N$ of equation~\eqref{Shape} as $(50/27)t_{NL}$. In the case of the local model this gives
\begin{eqnarray}
t_{NL}^{\rm{loc}A}=2.16 f_{NL}^2=1.5 \tau_{NL}\,,\qquad t_{NL}^{\rm{loc}B}=1.08 g_{NL},
\end{eqnarray}
where we note again that the relationship between $\tau_{NL}$ and $f_{NL}$ is only strictly true for single field inflation.
This approach assumes scale invariance and therefore will produce inconsistent results between models peaking or dipping at this central point. Also, this definition is not well-defined for models which are not scale-invariant, such as feature models, and it is simply not applicable to non-Gaussian signals created at late times, such as those induced by cosmic strings or secondary anisotropies. An alternative measure of the non-Gaussianity is given by comparison of the primordial trispectrum to the local primordial trispectrum but this approach is not well-defined and is essentially an order of magnitude estimation~\cite{Valen}.
\par
Therefore, we propose a universally defined trispectrum non-Gaussianity parameter $T_{NL}$ which (i) is a measure of the total observational signal expected for the trispectrum of the model in question and (ii) is normalised for direct comparison with the canonical local model (with $g_{NL}=0$). We define $\tilde{T}_{NL}$ from an adapted version of the estimator~\eqref{Estimator} with
\begin{eqnarray}
\tilde{T}_{NL}=\frac{1}{N_T \overline{N}_{T\rm{loc}A}}\sum_{l_i m_i}\frac{\langle a_{l_1m_1}a_{l_2m_2}a_{l_3m_3}a_{l_4m_4}\rangle_c \left(a_{l_1 m_1}^{\rm{obs}}a_{l_2 m_2}^{\rm{obs}}a_{l_3 m_3}^{\rm{obs}}a_{l_4 m_4}^{\rm{obs}}\right)_c}{C_{l_1}C_{l_2}C_{l_3}C_{l_4}}
\end{eqnarray}
where $N$ is the appropriate normalisation factor for the given model,
\begin{eqnarray}
N_T^2=\sum_{l_i, L}\frac{\left(T^{l_1 l_2}_{l_3 l_4}(L)\right)^2}{(2L+1)C_{l_1}C_{l_2}C_{l_3}C_{l_4}},
\end{eqnarray}
and $ \overline{N}_{T\rm{loc}A}$ is the normalisation for the local model with $\tau_{NL}=1, \,g_{NL}=0$.
\begin{eqnarray}
 \overline{N}_{T\rm{loc}A}^2&=&\sum_{l_i, L}\frac{\left(T^{l_1 l_2}_{l_3 l_4}(L)^{\rm{loc (\tau_{NL}=1, g_{NL}=0)} }\right)^{2}   }{(2L+1)C_{l_1}C_{l_2}C_{l_3}C_{l_4}}.
 \end{eqnarray}
The $\tilde{T}_{NL}$ estimator will recover $\tau_{NL}$ for the local model with $g_{NL}=0$, while it gives $( \overline{N}_{T\rm{loc}B}/ \overline{N}_{T\rm{loc}A})g_{NL}$ for the local model with $\tau_{NL}=0$ where $ \overline{N}_{T\rm{loc}B}$ is the normalisation for the local model with $\tau_{NL}=0, \,g_{NL}=1$. This coefficient is dependent on $l_{\rm{max}}$ but is a number of order unity.
\par
Results for primordial models should not depend strongly on the multipole cut-off $l_{\rm{max}}$. However, diffusion due to Silk damping in the transfer functions ensures that the primordial signal is exponentially suppressed for $l\gtrsim 2000$. Therefore, an appropriate choice for a canonical cut-off is $l_{\rm{max}}=2000$. Late-time anisotropies, such as cosmic strings, do not generically fall off exponentially for $l\gtrsim 2000$ but, nonetheless, in the domain $l\lesssim 2000$ we can make a meaningful comparison to the local $\tau_{NL}=1,g_{NL}=0$ model. Alternative measures must be proposed beyond this domain. As indicated in Appendix A the normalisation factor $N_T^2$ is a computationally intensive calculation. Instead we use the approximation $N_T\approx \overline{N}_{T\rm{loc}A} (N_{\mathcal{T}}/\overline{N}_{\mathcal{T}\rm{loc}A})$ where the subscript $\mathcal{T}$ instead of $T$ refers to using the reduced trispectrum instead of the full trispectrum in the above calculations. With these approximations we need only accurately calculate the full normalisation factor in the case of the model with $\tau_{NL}=1, g_{NL}=0$. Regardless of the accuracy we propose that given the vastly increased speed of the calculation we adopt this latter convention and define $T_{NL}$, i.e.
\begin{eqnarray}\label{TNL}
T_{NL}=\frac{ \overline{N}_{\mathcal{T}\rm{loc}A}}{N_\mathcal{T} \overline{N}_{T\rm{loc}A}^2}\sum_{l_i m_i}\frac{\langle a_{l_1m_1}a_{l_2m_2}a_{l_3m_3}a_{l_4m_4}\rangle_c \left(a_{l_1 m_1}^{\rm{obs}}a_{l_2 m_2}^{\rm{obs}}a_{l_3 m_3}^{\rm{obs}}a_{l_4 m_4}^{\rm{obs}}\right)_c}{C_{l_1}C_{l_2}C_{l_3}C_{l_4}}.
\end{eqnarray}
The relation between $T_{NL}$ and $\tilde{T}_{NL}$ as well as the accuracy of the above approximation for $N_T$ - which is only a conjecture at present- will be explored further in an upcoming paper. However we note here that the $T_{NL}$ estimator also recovers $\tau_{NL}$ in the case of the local model with $g_{NL}=0$.
\par
If the CMB trispectrum is not known precisely for the primordial model under study, then we can make an estimate for the normalisation factor $N_{\mathcal{T}}$ in~\eqref{TNL}  using the shape function for the reduced trispectrum $S_{\mathcal{T}}(k_1,k_2,k_3,k_4,K)$. One can obtain a fairly accurate approximation to the relative normalisations in~\eqref{TNL} from 
\begin{eqnarray}
\hat{N}^2=F(S_{\mathcal{T}},S_{\mathcal{T}})=\int d\mathcal{V}_k S_{\mathcal{T}}^2(k_1,k_2,k_3,k_4,K) \omega(k_1,k_2,k_3,k_4,K)
\end{eqnarray}
where the appropriate weight function was found in~\eqref{Weightk} and the domain $\mathcal{V}_k$ is given
 by \eqref{Domain1}, \eqref{Domain2}.  Using $N_{\mathcal{T}}/\overline{N}_{\mathcal{T}\rm{loc}A}\approx \hat{N}/\hat{N}_{\rm{loc}A}$ to approximate $N_{\mathcal{T}}$ we can make a fairly accurate estimate of the level of non-Gaussianity or can renormalise $\tau_{NL}$ constraints for different models into compatible constraints in a similar manner to the analysis  of $f_{NL}$ constraints in the case of the bispectrum in~\cite{Ferg3}.

\section{VII. Recovering the trispectrum}
\subsection{Recovering the primordial trispectrum}
The form of the estimator in~\eqref{EstimatorPrim1} suggests that further information may be extracted from the observed trispectrum beyond the $\tau_{NL}$ for one specific theoretical model. This is because, through the coefficients $\beta_{m}^{\mathcal{Q}}$ we have obtained some sort of mode decomposition of the trispectrum of the observational map. Consider the expectation value of $\beta_{m}^{\mathcal{Q}}$ obtained from an ensemble of maps generated for a particular theoretical model with shape function $S_{\mathcal{T}}=\sum_{m}\alpha_{m}^{\mathcal{Q}}\mathcal{Q}_m$. Since the shape function is in terms of the zeroth mode of the Legendre expansion of the primordial trispectrum, we can only hope to recover information about this mode via recovery of the shape function\footnote{This is to be somewhat expected since the primordial trispectrum is a six dimensional quantity whereas the CMB trispectrum is explicitly five dimensional.}. Using the expression
\begin{eqnarray}
\langle \beta_{m}^{\mathcal{Q}}\rangle&=&12\sum_{l_i m_i}\sum_{LM}\Bigg[\left(\int d\hat{n}_1 Y_{l_1 m_1}(\hat{n}_1)Y_{l_2 m_2}(\hat{n}_1)Y_{L M}^*(\hat{n}_1)\right)\left(\int d\hat{n}_2 Y_{l_3 m_3}(\hat{n}_2)Y_{l_4 m_4}(\hat{n}_2)Y_{L M}(\hat{n}_2)\right)\nonumber\\
&&\times \int r_1^2 dr_1 r_2^2 dr_2\mathcal{Q}^{l_1 l_2 l_3 l_4 L}_{m} (r_1,r_2)\Big] \frac{\langle a_{l_1 m_1} a_{l_2 m_2} a_{l_3 m_3} a_{l_4 m_4} \rangle_c}{C_{l_1}C_{l_2}C_{l_3}C_{l_4}},
\end{eqnarray}
as well as the identity for the Wigner $6$j symbol in Appendix A we find, after some algebra
\begin{eqnarray}
\langle \beta_{m}^{\mathcal{Q}}\rangle&=&12 \sum_{l_i,L}\int dr_1 dr_2 r_1^2 r_2^2 \mathcal{Q}_{m}^{l_1 l_2 l_3 l_4 L}(r_1, r_2)\sum_{m'}\alpha^{\mathcal{Q}}_{m'}\Bigg[h_{l_1 l_2 L}^2 h_{l_3 l_4 L}^2\int dr_1 dr_2 r_1^2 r_2^2 \mathcal{P}_{m'}^{l_1 l_2 l_3 l_4 L}(r_1, r_2)\nonumber\\
&&+\sum_{L'} h_{l_1 l_2 L}h_{l_3 l_4 L}h_{l_1 l_3 L'}h_{l_2 l_4 L'}(-1)^{l_2+l_3} \left\{ \begin{array}{ccc}
l_1 & l_2 & L \\
l_4 &l_3 & L' \end{array} \right\}\int dr_1 dr_2 r_1^2 r_2^2 \mathcal{P}_{m'}^{l_1 l_3 l_2 l_4 L'}(r_1, r_2)\nonumber\\
&&+\sum_{L'} h_{l_1 l_2 L}h_{l_3 l_4 L}h_{l_1 l_4 L'}h_{l_2 l_3 L'}(-1)^{L+L'} \left\{ \begin{array}{ccc}
l_1 & l_2 & L \\
l_3 &l_4 & L' \end{array} \right\}\int dr_1 dr_2 r_1^2 r_2^2 \mathcal{P}_{m'}^{l_1 l_4 l_2 l_3 L'}(r_1, r_2)\Bigg]\nonumber\\
&=&\sum_{m'}\Gamma_{m m'}\alpha^{\mathcal{Q}}_{m'},
\end{eqnarray}
where $\mathcal{P}_{m'}^{l_1 l_2 l_3 l_4 L}=\mathcal{Q}_{m'}^{l_1 l_2 l_3 l_4 L}+\mathcal{Q}_{m'}^{l_2 l_1 l_3 l_4 L}+\mathcal{Q}_{m'}^{l_1 l_2 l_4 l_3 L}+\mathcal{Q}_{m'}^{l_2 l_1 l_4 l_3 L}$. The quantity $\Gamma_{m m'}$ represents a matrix with positions labelled by $m,m'$ and can be inferred readily by the above equation. Inverting the relationship we can recover the $\alpha_{m}^{\mathcal{Q}}$ via
\begin{eqnarray}
\alpha_{m}^{\mathcal{Q}}=\sum_{m'}(\Gamma^{-1})_{m m'}\langle \beta_{m'}^{\mathcal{Q}}\rangle.
\end{eqnarray}
Therefore, if the decomposition coefficients are found with adequate significance, we can reconstruct the shape function through the expansion
\begin{eqnarray}
S_{\mathcal{T}}=\sum_{m}\sum_{ m'}(\Gamma^{-1})_{m m'} \beta_{m'}^{\mathcal{Q}}\mathcal{Q}_m.
\end{eqnarray}
This reconstruction will be sufficient to uniquely define the planar case ($\theta_4=0$) (as well as the 
general CMB case in the next section).  However, as already discussed the shape function only gives information about the zeroth Legendre mode of the primordial trispectrum. Therefore, recovery of the full primordial trispectrum is compromised by this degeneracy. However, as we have discussed in Section II, this degeneracy may be broken by using other probes of non-Gaussianity such as galaxy surveys and $21$ cm observations. We note also that the calculation of the matrix $\Gamma$ is computationally intensive due to the presence of the Wigner $6$j symbols. Nonetheless we include the discussion here for completeness. 

\subsection{Recovering the CMB trispectrum}
The recovery of the CMB bispectrum from a given observational map is more straightforward
(as for the bispectrum).  
In a similar fashion to the calculation in the case of the primordial trispectrum we find that
\begin{eqnarray}
\langle \overline{\beta}_{n}^{\mathcal{Q}}\rangle&=&12\sum_{l_i m_i}\sum_{LM}\Bigg[\left(\int d\hat{n}_1 Y_{l_1 m_1}(\hat{n}_1)Y_{l_2 m_2}(\hat{n}_1)Y_{L M}^*(\hat{n}_1)\right)\left(\int d\hat{n}_2 Y_{l_3 m_3}(\hat{n}_2)Y_{l_4 m_4}(\hat{n}_2)Y_{L M}(\hat{n}_2)\right)\nonumber\\
&&\times \int r_1^2 dr_1 r_2^2 dr_2\overline{\mathcal{Q}}^{l_1 l_2 l_3 l_4 L}_{n} (r_1,r_2)\Big] \frac{\langle a_{l_1 m_1} a_{l_2 m_2} a_{l_3 m_3} a_{l_4 m_4} \rangle_c}{v_{l_1}v_{l_2}v_{l_3}v_{l_4}v_{L}\sqrt{C_{l_1}C_{l_2}C_{l_3}C_{l_4}}}\nonumber\\
\implies \langle \overline{\beta}_{n}^{\mathcal{Q}}\rangle &=&12 \sum_{l_i,L}\overline{Q}^{l_1 l_2 l_3 l_4 L}_n\sum_{p} \overline{\alpha}^{\mathcal{Q}}_p\Bigg[ h_{l_1 l_2 L}^2 h_{l_3 l_4 L}^2 \mathcal{P}_{p}^{l_1 l_2 l_3 l_4 L}+\sum_{L'} h_{l_1 l_2 L}h_{l_3 l_4 L}h_{l_1 l_3 L'}h_{l_2 l_4 L'}(-1)^{l_2+l_3} \left\{ \begin{array}{ccc}
l_1 & l_2 & L \\
l_4 &l_3 & L' \end{array} \right\} \mathcal{P}_{p}^{l_1 l_3 l_2 l_4 L'}\nonumber\\
&&+\sum_{L'} h_{l_1 l_2 L}h_{l_3 l_4 L}h_{l_1 l_4 L'}h_{l_2 l_3 L'}(-1)^{L+L'} \left\{ \begin{array}{ccc}
l_1 & l_2 & L \\
l_3 &l_4 & L' \end{array} \right\} \mathcal{P}_{p}^{l_1 l_4 l_2 l_3 L'}    \Bigg]\nonumber\\
&=&\sum_{p}\overline{\Gamma}_{n p}\alpha^{\mathcal{Q}}_{p},
\end{eqnarray}
where $\overline{Q}^{l_1 l_2 l_3 l_4 L}=\overline{q}_p(l_1)\overline{q}_r(l_2)\overline{q}_s(l_3)\overline{q}_u(l_4)\overline{r}_v(L)$ and $\mathcal{P}_{p}^{l_1 l_2 l_3 l_4 L}=\mathcal{Q}_{p}^{l_1 l_2 l_3 l_4 L}+\mathcal{Q}_{p}^{l_2 l_1 l_3 l_4 L}+\mathcal{Q}_{p}^{l_1 l_2 l_4 l_3 L}+\mathcal{Q}_{p}^{l_2 l_1 l_4 l_3 L}$. Inverting the matrix $\overline{\Gamma}_{n p}$ we find 
\begin{eqnarray}
\alpha^{\mathcal{Q}}_{n}=\sum_p \overline{\Gamma}^{-1}_{n p} \langle \overline{\beta}_{p}^{\mathcal{Q}}\rangle.
\end{eqnarray}
Therefore, if we can measure the coefficients $\overline{\beta}_{p}^{\mathcal{Q}}$ with significance from a particular experiment, we can reconstruct the trispectrum map using~\eqref{DecompExtra}
\begin{eqnarray}
t^{l_1 l_2}_{l_3 l_4}(L)=\frac{\sqrt{C_{l_1}C_{l_2}C_{l_3}C_{l_4}}}{v_{l_1}v_{l_2}v_{l_3}v_{l_4}v_{L}}\sum_{n p}\overline{\Gamma}^{-1}_{n p}  \overline{\beta}_{p}^{\mathcal{Q}}\overline{\mathcal{Q}}_n.
\end{eqnarray}
The calculation of the matrix $\overline{\Gamma}$ remains computationally intensive but it is 
tractable.    We can, in principle, extract the full CMB trispectrum which, together with the extracted
CMB bispectrum~\cite{Ferg3},  will prove to be a key test of the Gaussianity of the Universe. 

\section{VIII. Map Making}
In this section we derive an algorithm for creating a non-Gaussian map with given trispectrum,
developing methods presented for the bispectrum in ref.~\cite{Smith} and generalised in ref.\cite{Ferg3}.
 This algorithm is valid in the limit of weak non-Gaussianity. 
\par
We define the function
\begin{eqnarray}
T_2[a^G]=\frac{1}{24}\sum_{l_i m_i}T_{l_1 m_1 l_2 m_2 l_3 m_3 l_4 m_4}a^G_{l_1 m_1}a^G_{l_2 m_2}a^G_{l_3 m_3}a^G_{l_4 m_4},
\end{eqnarray}
where $a^G_{lm}$ is the Gaussian part of the CMB multipoles, generated using the angular power spectrum $C_l$, while $T_{l_1 m_1 l_2 m_2 l_3 m_3 l_4 m_4}$ is the given trispectrum of the theoretical model for which simulations are required.
\par
Setting
\begin{eqnarray}
a_{l m}'&=&a_{l m}^G +\frac{1}{6}\sum_{l_i m_i}b_{l l_2 l_3}\mathcal{G}^{l l_2 l_3}_{m m_2 m_3}\frac{a_{l_2 m_2}^{* G}}{C_{l_2}} \frac{a_{l_3 m_3}^{* G}}{C_{l_3}}
+\frac{1}{4}\frac{\partial}{\partial a_{l m}^*}T_2[C^{-1} a^G]\nonumber\\
&=&a_{l m}^G +\frac{1}{6}\sum_{l_i m_i}b_{l l_2 l_3}\mathcal{G}^{l l_2 l_3}_{m m_2 m_3}\frac{a_{l_2 m_2}^{* G}}{C_{l_2}} \frac{a_{l_3 m_3}^{* G}}{C_{l_3}}
+\frac{1}{24}\sum_{l_i m_i}T_{l m l_2 m_2 l_3 m_3 l_4 m_4}\frac{a_{l_2 m_2}^{* G}}{C_{l_2}}\frac{a_{l_3 m_3}^{* G}}{C_{l_3}}\frac{a_{l_4 m_4}^{* G}}{C_{l_4}},
\end{eqnarray}
we recover the bispectrum from $\langle a_{l_1 m_1}' a_{l_2 m_2}' a_{l_3 m_3}'\rangle$ (as described in \cite{Smith}). Next, we calculate the four point correlator of the $a_{lm}'$'s and find
\begin{eqnarray}
\langle a_{l_1 m_1}' a_{l_2 m_2}' a_{l_3 m_3}'a_{l_4 m_4}'\rangle&=&\langle a_{l_1 m_1}^G a_{l_2 m_2}^G a_{l_3 m_3}^G a_{l_4 m_4}^G \rangle+\langle\frac{1}{24}\sum_{l_j m_j}T_{l_1 m_1 l_b m_b l_c m_c l_d m_d}\frac{a_{l_b m_b}^{* G}}{C_{l_b}}\frac{a_{l_c m_c}^{* G}}{C_{l_c}}\frac{a_{l_d m_d}^{* G}}{C_{l_d}}a_{l_2 m_2}^G a_{l_3 m_3}^G a_{l_4 m_4}^G\rangle\nonumber\\
&&+\rm{permutations},
\end{eqnarray}
where $j\in (b,c,d)$. The first term clearly gives the unconnected component of the four-point correlator. We note that the contribution from the bispectrum is zero
since the correlator over an odd number of $a_{l m}^G$ vanishes. Evaluating the correlators in the second term on the right hand side and adding up the different permutations we find
\begin{eqnarray}
\langle a_{l_1 m_1}' a_{l_2 m_2}' a_{l_3 m_3}'a_{l_4 m_4}'\rangle=
\langle a_{l_1 m_1}^G a_{l_2 m_2}^G a_{l_3 m_3}^G a_{l_4 m_4}^G \rangle+ T_{l_1 m_1 l_2 m_2 l_3 m_3 l_4 m_4}.
\end{eqnarray}
This verifies the validity of the use of $a_{lm}'$ to make maps including the Gaussian, bispectrum and trispectrum contributions to the model under study.
\par
We observe, using \eqref{TotalRedTrisp}, that we may write $T_2[a^G]$ in the form
\begin{eqnarray}
T_2[a^G]=\frac{1}{2}\sum_{l_i m_i}\mathcal{T}_{l_1 m_1 l_2 m_2 l_3 m_3 l_4 m_4}a^G_{l_1 m_1} a^G_{l_2 m_2}a^G_{l_3 m_3}a^G_{l_4 m_4}.
\end{eqnarray}
Using this formula we may also rewrite the trispectrum contribution to $a_{l m}'$, which we denote $a_{l m}^{NG,T}$ as
\begin{eqnarray}\label{almT}
a_{l m}^{NG,T}&=&\frac{1}{24}\sum_{l_i m_i}T_{l m l_2 m_2 l_3 m_3 l_4 m_4}\frac{a_{l_2 m_2}^{* G}}{C_{l_2}}\frac{a_{l_3 m_3}^{* G}}{C_{l_3}}\frac{a_{l_4 m_4}^{* G}}{C_{l_4}}\\
&=&\frac{1}{8}\sum_{l_i m_i}\left(\mathcal{T}_{l m l_2 m_2 l_3 m_3 l_4 m_4}+\mathcal{T}_{l_2 m_2 l m l_3 m_3 l_4 m_4}+\mathcal{T}_{l_2 m_2 l_3 m_3 l m l_4 m_4}+\mathcal{T}_{l_2 m_2 l_3 m_3 l_4 m_4 l m}\right)\frac{a_{l_2 m_2}^{* G}}{C_{l_2}}\frac{a_{l_3 m_3}^{* G}}{C_{l_3}}\frac{a_{l_4 m_4}^{* G}}{C_{l_4}}\nonumber.
\end{eqnarray}
Using the formulae for the extra-reduced trispectrum \eqref{ExtraTrispRed} and the Gaunt integral \eqref{Gaunt}, we note that
\begin{eqnarray}
\mathcal{T}_{l_1 m_1 l_2 m_2 l_3 m_3 l_4 m_4}&=&\sum_{L M}\left(\int d\Omega_{\hat{n}_1}Y_{l_1 m_1}(\hat{\mathbf{n}}_1)Y_{l_2 m_2}(\hat{\mathbf{n}}_1)Y_{L M}^*(\hat{\mathbf{n}}_1)\right)\left(\int d\Omega_{\hat{n}_2}Y_{l_3 m_3}(\hat{\mathbf{n}}_2)Y_{l_4 m_4}(\hat{\mathbf{n}}_2)Y_{L M}(\hat{\mathbf{n}}_2)\right)t^{l_1 l_2}_{l_3 l_4}(L).\nonumber\\
&&
\end{eqnarray}
As an aside, we note that if $t^{l_1 l_2}_{l_3 l_4}(L)$ is independent of the diagonal $L$ then we can use equation~\eqref{UsefulIdentity} to write
\begin{eqnarray}\label{SpecialId}
\mathcal{T}_{l_1 m_1 l_2 m_2 l_3 m_3 l_4 m_4}&=&\int d\Omega_{\hat{n}}Y_{l_1 m_1}(\hat{\mathbf{n}})Y_{l_2 m_2}(\hat{\mathbf{n}})Y_{l_3 m_3}(\hat{\mathbf{n}})Y_{l_4 m_4}(\hat{\mathbf{n}})t^{l_1 l_2}_{l_3 l_4},
\end{eqnarray}
where we drop the label $L$ from the extra-reduced trispectrum. This special class of trispectra is explored further in Appendix D.
\par
Denoting the bispectrum contribution to $a_{lm}'$ as $a_{l m}^{NG, B}$ we have verified the following prescription for forming maps including the bispectrum and trispectrum contributions
\begin{eqnarray}
a_{l m}'=a_{lm}^G+f_{NL }\tilde{a}_{lm}^{NG, B}+\tau_{NL}\tilde{a}_{lm}^{NG, T}
\end{eqnarray}
where we have written $a_{lm}^{NG, B}=f_{NL }\tilde{a}_{lm}^{NG, B}$ and $a_{lm}^{NG, T}=\tau_{NL }\tilde{a}_{lm}^{NG, T}$ to make the size of the respective non-Gaussian components more explicit.
\par
Since the computation of the reduced trispectrum is more efficient using the late time expression (due to the absence of the line of sight integrals) we write out the formula for $a_{l m}^{NG,T}$ using the late-time mode decomposition. It is straightforward to find the equivalent formula using the primordial expression. Later in the section we present a particular application using the primordial local model of this formalism.
\par
The late-time mode decomposition of the extra-reduced trispectrum $t^{l_1 l_2}_{l_3 l_4}(L)$ - as detailed at the end of section V - may be written as
\begin{eqnarray}\label{Special}
t^{l_1 l_2}_{l_3 l_4}(L)=\sum_n \overline{\alpha}_n^{\mathcal{Q}}\overline{q}_p(l_1)\overline{q}_r(l_2)\overline{q}_s(l_3)\overline{q}_u(l_4)\overline{r}_v(L).
\end{eqnarray}
Using these expressions we have
\begin{eqnarray}
\sum_{l_i m_i}\mathcal{T}_{l m l_2 m_2 l_3 m_3 l_4 m_4}\frac{a_{l_2 m_2}^{* G}}{C_{l_2}}\frac{a_{l_3 m_3}^{* G}}{C_{l_3}}\frac{a_{l_4 m_4}^{* G}}{C_{l_4}}&=&\sum_n \overline{\alpha}_n^{\mathcal{Q}}\int d\Omega_{\hat{\mathbf{n}}_1}d\Omega_{\hat{\mathbf{n}}_2}Y_{l m}(\hat{\mathbf{n}}_1)\overline{q}_p(l)\overline{\mathcal{M}}_r(\hat{\mathbf{n}}_1)\overline{\mathcal{M}}_s(\hat{\mathbf{n}}_2)\overline{\mathcal{M}}_u(\hat{\mathbf{n}}_2)\overline{\mathcal{N}}_v(\hat{\mathbf{n}}_1,\hat{\mathbf{n}}_2),\nonumber\\
&&
\end{eqnarray}
where
\begin{eqnarray}
\overline{\mathcal{M}}_r(\hat{\mathbf{n}}_1)&=&\sum_{l_2 m_2}\frac{Y_{l_2 m_2}(\hat{\mathbf{n}}_1)a_{l_2 m_2}^{* G}}{C_{l_2}}\overline{q}_r(l_2),\nonumber\\
\overline{\mathcal{N}}_v(\hat{\mathbf{n}}_1,\hat{\mathbf{n}}_2)&=&\overline{\mathcal{N}}_v(\hat{\mathbf{n}}_2,\hat{\mathbf{n}}_1)=\sum_{L M}Y_{L M}^*(\hat{\mathbf{n}}_1)Y_{L M}(\hat{\mathbf{n}}_2)\overline{r}_v(L).
\end{eqnarray}
Evaluating, in a similar way, the other terms in equation~\eqref{almT} we find
\begin{eqnarray}
a_{l m}^{NG,T}&=&\frac{1}{8}\sum_n \overline{\alpha}_n^{\mathcal{Q}}\int d\Omega_{\hat{\mathbf{n}}_1}d\Omega_{\hat{\mathbf{n}}_2}Y_{l m}(\hat{\mathbf{n}}_1)\Big[\left(\overline{q}_p(l)\overline{\mathcal{M}}_r(\hat{\mathbf{n}}_1)+\overline{q}_r(l)\overline{\mathcal{M}}_p(\hat{\mathbf{n}}_1) \right)\overline{\mathcal{M}}_s(\hat{\mathbf{n}}_2) \overline{\mathcal{M}}_u(\hat{\mathbf{n}}_2) \nonumber\\
&&+\left(\overline{q}_s(l)\overline{\mathcal{M}}_u(\hat{\mathbf{n}}_1)+\overline{q}_u(l)\overline{\mathcal{M}}_s(\hat{\mathbf{n}}_1) \right)\overline{\mathcal{M}}_p(\hat{\mathbf{n}}_2) \overline{\mathcal{M}}_r(\hat{\mathbf{n}}_2) 
\Big]\overline{\mathcal{N}}_v(\hat{\mathbf{n}}_1,\hat{\mathbf{n}}_2).
\end{eqnarray}
As emphasised in~\cite{Ferg3} the condition that the map has the power spectrum $C_l$ specified in the imput will only be satisfied if the power spectrum of the non-Gaussian components $C_l^{NG}$ is small. Therefore, one has to ascertain that spuriously large $C_l^{NG}$ contributions do not affect the overall power spectrum significantly. We will discuss the implementation of the algorithm presented here in an upcoming paper \cite{Regan2}.

\subsection{Application to the Local Model}
The reduced trispectrum for the local model, as shown in Section IV, is made up of two terms which we denoted $\rm{locA}$ and $\rm{locB}$. As shown in \cite{Okamoto} - and can be deduced from Section IV - the extra reduced trispectra may be written as
\begin{eqnarray}
{t^{l_1 l_2}_{l_3 l_4}(L)}^{\rm{locA}}&=&\frac{25}{9}\tau_{NL}\int d r_1 d r_2 r_1^2 r_2^2 F_L (r_1,r_2) \alpha_{l_1}(r_1)\beta_{l_2}(r_1)\alpha_{l_3}(r_2)\beta_{l_4}(r_2),\\
{t^{l_1 l_2}_{l_3 l_4}(L)}^{\rm{locB}}&=&g_{NL}\int d r r^2 \beta_{l_2}(r) \beta_{l_4}(r)\left(\mu_{l_1}(r)\beta_{l_3}(r)+\beta_{l_1}(r)\mu_{l_3}(r)\right),
\end{eqnarray}
where
\begin{eqnarray}
F_L (r_1,r_2)&=&\frac{2}{\pi}\int K^2 dK P_{\Phi}(K)j_L(K r_1)j_L(K r_2),\nonumber\\
\alpha_{l}(r)&=&\mu_{l}(r)=\frac{2}{\pi}\int k^2 dk \Delta_{l}(k) j_l(k r),\nonumber\\
\beta_{l}(r)&=&\frac{2}{\pi}\int k^2 dk P_{\Phi}(k) \Delta_{l}(k) j_l(k r).
\end{eqnarray}
Using these formulae, and exploiting that the $\rm{locB}$ trispectrum is independent of the diagonal $L$ with equation~\eqref{SpecialId}, we find
\begin{eqnarray}
\sum_{l_i m_i}\mathcal{T}_{l m l_2 m_2 l_3 m_3 l_4 m_4}^{\rm{locA}}\frac{a_{l_2 m_2}^{* G}}{C_{l_2}}\frac{a_{l_3 m_3}^{* G}}{C_{l_3}}\frac{a_{l_4 m_4}^{* G}}{C_{l_4}}&=&\frac{25}{9}\tau_{NL}\int d r_1 dr_2 r_1^2 r_2^2 \alpha_l(r_1)\int d\Omega_{\hat{\mathbf{n}}_1}d\Omega_{\hat{\mathbf{n}}_2}Y_{l m}(\hat{\mathbf{n}}_1)\mathcal{M}_F(\hat{\mathbf{n}}_1,\hat{\mathbf{n}}_2,r_1,r_2)\mathcal{M}_{\beta}(\hat{\mathbf{n}}_1,r_1)\nonumber\\
&&\times\mathcal{M}_{\alpha}(\hat{\mathbf{n}}_2,r_2)\mathcal{M}_{\beta}(\hat{\mathbf{n}}_2,r_2),\\
\sum_{l_i m_i}\mathcal{T}_{l m l_2 m_2 l_3 m_3 l_4 m_4}^{\rm{locB}}\frac{a_{l_2 m_2}^{* G}}{C_{l_2}}\frac{a_{l_3 m_3}^{* G}}{C_{l_3}}\frac{a_{l_4 m_4}^{* G}}{C_{l_4}}&=&g_{NL}\int d r r^2\int d \Omega_{\hat{\mathbf{n}}}Y_{l m}(\hat{\mathbf{n}})\Bigg[ \mu_l (r)\mathcal{M}_{\beta}(\hat{\mathbf{n}},r)+ \beta_l (r)\mathcal{M}_{\mu}(\hat{\mathbf{n}},r)\Bigg]\mathcal{M}_{\beta}(\hat{\mathbf{n}},r)\mathcal{M}_{\beta}(\hat{\mathbf{n}},r),\nonumber\\
&&
\end{eqnarray}
where
\begin{eqnarray}
\mathcal{M}_{\alpha}(\hat{\mathbf{n}},r)&=&\mathcal{M}_{\mu}(\hat{\mathbf{n}},r)=\sum_{l m}\alpha_l(r)\frac{Y_{l m}(\hat{\mathbf{n}}) a_{l m}^{* G}}{C_{l}},\nonumber\\
\mathcal{M}_{\beta}(\hat{\mathbf{n}},r)&=&\sum_{l m}\beta_l(r)\frac{Y_{l m}(\hat{\mathbf{n}}) a_{l m}^{* G}}{C_{l}},\nonumber\\
\mathcal{M}_F(\hat{\mathbf{n}}_1,\hat{\mathbf{n}}_2,r_1,r_2)&=& \sum_{L M}Y_{L M}^*(\hat{\mathbf{n}}_1)Y_{L M}(\hat{\mathbf{n}}_2)F_L(r_1,r_2).
\end{eqnarray}
We similarly find the other terms in~\eqref{almT} to get
\begin{eqnarray}
(a_{l m}^{NG,T})_{\rm{locA}}&=&\frac{25}{36}\tau_{NL}\int d r_1 dr_2 r_1^2 r_2^2 \left[\alpha_l(r_1)\int d\Omega_{\hat{\mathbf{n}}_1}d\Omega_{\hat{\mathbf{n}}_2}Y_{l m}(\hat{\mathbf{n}}_1)\mathcal{M}_{\beta}(\hat{\mathbf{n}}_1,r_1)+\beta_l(r_1)\int d\Omega_{\hat{\mathbf{n}}_1}d\Omega_{\hat{\mathbf{n}}_2}Y_{l m}(\hat{\mathbf{n}}_1)\mathcal{M}_{\alpha}(\hat{\mathbf{n}}_1,r_1)\right]\nonumber\\
&&\times\mathcal{M}_{\alpha}(\hat{\mathbf{n}}_2,r_2)\mathcal{M}_{\beta}(\hat{\mathbf{n}}_2,r_2)\mathcal{M}_{F}(\hat{\mathbf{n}}_1,\hat{\mathbf{n}}_2,r_1,r_2),\\
(a_{l m}^{NG,T})_{\rm{locB}}&=&\frac{1}{4}g_{NL}\int d r r^2 \Bigg[\mu_l(r)\int d\Omega_{\hat{\mathbf{n}}}Y_{l m}(\hat{\mathbf{n}})\mathcal{M}_{\beta}(\hat{\mathbf{n}},r)\mathcal{M}_{\beta}(\hat{\mathbf{n}},r)\mathcal{M}_{\beta}(\hat{\mathbf{n}},r)+         \beta_l(r)\int d\Omega_{\hat{\mathbf{n}}}   Y_{l m}(\hat{\mathbf{n}})\nonumber
\\&&\times\Bigg(3 \mathcal{M}_{\beta}(\hat{\mathbf{n}},r)\mathcal{M}_{\mu}(\hat{\mathbf{n}},r)\mathcal{M}_{\beta}(\hat{\mathbf{n}},r)\Bigg)\Bigg].
\end{eqnarray}
In the case of the bispectrum, direct implementation of the explicitly separable local shape results in spuriously large $C_l^{NG}$ contributions. However, it was found that using the eigenmode expansion in ref.~\cite{Ferg3} was much more robust circumventing such effects because of the bounded nature
of the polynomial eigenmodes. This improvement is expected to occur for the trispectrum. An alternative method is to regularise the expressions given here by eliminating pathological terms, while leaving the final trispectrum of the map unchanged.  For arbitrary separable trispectra (unlike the eigenmode
expansion),  convergence must be achieved by hand on a case-by-case basis.

\section{IX. Conclusions}
We have described in this paper two comprehensive pipelines for the analysis of general primordial or CMB trispectra. The methods are based on mode expansions, exploiting a complete orthonormal eigenmode basis to efficiently decompose arbitrary trispectra into a separable polynomial expansion. These separable mode expansions allow for a reduction of the computational overhead to tractable levels, regardless of whether the reduced trispectrum is being computed at Planck resolution, or we are directly finding an estimator for the size of the trispectrum from a real data set. A shape decomposition has been described allowing for a visualisation of a scale invariant reduced trispectrum on particular slices.
\par
We have presented a correlator for comparing trispectra. We have also defined a correlator for comparing the shape functions that is expected to closely approximate the former. However, the main purpose of this paper was to present a detailed theoretical framework for finding an estimator for the size of the trispectrum using separable eigenmode expansions. Using this efficient method for finding an estimator for the trispectrum we have defined a universal measure $T_{NL}$ which will allow for consistent comparison between theoretical models. This measure can be calculated for both primordial models and late-time models, e.g. due to active models such as cosmic strings. The completeness of the orthogonal eigenmodes should allow for a reconstruction of the full CMB trispectrum from the data, assuming the presence of a sufficiently significant non-Gaussian signal. We have also detailed an algorithm for producing non-Gaussian simulations with a given power spectrum, bispectrum and trispectrum. The implementation of these methods will be discussed in a future publication~\cite{Regan2}. Clearly, the full implementation of the primordial and late-time pipelines represents a significant challenge. However the generality and robustness of the methodology described here indicates that there is an intriguing possibility of exploring and constraining a wide class of non-Gaussian models using the trispectrum.

\section{Acknowledgements}

We are very grateful for many informative and illuminating discussions with 
Xingang Chen and Michele Liguori. We also thank S\'ebastien Renaux-Petel for comments on an earlier version of this paper. EPS and JRF was supported by STFC grant ST/F002998/1 and the 
Centre for Theoretical Cosmology.  DMR was supported by EPSRC, the Isaac Newton Trust and the Cambridge European Trust.

\bigskip

\section{Appendix A: Normalisation factor}
Clearly the appropriate normalisation factor for the estimator~\eqref{Estimator} is of the form
\begin{eqnarray*}
N_{T'}&=&\sum_{l_i m_i}\frac{\langle a_{l_1m_1}a_{l_2m_2}a_{l_3m_3}a_{l_4m_4}\rangle_c \langle a_{l_1m_1}a_{l_2m_2}a_{l_3m_3}a_{l_4m_4}\rangle_c}{C_{l_1}C_{l_2}C_{l_3}C_{l_4}}. 
\end{eqnarray*}
In order to relate this to $N_T$ expressed in~\eqref{Normal} we use equation~\eqref{Ttot} to expand $N_{T'}$ in the form
\begin{eqnarray}
N_{T'}&=&\sum_{l_i L L'}\sum_{m_i M M'} \frac{ 1      }{C_{l_1}C_{l_2}C_{l_3}C_{l_4}}\nonumber\\
&&\times (-1)^M \left( \begin{array}{ccc}
l_1 & l_2 & L \\
m_1 & m_2 & -M \end{array} \right) \left( \begin{array}{ccc}
l_3 & l_4 & L \\
m_3 & m_4 & M \end{array} \right) T^{l_1 l_2}_{l_3 l_4}(L)   (-1)^{M'} 
\left( \begin{array}{ccc}
l_1 & l_2 & L' \\
m_1 & m_2 & -M' \end{array} \right) \left( \begin{array}{ccc}
l_3 & l_4 & L' \\
m_3 & m_4 & M' \end{array} \right) T^{l_1 l_2}_{l_3 l_4}(L').\nonumber    
\end{eqnarray}
Now, using
\begin{eqnarray}\label{Orthog3}
\sum_{m_1 m_2} \left( \begin{array}{ccc}
l_1 & l_2 & L \\
m_1 & m_2 & -M \end{array} \right)  \left( \begin{array}{ccc}
l_1 & l_2 & L' \\
m_1 & m_2 & -M' \end{array} \right)&=&\frac{\delta_{L,L'} \delta_{M,M'} }{2L+1},\nonumber\\
\sum_M \sum_{m_3 m_4} \left( \begin{array}{ccc}
l_3 & l_4 & L \\
m_3 & m_4 & M \end{array} \right)  \left( \begin{array}{ccc}
l_3 & l_4 & L \\
m_3 & m_4 & M \end{array} \right)&=&1,
\end{eqnarray}
we find that
\begin{eqnarray*}
N_{T'}=\sum_{l_i, L}\frac{T^{l_1 l_2}_{l_3 l_4}(L)T^{l_1 l_2}_{l_3 l_4}(L)}{(2L+1)C_{l_1}C_{l_2}C_{l_3}C_{l_4}}=N_T.
\end{eqnarray*}
This verifies the use of equation~\eqref{Normal} to normalise the estimator~\eqref{Estimator}.
\par
We can expand $N_T$ in terms of the reduced trispectrum using
\begin{eqnarray}
N_T=12\sum_{l_i m_i}\frac{\mathcal{T}_{l_1 m_1 l_2 m_2 l_3 m_3 l_4 m_4}T_{l_1 m_1 l_2 m_2 l_3 m_3 l_4 m_4}}{C_{l_1}C_{l_2}C_{l_3}C_{l_4}}.
\end{eqnarray}
Then with the identity for the Wigner $6$j symbol (see Appendix in \cite{Hu})
\begin{eqnarray}
\left\{ \begin{array}{ccc}
a & b & e \\
c & d & f \end{array}\right\}=\sum_{\alpha \beta \gamma}\sum_{\delta \epsilon \phi}(-1)^{e+f+\epsilon+\phi} \left( \begin{array}{ccc}
a & b & e \\
\alpha & \beta & \epsilon \end{array} \right) \left( \begin{array}{ccc}
c & d & e \\
\gamma & \delta & -\epsilon \end{array} \right) \left( \begin{array}{ccc}
a & d & f \\
\alpha & \delta & -\phi \end{array} \right) \left( \begin{array}{ccc}
c & b & f \\
\gamma & \beta & \phi \end{array} \right),
\end{eqnarray}
the identities \eqref{Orthog3} and relations for $P$ in \eqref{Prel1} and \eqref{Prel2} we find that
\begin{eqnarray}
N_T=12\sum_{l_i,L}\frac{\mathcal{T}^{l_1 l_2}_{l_3 l_4}(L)}{C_{l_1}C_{l_2}C_{l_3}C_{l_4}}\left(\frac{P^{l_1 l_2}_{l_3 l_4}(L)}{2L+1}+\sum_{L'}(-1)^{l_2+l_3}\left\{ \begin{array}{ccc}
l_1 & l_2 & L \\
l_4 & l_3 & L' \end{array}\right\} P^{l_1 l_3}_{l_2 l_4}(L')+\sum_{L'}(-1)^{L+L'}\left\{ \begin{array}{ccc}
l_1 & l_2 & L \\
l_3 & l_4 & L' \end{array}\right\} P^{l_1 l_4}_{l_3 l_2}(L')\right).\nonumber\\
\end{eqnarray}
Due to the presence of the $6$j symbols the calculation of $N_T$ is computationally very expensive in general.

\section{Appendix B: Optimal Estimator}
When non-Gaussianity is weak we can exploit the multivariate Edgeworth expansion~\cite{Amendola} around the Gaussian probability distribution function (PDF), $P^G(a)$, i.e.
\begin{eqnarray}
P(a)&=&\Bigg[1-\sum_{l_i m_i}\langle a_{l_1 m_1}a_{l_2 m_2}a_{l_3 m_3}\rangle \frac{\partial}{\partial a_{l_1 m_1}}	\frac{\partial}{\partial a_{l_2 m_2}}	\frac{\partial}{\partial a_{l_3 m_3}}	\nonumber\\
&&+\sum_{l_i m_i}\langle a_{l_1 m_1}a_{l_2 m_2}a_{l_3 m_3}a_{l_4 m_4}\rangle_c \frac{\partial}{\partial a_{l_1 m_1}}	\frac{\partial}{\partial a_{l_2 m_2}}	\frac{\partial}{\partial a_{l_3 m_3}}\frac{\partial}{\partial a_{l_4 m_4}}+\dots	\Bigg]P^G(a),
\end{eqnarray}
where the Gaussian PDF is given by
\begin{eqnarray}
P^G(a)=\frac{e^{-\frac{1}{2}\sum_{l m}\sum_{l' m'}a_{l m} (C^{-1})_{l m, l' m'} a_{l' m'}
 }}{(2\pi)^{N/2}|C|^{1/2}}
\end{eqnarray}
with $C_{l m, l' m'}=\langle a_{l m} a_{l' m'}\rangle$ and $N$ the number of $l$ and $m$. Maximising over the three point correlator results in the optimal bispectrum estimator. Here we will ignore this term (setting it to zero for convenience) and concentrate on the four-point correlator. We find
\begin{eqnarray}
P(a)&=&\Bigg[1+\sum_{l_i m_i}\langle a_{l_1 m_1}a_{l_2 m_2}a_{l_3 m_3}a_{l_4 m_4}\rangle_c\Bigg((C^{-1} a^{ })_{l_1 m_1}(C^{-1} a^{ })_{l_2 m_2} (C^{-1} a^{ })_{l_3 m_3} (C^{-1} a^{ })_{l_4 m_4}\nonumber\\
&&-6(C^{-1})_{l_2 m_2,l_1 m_1} (C^{-1} a^{ })_{l_3 m_3} (C^{-1} a^{ })_{l_4 m_4}+3 (C^{-1})_{l_1 m_1,l_2 m_2}(C^{-1})_{l_3 m_3,l_4 m_4}		\Bigg)	\Bigg]P^G(a),
\end{eqnarray}
where $(C^{-1} a)_{l m}=\sum_{l' m'}C^{-1}_{l' m', l m}a_{l' m'}$.
Parametrising the size of the trispectrum by $\mathcal{E}$ we wish to maximise the PDF with respect to this. We assume that $ (a_{l_1 m_1}a_{l_2 m_2}a_{l_3 m_3}a_{l_4 m_4})_c\propto \mathcal{E}$ such that the second term is proportional to $\mathcal{E}$. Maximising the PDF means that we wish to set $(d P/d\mathcal{E})=0$, such that the Taylor expansion about $\mathcal{E}=0$ reads
\begin{eqnarray}
P(a)= \Bigg[1+\frac{d (P/P^G)}{d\mathcal{E}}\mathcal{E} +\frac{1}{2}\frac{d^2  (P/P^G)}{d\mathcal{E}^2} \mathcal{E}^2+\dots\Bigg]P^G(a) \approx  \Bigg[1+\frac{1}{2}\frac{d^2  (P/P^G)}{d\mathcal{E}^2} \mathcal{E}^2 \Bigg]P^G(a),
\end{eqnarray}
Since 
\begin{eqnarray*}
\frac{d^2 P}{d\mathcal{E}^2}\propto 2 \sum_{l_i m_i}  \langle a_{l_1m_1}a_{l_2m_2}a_{l_3m_3}a_{l_4m_4}\rangle_c  (C^{-1})_{l_1 m_1,l_1' m_1'} (C^{-1})_{l_2 m_2,l_2' m_2'} (C^{-1})_{l_3 m_3,l_3' m_3'} (C^{-1})_{l_4 m_4,l_4' m_4'} ( a_{l_1' m_1'}a_{l_2' m_2'}a_{l_3' m_3'}a_{l_4' m_4'})_c
\end{eqnarray*}
we find that the estimator is maximised by setting (with appropriate choice of proportionality constant)
\begin{eqnarray}
\mathcal{E}&=&\frac{f_{\rm{sky}}}{\tilde{N}}\sum_{l_i m_i}\langle a_{l_1 m_1}a_{l_2 m_2}a_{l_3 m_3}a_{l_4 m_4}\rangle_c\Bigg((C^{-1} a^{ })_{l_1 m_1}(C^{-1} a^{ })_{l_2 m_2} (C^{-1} a^{ })_{l_3 m_3} (C^{-1} a^{ })_{l_4 m_4}\nonumber\\
&&-6(C^{-1})_{l_1 m_1,l_2 m_2} (C^{-1} a)_{l_3 m_3} (C^{-1} a^{ })_{l_4 m_4}+3  (C^{-1})_{l_1 m_1,l_2 m_2}(C^{-1})_{l_3 m_3,l_4 m_4}		\Bigg)
\end{eqnarray}
where 
\begin{eqnarray*}
\tilde{N}= \sum_{l_i m_i}  \langle a_{l_1m_1}a_{l_2m_2}a_{l_3m_3}a_{l_4m_4}\rangle_c  (C^{-1})_{l_1 m_1,l_1' m_1'} (C^{-1})_{l_2 m_2,l_2' m_2'} (C^{-1})_{l_3 m_3,l_3' m_3'} (C^{-1})_{l_4 m_4,l_4' m_4'} ( a_{l_1' m_1'}a_{l_2' m_2'}a_{l_3' m_3'}a_{l_4' m_4'})_c.
\end{eqnarray*}

\section{Appendix C: Non-Optimal Estimators - Skewness and Kurtosis}
The deviation from non-Gaussianity may be measured in a non-optimal way by estimating the departure of the one-point PDF from Gaussian behaviour. This deviation may be measured in terms of the skewness and kurtosis. The skewness is given by
\begin{eqnarray}
g_1=\frac{\Bigg\langle\left(\frac{\Delta T}{T}(\hat{n}) \right)^3\Bigg\rangle}{\left(\Bigg\langle \left(\frac{\Delta T}{T}(\hat{n}) \right)^2\Bigg\rangle\right)^{3/2}},
\end{eqnarray}
while the kurtosis is given by \eqref{Kurtosis1}. Using \eqref{DeltaT} we evaluate the variance as
\begin{eqnarray}
\Bigg\langle\left(\frac{\Delta T}{T}(\hat{n}) \right)^2\Bigg\rangle=\sum_{l' m'}\sum_{l m}\int \frac{d\Omega_{\hat{n}}}{4\pi}\langle a_{l' m'}^* a_{l m} \rangle Y_{l' m'}^*(\hat{\mathbf{n}})Y_{l m}(\hat{\mathbf{n}})=\frac{1}{4\pi}\sum_{l m}\langle a_{l m}^* a_{l m} \rangle =\frac{\sum_l (2l+1)C_l}{4\pi}.
\end{eqnarray}
The three-point temperature correlator is similarly given by
\begin{eqnarray}
\Bigg\langle\left(\frac{\Delta T}{T}(\hat{n}) \right)^3\Bigg\rangle&=&\sum_{l_i m_i}\langle a_{l_1 m_1}a_{l_2 m_2}a_{l_3 m_3}\rangle \int \frac{d\Omega_{\hat{n}}}{4\pi} Y_{l_1 m_1}(\hat{\mathbf{n}})Y_{l_2 m_2}(\hat{\mathbf{n}})Y_{l_3 m_3}(\hat{\mathbf{n}})\nonumber\\
&=&\frac{1}{4\pi}\sum_{l_i m_i}\left(\mathcal{G}^{l_1 l_2 l_3}_{m_1 m_2 m_3}\right)^2 b_{l_1 l_2 l_3}
\end{eqnarray}
where $\mathcal{G}^{l_1 l_2 l_3}_{m_1 m_2 m_3}$ is the Gaunt integral and $b_{l_1 l_2 l_3}$ is the reduced bispectrum. The Gaunt integral is given by
\begin{eqnarray}
\mathcal{G}^{l_1 l_2 l_3}_{m_1 m_2 m_3}=h_{l_1 l_2 l_3}\left( \begin{array}{ccc}
l_1 & l_2 & l_3 \\
m_1 & m_2 & m_3 \end{array} \right).
\end{eqnarray}
Using equation~\eqref{Orthog3} we can simplify this expression to get
\begin{eqnarray}
\Bigg\langle\left(\frac{\Delta T}{T}(\hat{n}) \right)^3\Bigg\rangle&=&\frac{1}{4\pi}\sum_{l_i} h_{l_1 l_2 l_3}^2 b_{l_1 l_2 l_3}.
\end{eqnarray}
Next in order to evaluate the kurtosis we calculate the quantity
\begin{eqnarray}
K&=&\Bigg\langle\left(\frac{\Delta T}{T}(\hat{n}) \right)^4\Bigg\rangle-3\left(\Bigg\langle \left(\frac{\Delta T}{T}(\hat{n}) \right)^2\Bigg\rangle\right)^{2}\nonumber\\
&=&\sum_{l_i m_i}\langle a_{l_1 m_1}a_{l_2 m_2}a_{l_3 m_3}a_{l_4 m_4}\rangle \int \frac{d\Omega_{\hat{n}}}{4\pi} Y_{l_1 m_1}(\hat{\mathbf{n}})Y_{l_2 m_2}(\hat{\mathbf{n}})Y_{l_3 m_3}(\hat{\mathbf{n}})Y_{l_4 m_4}(\hat{\mathbf{n}})\nonumber\\
&&-\sum_{l_i m_i}\int \frac{d\Omega_{\hat{n}_1}}{4\pi}  \frac{d\Omega_{\hat{n}_2}}{4\pi}\langle a_{l_1 m_1}a_{l_2 m_2}\rangle \langle a_{l_3 m_3}a_{l_4 m_4}\rangle Y_{l_1 m_1}(\hat{\mathbf{n}}_1)Y_{l_2 m_2}(\hat{\mathbf{n}}_1)Y_{l_3 m_3}(\hat{\mathbf{n}}_2)Y_{l_4 m_4}(\hat{\mathbf{n}}_2)\nonumber\\
&&-\sum_{l_i m_i}\int \frac{d\Omega_{\hat{n}_1}}{4\pi}  \frac{d\Omega_{\hat{n}_2}}{4\pi}\langle a_{l_1 m_1}a_{l_3 m_3}\rangle \langle a_{l_2 m_2}a_{l_4 m_4}\rangle Y_{l_1 m_1}(\hat{\mathbf{n}}_1)Y_{l_3 m_3}(\hat{\mathbf{n}}_1)Y_{l_2 m_2}(\hat{\mathbf{n}}_2)Y_{l_4 m_4}(\hat{\mathbf{n}}_2)
\nonumber\\
&&-\sum_{l_i m_i}\int \frac{d\Omega_{\hat{n}_1}}{4\pi}  \frac{d\Omega_{\hat{n}_2}}{4\pi}\langle a_{l_1 m_1}a_{l_4 m_4}\rangle \langle a_{l_2 m_2}a_{l_3 m_3}\rangle Y_{l_1 m_1}(\hat{\mathbf{n}}_1)Y_{l_4 m_4}(\hat{\mathbf{n}}_1)Y_{l_2 m_2}(\hat{\mathbf{n}}_2)Y_{l_3 m_3}(\hat{\mathbf{n}}_2),
\end{eqnarray}
where the final three terms are clearly equivalent. Using
\begin{eqnarray}
\langle a_{l_1 m_1}a_{l_2 m_2}a_{l_3 m_3}a_{l_4 m_4}\rangle&=&\langle a_{l_1 m_1}a_{l_2 m_2} a_{l_3 m_3}a_{l_4 m_4}\rangle_c+\langle a_{l_1 m_1}a_{l_2 m_2}\rangle \langle a_{l_3 m_3}a_{l_4 m_4}\rangle+\langle a_{l_1 m_1}a_{l_3 m_3}\rangle \langle a_{l_2 m_2}a_{l_4 m_4}\rangle\nonumber\\
&&+\langle a_{l_1 m_1}a_{l_4 m_4}\rangle \langle a_{l_2 m_2}a_{l_3 m_3}\rangle,
\end{eqnarray}
it can be shown after some algebra that
\begin{eqnarray}
K=\frac{1}{4\pi}\sum_{l_i m_i}\langle a_{l_1 m_1}a_{l_2 m_2}a_{l_3 m_3}a_{l_4 m_4}\rangle_c \int d\Omega_{\hat{n}} Y_{l_1 m_1}(\hat{\mathbf{n}})Y_{l_2 m_2}(\hat{\mathbf{n}})Y_{l_3 m_3}(\hat{\mathbf{n}})Y_{l_4 m_4}(\hat{\mathbf{n}}).
\end{eqnarray}
From equation~\eqref{TotalRedTrisp} this may be written in terms of the reduced trispectrum as
\begin{eqnarray}
K=\frac{12}{4\pi}\sum_{l_i m_i}\mathcal{T}_{l_1 m_1 l_2 m_2 l_3 m_3 l_4 m_4} \int d\Omega_{\hat{n}} Y_{l_1 m_1}(\hat{\mathbf{n}})Y_{l_2 m_2}(\hat{\mathbf{n}})Y_{l_3 m_3}(\hat{\mathbf{n}})Y_{l_4 m_4}(\hat{\mathbf{n}}).
\end{eqnarray}
Next, noting that the product of two spherical harmonics can be written as
\begin{eqnarray}
Y_{l_1 m_1}(\hat{n})Y_{l_2 m_2}(\hat{n})=\sum_{L' M'}h_{l_1 l_2 L'}\left( \begin{array}{ccc}
l_1 & l_2 & L' \\
m_1 & m_2 & -M' \end{array} \right)(-1)^{M'}Y_{L'  M'}(\hat{n})
\end{eqnarray}
and equation~\eqref{Gaunt}, we find
\begin{eqnarray}\label{UsefulIdentity}
\int d\Omega_{\hat{n}} Y_{l_1 m_1}(\hat{\mathbf{n}})Y_{l_2 m_2}(\hat{\mathbf{n}})Y_{l_3 m_3}(\hat{\mathbf{n}})Y_{l_4 m_4}(\hat{\mathbf{n}})=\sum_{L' M'}(-1)^{M'}h_{l_1 l_2 L'}h_{l_3 l_4 L'}\left( \begin{array}{ccc}
l_1 & l_2 & L' \\
m_1 & m_2 & -M' \end{array}\right) \left( \begin{array}{ccc}
l_3 & l_4 & L' \\
m_3 & m_4 & M' \end{array} \right).
\end{eqnarray}
Finally, using equation~\eqref{RedTrisp} for the reduced trispectrum, the orthogonality relation between Wigner $3$j functions as expressed in~\eqref{Orthog1} and the `extra'-reduced trispectrum~\eqref{ExtraTrispRed}, we find
\begin{eqnarray}
K=\frac{12}{4\pi}\sum_{l_i, L} \frac{h_{l_1 l_2 L}h_{l_3 l_4 L}}{2L+1}\mathcal{T}^{l_1 l_2}_{l_3 l_4}(L)=\frac{12}{4\pi}\sum_{l_i, L} \frac{h_{l_1 l_2 L}^2 h_{l_3 l_4 L}^2}{2L+1}t^{l_1 l_2}_{l_3 l_4}(L).
\end{eqnarray}
In summary, the skewness and kurtosis are given respectively by
\begin{eqnarray}
g_1&=&\sqrt{4\pi}\frac{\sum_{l_i} h_{l_1 l_2 l_3}^2 b_{l_1 l_2 l_3}}{\left( \sum_l(2l+1)C_l\right)^{3/2}},\\
g_2&=&\frac{48\pi\sum_{l_i, L} h_{l_1 l_2 L}^2 h_{l_3 l_4 L}^2 t^{l_1 l_2}_{l_3 l_4}(L)/(2L+1)}{\left( \sum_l(2l+1)C_l\right)^2}.
\end{eqnarray}

\section{Appendix D: Special case of trispectrum independent of diagonal}
Suppose that the primordial reduced trispectrum is independent of the diagonal $K$. In particular, we write $\mathcal{T}_{\Phi,0}(k_1,k_2,k_3,k_4;K)=\mathcal{T}_{\Phi,0}(k_1,k_2,k_3,k_4)$. In that case the `extra'-reduced trispectrum (see \eqref{TrispRed2} and \eqref{ExtraTrispRed}) becomes
\begin{eqnarray}
t^{l_1 l_2}_{l_3 l_4}(L)&=&\left(\frac{2}{\pi}\right)^5 \int (k_1 k_2 k_3 k_4 K)^2 dk_1 dk_2 dk_3 dk_4 dK r_1^2 dr_1 r_2^2 dr_2 j_L(K r_1) j_L(K r_2)\nonumber\\
&&\times[j_{l_1}(k_1 r_1)\Delta_{l_1}(k_1)][j_{l_2}(k_2 r_1)\Delta_{l_2}(k_2)][j_{l_3}(k_3 r_2)\Delta_{l_3}(k_3)][j_{l_4}(k_4 r_2)\Delta_{l_4}(k_4)] \mathcal{T}_{\Phi,0}(k_1,k_2, k_3,k_4).
\end{eqnarray}
Next, using equation~\eqref{IdentitySpecial} we find
\begin{eqnarray}
\int dK K^2 j_L(K r_1) j_L(K r_2)=\frac{\pi}{2 r_2^2}\delta(r_2-r_1).
\end{eqnarray}
This implies that
\begin{eqnarray}
t^{l_1 l_2}_{l_3 l_4}(L)&=&\left(\frac{2}{\pi}\right)^4 \int (k_1 k_2 k_3 k_4)^2 dk_1 dk_2 dk_3 dk_4 r_1^2 dr_1 \nonumber\\
&&\times[j_{l_1}(k_1 r_1)\Delta_{l_1}(k_1)][j_{l_2}(k_2 r_1)\Delta_{l_2}(k_2)][j_{l_3}(k_3 r_1)\Delta_{l_3}(k_3)][j_{l_4}(k_4 r_1)\Delta_{l_4}(k_4)] \mathcal{T}_{\Phi,0}(k_1,k_2, k_3,k_4),
\end{eqnarray}
i.e. we only have one line of sight integral. This expression also shows that, if the primordial trispectrum is independent of the diagonal $K$, then $t^{l_1 l_2}_{l_3 l_4}(L)$ is independent of $L$. We can exploit this property in our estimators. From equations~\eqref{RedTrisp},~\eqref{ExtraTrispRed} and equation~\eqref{Gaunt} for the Gaunt integral (which we denote here in the form $\mathcal{G}^{l_1 l_2 l_3}_{m_1 m_2 m_3}$) we have
\begin{eqnarray}
\mathcal{T}_{l_1 m_1 l_2 m_2 l_3 m_3 l_4 m_4}=\sum_{L M}(-1)^M \mathcal{G}^{l_1 l_2 L}_{m_1 m_2 -M} \mathcal{G}^{l_3 l_4 L}_{m_3 m_4 M} t^{l_1 l_2}_{l_3 l_4}(L).
\end{eqnarray}
If the extra-reduced trispectrum is independent of $L$ we can use equation~\eqref{UsefulIdentity} to write
\begin{eqnarray}
\mathcal{T}_{l_1 m_1 l_2 m_2 l_3 m_3 l_4 m_4}=\int d\Omega_{\hat{n}} Y_{l_1 m_1}(\hat{\mathbf{n}})Y_{l_2 m_2}(\hat{\mathbf{n}})Y_{l_3 m_3}(\hat{\mathbf{n}})Y_{l_4 m_4}(\hat{\mathbf{n}}) t^{l_1 l_2}_{l_3 l_4}.
\end{eqnarray}
where we drop the label $L$ from the extra-reduced trispectrum. The extra reduced trispectrum now has the following mode expansion
\begin{eqnarray}
t^{l_1 l_2}_{l_3 l_4}=N\Delta_{\Phi}^3\sum_{m}\alpha_{m}^{\mathcal{Q}}\int dr r^2 \mathcal{Q}_m^{l_1 l_2 l_3 l_4}(r),
\end{eqnarray}
where now we have
\begin{eqnarray}
\mathcal{Q}^{l_1 l_2 l_3 l_4}_{m} (r)=     q_p^{l_1}(r)q_r^{l_2}(r)q_s^{l_3}(r)q_{u}^{l_4}(r)
\end{eqnarray}
with
\begin{eqnarray}
q_p^{l}(r)&=&\frac{2}{\pi}\int dk q_p(k) \Delta_{l}(k)j_{l}(k r).
\end{eqnarray}
The mode decomposition is similar to that described in section V with $r_v = \rm{constant}$. In this case we use the following primordial decomposition
\begin{eqnarray}
(k_1 k_2 k_3 k_4)^2\mathcal{T}_{\Phi,0}(k_1,k_2,k_3,k_4)=\sum_{m}\alpha_{m}^{\mathcal{Q}}q_p(k_1)q_r(k_2)q_s(k_3)q_t(k_4).
\end{eqnarray}
Using this in the expression for the general estimator~\eqref{Estimator2}, which can be re-expressed as
\begin{eqnarray}
\mathcal{E}=\frac{12}{N_T}\sum_{l_i m_i}\mathcal{T}_{l_1 m_1 l_2 m_2 l_3 m_3 l_4 m_4} \left(a_{l_1 m_1}	a_{l_2 m_2}	a_{l_3 m_3}	a_{l_4 m_4}		\right)^{\rm{obs}}_c
\end{eqnarray}
with
\begin{eqnarray}
 \left(a_{l_1 m_1}	a_{l_2 m_2}	a_{l_3 m_3}	a_{l_4 m_4}		\right)^{\rm{obs}}_c&=& a_{l_1 m_1}^{\rm{obs}}	a_{l_2 m_2}^{\rm{obs}}	a_{l_3 m_3}^{\rm{obs}}	a_{l_4 m_4}^{\rm{obs}}	- \Big((-1)^{m_1}C_{l_1}\delta_{l_1 l_2}\delta_{m_1 -m_2}a^{\rm{obs}}_{l_3 m_3}a^{\rm{obs}}_{l_4 m_4}+\rm{5\,perms}\Big)\nonumber\\
&&+\Big((-1)^{m_1+m_3}\delta_{l_1 l_2}\delta_{m_1 -m_2} \delta_{l_3 l_4}\delta_{m_3 -m_4}C_{l_1}C_{l_3}	+\rm{2\,perms}\Big),
\end{eqnarray}
we find
\begin{eqnarray}
\mathcal{E}=\frac{12N \Delta_{\Phi}^3}{N_T}\sum_{m}\alpha_{m}^{\mathcal{Q}} \int d \hat{n}\int dr &r^2& \Big[  M_p(\hat{n},r) M_r(\hat{n},r) M_s(\hat{n},r) M_t(\hat{n},r)-\Big(M^{\rm{uc}}_{pr}(\hat{n},r) M_s(\hat{n},r)M_t(\hat{n},r)+\rm{5\,perms}\Big)   \nonumber\\
&&+ \Big(M^{\rm{uc}}_{pr}(\hat{n},r) M_{s t}^{\rm{uc}}(\hat{n},r)+\rm{2\,perms}\Big)   \Big],
\end{eqnarray}
where
\begin{eqnarray}
M_p(\hat{n},r)&=&\sum_{l m}\frac{a_{l m}Y_{l m}(\hat{n})}{C_l} q_p^{l}(r),\nonumber\\
M_{p r}^{\rm{uc}}(\hat{n},r)&=&\sum_{l m}\frac{Y^*_{l m}(\hat{n})Y_{l m}(\hat{n})}{C_l} q_p^{l}(r)q_r^{l}(r).
\end{eqnarray}
\par
We again can estimate the computational time needed to find this estimator. Since we now have only one line of sight integral and one integral over the sky $\sim \int d\hat{n}$ we use the prescription outline in Section VI to estimate the complexity conservatively as $\mathcal{O}(100)\times \mathcal{O}(l_{\rm{max}}^3)$.
\par
The implementation in the case of the late-time estimator for which the extra-reduced trispectrum is independent of $L$ can be found similarly. Since this estimator no longer requires a line of sight integral the complexity of the calculation can be estimated as  $\mathcal{O}(l_{\rm{max}}^3)$.

\bibliographystyle{hunsrt}
\bibliography{Trispectrum2}

\end{document}